%% file: acta_tekel.tex
\documentclass[twoside,fleqn]{ActaStyle}
\usepackage[colorlinks=true, pdfstartview=FitV, linkcolor=black,
            citecolor=black, urlcolor=black]{hyperref}
\usepackage{graphicx}
\usepackage{wrapfig}
\usepackage{times}
\usepackage{cite}
\usepackage{amsmath}
\usepackage{amssymb}
\usepackage{fancyhdr}
\usepackage{latexsym}
\usepackage{amsfonts}
\usepackage{amsthm}
\usepackage{simplewick}

\renewcommand{\subsectionmark}[1]{}
\renewcommand{\thesection}{\arabic{section}}
\renewcommand{\thesubsection}{\thesection.\arabic{subsection}}

\renewcommand{\thefigure}{\thesection.\arabic{figure}}
\renewcommand{\thetable}{\thesection.\arabic{table}}

\def\ep{\varepsilon}
\def\half{\frac{1}{2}}

\def\K{\mathcal{K}}
\def\C{\mathcal{C}}
\def\A{\mathcal{A}}

\def\O{\mathcal{O}}
\def\H{\mathcal{H}}
\def\D{\mathcal{D}}
\def\C{\mathcal{C}}
\def\K{\mathcal{K}}
\def\A{\mathcal{A}}
\def\I{\mathcal{I}}
\def\F{\mathcal{F}}
\def\bra#1{\left\langle#1\right|}
\def\ket#1{\left|#1\right\rangle}
\def \bear {\begin{eqnarray}}
\def \eear {\end{eqnarray}}
\def\lr#1{\left(#1\right)}
\def\slr#1{\left[#1\right]}

\def\trl#1{\textrm{Tr}\lr{#1}}
\def\trsl#1{\textrm{Tr}\slr{#1}}
\def\avg#1{\left\langle #1\right\rangle}

\def\CPn{\mathbb C P^n}
\def\CPFn{\mathbb C P^n_F}
\def\RS2{\mathbb R\times S^2_F}
\def\Rnn{\mathbb R ^{2n}}
\def\R{\mathbb R}

\def\bse{\begin{subequations}}
\def\ese{\end{subequations}}

\def \be  {\begin{equation}}
\def \ee  {\end{equation}}
\def \bex  {\begin{equation*}}
\def \eex  {\end{equation*}}
\def \bea {\begin{eqnarray}}
\def \bal {\begin{align}}
\def \eea {\end{align}}
\def\quater{\frac{1}{4}}
\def\tFo{\left._2 F\right._1}
\def\no{\nonumber\\}
\def\Tr{\textrm{Tr}}
\def\pd#1#2{\frac{\partial #1}{\partial #2}}

\def\reff{{r_{eff}}}
\def\geff{{g_{eff}}}

\def\podnadpis#1{\bigskip {\bf #1}}

\def \PRD {{Phys. Rev. D\ }}
\def \JHEP {{JHEP\ }}
\def \RMP {{Rev. Mod. Phys.\ }}

\allowdisplaybreaks
\def\authorheading{J. Tekel}
\def\shorttitle{Phase structure of fuzzy field theories and matrix models}
	
\begin{document}

\pagerange{369}{468}

\title{PHASE STRUCTURE OF FUZZY FIELD THEORIES\\ AND MULTITRACE MATRIX MODELS}

\author{Juraj Tekel\email{tekel@fmph.uniba.sk}}
{Department of Theoretical Physics, Faculty of Mathematics, Physics and Informatics, Comenius University, Bratislava, Slovakia}

\abstract{We review the interplay of fuzzy field theories and matrix models, with an emphasis on
the phase structure of fuzzy scalar field theories. We give a self-contained introduction to
these topics and give the details concerning the saddle point approach for the usual single
trace and multitrace matrix models. We then review the attempts to explain the phase
structure of the fuzzy field theory using a corresponding random matrix ensemble, showing
the strength and weaknesses of this approach. We conclude with a list of challenges one needs
to overcome and the most interesting open problems one can try to solve.}


\begin{minipage}{2.5cm}
\quad{\small {\sf KEYWORDS:}}
\end{minipage}
\begin{minipage}{10.0cm}
Multitrace matrix models, Noncommutative geometry, Fuzzy field theory, Phase diagram of fuzzy field theory
\end{minipage}

%%%%%%%%%%%%%%%%%%%%%%%%%%%%%%%%%%%%%%%%%%%%%%%%%%
%%%%%%%%%%%%%%%%%%%%%%%%%%%%%%%%%%%%%%%%%%%%%%%%%%

\setcounter{tocdepth}{3}
\tableofcontents

\setcounter{equation}{0} \setcounter{figure}{0} \setcounter{table}{0}\newpage
\section{Introduction}
\input{0_intro}

%%%%%%%%%%%%%%%%%%%%%%%%%%%%%%%%%%%%%%%%%%%%%%%%%%
%%%%%%%%%%%%%%%%%%%%%%%%%%%%%%%%%%%%%%%%%%%%%%%%%%
\renewcommand{\thefootnote}{\thesection.\arabic{footnote}}

\setcounter{equation}{0} \setcounter{figure}{0} \setcounter{table}{0} \setcounter{footnote}{0}\newpage
\section{Fuzzy scalar field theory}\label{sec1}

\subsection{Fuzzy and noncommutative spaces}
\input{1_1spaces}

\subsection{Scalar field theory on noncommutative spaces}\label{sec12}
\input{1_2fields}

\subsection{Phases of noncommutative scalar fields}\label{sec13}
\input{1_3phases}

%%%%%%%%%%%%%%%%%%%%%%%%%%%%%%%%%%%%%%%%%%%%%%%%%%
%%%%%%%%%%%%%%%%%%%%%%%%%%%%%%%%%%%%%%%%%%%%%%%%%%

\setcounter{equation}{0} \setcounter{figure}{0} \setcounter{table}{0} \setcounter{footnote}{0}\newpage
\section{Matrix models}\label{sec2}
\input{2_0intro}

\subsection{General aspects of the matrix models}
\input{2_1matrixmodels}

\subsection{Saddle point approximation (for symmetric quartic potential)}\label{sec22}
\input{2_21sadlepoint}
\input{2_22wigner}
\input{2_23quartic}
\input{2_24lesons}

\subsection{Saddle point approximation for asymmetric quartic potential}\label{sec23}
\input{2_3asymquartic}

\subsection{Saddle point approximation for multitrace matrix models}\label{sec24}
\input{2_4multitrace}

%%%%%%%%%%%%%%%%%%%%%%%%%%%%%%%%%%%%%%%%%%%%%%%%%%
%%%%%%%%%%%%%%%%%%%%%%%%%%%%%%%%%%%%%%%%%%%%%%%%%%

\setcounter{equation}{0} \setcounter{figure}{0} \setcounter{table}{0} \setcounter{footnote}{0}\newpage
\section{Matrix models of fuzzy scalar field theory}\label{sec3}
\input{3_intro}

\subsection{Fuzzy scalar field theory as a matrix model}\label{sec31}
\input{3_1matrix}

\subsection{Fuzzy scalar field theory as a multitrace matrix model}\label{sec32}
\input{3_2multitrace}

\subsection{Phase structure of the second moment multitrace model}\label{sec33simplphase}
\input{3_3phasesimple}

\subsection{Phase structure of the fourth moment multitrace model}
\input{3_4phasefull}

%%%%%%%%%%%%%%%%%%%%%%%%%%%%%%%%%%%%%%%%%%%%%%%%%%
%%%%%%%%%%%%%%%%%%%%%%%%%%%%%%%%%%%%%%%%%%%%%%%%%%

\setcounter{equation}{0} \setcounter{figure}{0} \setcounter{table}{0} \setcounter{footnote}{0}\newpage
\section{Conclusions and outlook}
\input{4_concl}

%%%%%%%%%%%%%%%%%%%%%%%%%%%%%%%%%%%%%%%%%%%%%%%%%%
%%%%%%%%%%%%%%%%%%%%%%%%%%%%%%%%%%%%%%%%%%%%%%%%%%
\bigskip
\bigskip

\begin{ack}
I would like to thank Denjoe O'Connor for useful discussions. This work was supported by the \emph{Alumni FMFI} foundation as a part of the \emph{N\'{a}vrat teoretikov} project.
\end{ack}
\addcontentsline{toc}{section}{Acknowledgement}

%\section{Some plots of the distributions}\label{ap1}
%\input{appendix1}

\setcounter{equation}{0} \setcounter{figure}{0} \setcounter{table}{0} \setcounter{footnote}{0}\newpage
\appendix
\section{Description of the numerical algorithm}\label{ap2}
\input{appendix2}

\newpage
\fancyhead[LO]{References}

\end{document}

%% file: 0_intro.tex
This text tells the story of the connection between fuzzy field theories and matrix models,
with an emphasis on the phase structure of both. It is an interesting connection, since both
these fields have a long standing place among the concepts in the theoretical physics. Their
connection provides an interesting bridge for ideas to migrate from one side to the other and
help to provide insight. We will investigate, how such migration can help to understand the
phase structure of fuzzy field theories by looking at the phase structure of a particular
matrix model.

In the rest of this introduction we briefly summarize the appearance of matrix models in the fuzzy field theory and give some very basic feeling for the role of noncommutative spaces and matrix models in physics.

Section \ref{sec1} then gives a more thorough and complete overview of the construction of the fuzzy and noncommutative spaces, of the scalar field theory defined on them and of the current understanding of their phase structure.

In the section \ref{sec2} we review the matrix models, with the basic notions and the technique of the saddle point approximation. We also review the multitrace matrix models, a more complicated matrix models which will turn out to be essential in the last section. We elaborate on several examples of matrix models, so that the reader can get good grip on the basic techniques, which will be used as well as hopefully better understand the concepts.

In the last section \ref{sec3} we first review and describe more thoroughly how the matrix models arise in the study of the fuzzy field theory. We then put the machinery of matrix models to work in a study of two different models which approximate the fuzzy field theory on the fuzzy sphere, in an effort to explain analytically the phase structure obtained by numerical computations.

\subsection{Fuzzy field theory and random matrices}

Both fuzzy spaces and random matrices have a firm place in modern theoretical physics. They arise as object of interest in many areas or are often used as a very useful computational tool. Since fields on fuzzy spaces have a finite number of modes, observables are naturally defined as matrices and there is a straightforward connection to the random matrix theory. Computing expectation values of observables in the fuzzy field theory and in the random matrix theory with certain probability measure is essentially the same calculation. This was first observed in \cite{stein05_1,stein05_2,stein04_1}.

Indeed, the fuzzy fields very naturally define a wide class of new random matrix ensembles, as was pointed out in \cite{our1}. The new feature is the presence of a kinetic term in the measure. This term couples the random matrix to a set of fixed external matrices, which are related to the underlying fuzzy space of the corresponding field theory. The new term also introduces new observables of interest, given by various commutators of the external matrices and the random matrix, and their products.

The computational problem is clear. The new term in the measure does not reduce upon diagonalization to an expression involving only the eigenvalues of the random matrix and the integral over the angular degrees of freedom, which has to be done in order to compute the averages, is not trivial anymore. Therefore the standard approach of saddle point approximation which is used to obtain the results in the limit of large matrices is not usable. Some further investigation into the structure of angular integral revealed that the fuzzy field theory is described by a particular multitrace model \cite{ocon}. The generalization of the saddle point approximation to multitrace models is known, so we have hope to analyze the fuzzy field theory via this matrix model.

But to be able to do that, we have a long way to cover.

\subsection{Non-commutative spaces in physics}

The idea of the non-commutative geometry arises in the correspondence of commutative $C^*$ algebras and differentiable manifolds. Every manifold comes with a naturally defined algebra of functions and every $C^*$ algebra is an algebra of functions on some manifold. One then defines non-commutative spaces as spaces that correspond to a non-commutative algebra \cite{connes1,landi}. One introduces a spectral triple of a $C^*$ algebra $\A$, a Hilbert space $\H$ on which this algebra can be realized as bounded operators and a special Dirac operator $\D$, which will characterize the geometry. Using these three ingredients, it is possible to define differential calculus on a manifold.

More specifically, in a finite dimensional case one arrives at a notion of a fuzzy space, which will be central for this presentation. Here, the Hilbert space $\H$ is finite dimensional and algebra $\A$ can be realized as an algebra of ${N\times N}$ matrices. This can be thought of as a generalization of a well-known notion from quantum mechanics. With large number of quanta the quantum theory is well approximated by the classical theory, with the functions on the classical phase space representing the linear Hermitian operators. The fuzzy space defined by $(\A,\H,\Delta_N)$, with $\Delta_N$ the matrix Laplacian, are finite state-approximation to the classical phase space manifold. The finite dimensionality of $\H$ amounts for compactness of the classical version of the manifold.

The original motivation for considering non-commutative manifolds in physics dates back to the early days of quantum field theories. It was suggested by Heisenberg and later formalized by Snyder \cite{snyder} that the divergences which plague the qft's can be regularized by the space-time noncommutativity. Opposing to other methods, non-commutative space-time keeps its Lorentz invariance. However, since renormalization proved to be effective in providing accurate numerical results, this idea was abandoned.

More recently, it has been shown that putting the quantum theory and gravity together introduces some kind of nontrivial short distance structure to the theory \cite{doplicher}. Noncommutative spaces, as we will see in the next section, yield such structure without breaking the symmetries of the space.

In the spirit of the original motivation for non-commutative spaces, there has been work done to formulate the standard model of particle physics solely in the terms of the spectral triple and the framework of the non-commutative geometry \cite{ncsm1,ncsm2,ncsm3}, for a recent review see \cite{walter}. There have also been attempts to formulate the theory of gravity purely in the terms of the spectral triple \cite{grav1,grav2}.

Fuzzy spaces arise also in the description of the Quantum Hall Effect, i.e. dynamics of charged particles in the presence of magnetic field. Originally, the problem was considered in the plane, but fuzzy spaces allow for a natural generalization to higher dimensional curved spaces. When the field is strong enough, particles are confined to the lowest Landau level and the projections of the Hermitian operators on this one level no longer commute, with the magnitude of the magnetic field being related to the noncommutativity parameter \cite{qhe1}. The lowest Landau level states form the Hilbert space $\H$ and considering particle dynamics on a general manifold reproduces the fuzzy version, with the corresponding Hilbert space. We can explore the geometry of the fuzzy space considering the behavior of the electron liquid and even study the case of non-Abelian background field \cite{qhe2,qhe3,fuzzygauge}.

Something similar happens when considering D-branes in string theory and the effective dynamics of opened strings is described by the non-commutative gauge theory and the D-brane worldvolume becomes a non-commutative space \cite{ppsug,witten}, see \cite{noncom1,noncom2} for a review. Finally, fuzzy spaces arise in the $M$-theory and related matrix formulations of the string theory as brane solutions \cite{branekabat,branenair}, see \cite{taylor} for a review of $M$-theory and \cite{abedis} for a review of fuzzy spaces as brane solutions.

And last, but not least, fuzzy spaces are useful as a regularization method for lattice calculations, their main advantage being the preservation of the symmetries of the underlying space. See \cite{fuzzy3,num14panero1} for a review.

\subsection{Random matrices in physics and matrix models}

Random matrices, as the name suggests, are matrices with random entries given by some overall probability measure, usually with some symmetry. Computing the averages in this theory then amounts to computing matrix integrals with this probability weight.

For a very thorough overview of properties and applications of random matrices, both in physics and beyond, see \cite{revmat}.

Historically, the random matrices arose in the work of Wigner \cite{wigner1,wigner2}, in an attempt to describe the energy levels of heavy nuclei. These are too dense to be described individually and Wigner suggested an statistical approach, where he showed that the level distribution is in a good agreement with the the eigenvalue distribution of a random matrix. This success lies in the universality of the statistical properties of the random matrices. See \cite{metha} for overview.

On a very different front, t'Hooft suggested an $1/N_c$ expansion of QCD \cite{thooft}. The gluons of the theory are $N_c\times N_c$ matrices, here $N_c$ is the number of colors, and since the fluctuations of these need to be integrated out, this is a theory of random matrices. t'Hooft showed, that the $N_c\to\infty$ limit simplifies the theory greatly, since only planar diagrams survive, and corrections of higher orders in $1/N_c$ correspond to topologically more complicated contributions.

This is related to the possibility of using the random matrix theory to discretize two dimensional surfaces. Diagrams are generated by the matrix integrals and when computing large matrix limit of these integrals, we can count the possible discretizations, incorporate more complicated structures or obtain the continuum limit by a suitable rescaling of the parameters of the theory. This is then important in the considerations of conformal field theories and extraction of critical exponents. Similarly in string theory, the world sheet of the string is two dimensional and considering a rescaling of parameters of the theory, one can generate two dimensional surfaces of all genera, i.e. string interactions. See \cite{gaume,matrix2,matrix3,string}.

Random matrix theory has a wide application in the condensed matter physics, again mostly due to the universality properties. The large number of electrons or other constituents that come into play corresponds to the limit of a large dimension of the matrix. One can either study thermodynamical properties of some closed system, then the random matrix is the Hamiltonian $H$. Or one can study transport properties, where the matrix of interest is the scattering matrix $S$. See \cite{revmat1,revmat2} for review. Since level repulsion is a characteristic for both random matrices and chaotic systems, it is believed that random matrix theory will lead to the theory of quantum chaos \cite{revmat2,revmat3,chaos}.

From a pure mathematical point of view random matrices and their statistical properties have applications in fields as combinatorics, graph theory, theory of knots and many more. Let us mention one quite remarkable and unexpected appearance. Random matrices have a clear connection to Riemann conjecture. Namely, the two-point correlation function of the zeros of the zeta function $\zeta(z)$ on the critical line seems to be the same as the two-point correlation function of the eigenvalues of a random Hermitian matrix from a simple Gaussian ensemble \cite{rieman} and also share other statistical properties with random matrix ensembles \cite{rieman2}. However, there is no true understanding of the reason behind these observations.

%% file: 1_1spaces.tex
In this section, we will explain what the fuzzy and noncommutative spaces are, how they are constructed and what are their basic properties. The notions are rather abstract, so before we proceed with the full treatment, let us give the reader a taste of what is coming and what it means.

\subsubsection{An appetizer - The Fuzzy Sphere}

We will start with giving a flavor of what the idea behind the non-commutative spaces is and what the main features are, with emphasis on the non-commutative version of the two sphere $S^2_F$. More details will follow in later sections.

Every manifold comes with a naturally defined associative algebra of functions with point-wise multiplication. This algebra is generated by the coordinate functions of the manifold and is from the definition commutative. As it turns out, this algebra contains all the information about the original manifold and we can describe geometry of the manifold purely in terms of the algebra. Also, every commutative algebra is an algebra of functions on some manifold. Therefore, what we get is
\[\textrm{commutative algebras }\longleftrightarrow\textrm{ differentiable manifolds .}\]
See \cite{ncgeo} for details. A natural question to ask is whether there is a similar expression for non-commutative algebras, or
\[\textrm{non-commutative algebras }\longleftrightarrow\textrm{ ???}\]
Quite obvious answer is no, there is no space to put on the other side of the expression. Coordinates on all the manifolds commute and that is the end of the story. So, as is often the case, we define new objects, called non-commutative manifolds that are going to fit on the right hand side. Namely we look how aspects of the regular commutative manifolds are encoded into their corresponding algebras and we call the non-commutative manifold object that would be encoded in the same way in a non-commutative algebra.

This is going to introduce noncommutativity among the coordinates. This notion is not completely new, as one recalls the commutation relations of the quantum mechanics $[x^i,p_j]=i\hbar\delta^i_j$. In classical physics, the phase space of the theory was a regular manifold. However in quantum theory we introduce noncommutativity between (some) of the coordinates and therefore the phase space of the theory becomes non-commutative. One of the most fundamental consequences of the commutation relations is the uncertainty principle. The exact position and momentum of the particle cannot be measured and therefore we cannot specify a single particular point of the phase space. Similarly, if there is noncommutativity between the coordinates, there is a corresponding uncertainty principle in measurement of coordinates. The notion of a space-time point stops to make sense, since we cannot exactly say, where we are. This is the motivation behind the name of the fuzzy space.

In practice, we often `deform' a commutative space into its non-commutative analogue. In this way we get non-commutative spaces that give a desired commutative limit and also writing such a space from scratch is very difficult. An example of such deformation is already mentioned phase space of quantum mechanics or the more general case of non-commutative flat space $\mathbb R^2_\theta$, given by
\be\label{11qm}
[x_i,x_j]\equiv x_i x_j-x_j x_i=i\theta_{ij},
\ee
for some constant, anti-symmetric tensor $\theta^{ij}$. To illustrate the procedure of deformation better let us consider a different example and show how this works for a two-sphere \cite{sf21,sf22,sf23}.

The regular two sphere is defined as the set of points with a given distance from the origin, i.e. $\sum_{i=1}^3x_i^2=R^2$. This comes with an understood condition on commutativity of the coordinates $x_ix_j-x_jx_i=0$. Coordinate functions constrained in this way generate the algebra of all the functions on the sphere.\footnote{Note that this is technically not the easiest way to do so. It is easier to introduce only two coordinates $\theta,\varphi$ on the sphere and define the algebra of functions not by the generators, but by the basis, e.g. the spherical harmonics. However the two sphere defined in our way is easier deformed into the non-commutative analogue.}

Now we define the fuzzy two sphere by the coordinates $\hat x_i$, which obey the following conditions
\be
\sum_{i=1}^3\hat x_i^2=\rho^2 \ \ \ , \ \ \ \hat x_i\hat x_j-\hat x_j\hat x_i=i\theta\ep_{ijk}\hat x_k,
\label{fuzs2}
\ee
where $\rho,\theta$ are parameters describing the fuzzy sphere, in a similar way as $R$ did describe the regular sphere. The radius of the original sphere was encoded in the sum of the squares of the coordinates, so we will call $\rho$ the 'radius' of the non-commutative sphere. We see, that such $\hat x$'s are achieved by a spin-$j$ representation of the $SU(2)$. If we chose
\be
\hat x_i=\frac{2r}{\sqrt{(2j+1)^2-1}}L_i \ \ \ , \ \ \ [L_i,L_j]=i\ep_{ijk}L_k \ \ \ , \ \ \ \sum_{i=1}^3 L_i^2=j(j+1) ,
\label{2.2}
\ee
where $L_i$'s are the generators of $SU(2)$, we get
\be\label{11tehta}
\sum_{i=1}^3\hat x_i^2=\frac{4\rho^2}{N^2-1}\lr{\frac{N-1}{2}}\lr{\frac{N-1}{2}+1}=\rho^2 \ \ \ , \ \ \ \theta=\frac{2r}{\sqrt{N^2-1}},
\ee
with $N=2j+1$ the dimension of the representation. Matrices $\hat x_i$ become coordinates on the non-commutative sphere. Note, that the limit $N\to\infty$ removes the noncommutativity, since $\theta\to0$, and we recover a regular sphere with radius $r$. This explains a rather strange choice of parametrization in (\ref{2.2}). Also note, that this way we got a series of spaces, one for each $j$ (or $N$). The important fact is that the coordinates still do have the $SU(2)$ symmetry and therefore it makes sense to talk about this object as spherically symmetric. This explains the particular choice of deformation in (\ref{fuzs2}).

The non-commutative analogue of the derivative is the $L$-commutator, since it captures the change under a small translation, which is rotation in the case of the sphere. The integral of a function becomes a trace, since it is a scalar product on the space of matrices. As we sill see in the next section, both of these have the correct commutative limits.

Spherical harmonic functions $Y_l^m$ form a basis of the algebra of functions on the regular sphere. These are labeled by $l=0,1,2,\ldots$ and by $m=-l,-l+1,\ldots,l-1,l$ and this basis is infinite. If we truncate this algebra, namely take the following set of functions
\be
Y_l^m \ \ \ , \ \ \ m=-l,-l+1,\ldots,l-1,l \ \ \ \& \ \ \ l=0,1,\ldots,N-1 \ ,
\ee
we recover a different algebra. This is obviously not the algebra of functions on the regular sphere and also to make this algebra closed, it can be seen that we need to introduce some non-trivial commutation rules \cite{sf22}. And one can check, that the $N^2$ independent matrices generated by (\ref{2.2}) are in one-to-one correspondence with this truncated set of spherical harmonics. This means, that in the limit of large $N$ the algebra we recover is truly the algebra of functions on regular commutative sphere $S^2$.

Here we can see in a different way why the fuzzy-ness introduces short distance structure. $l$ measures the momentum of the mode and cutting off the modes we have introduced the highest possible momentum. This in turn introduces the shortest possible distance to measure.

%%%%%%%%%%%%%%%%%%%%%%%%%%%%%%%%%%%%%%%%%%%%%%%%%%%%%%%%%%%%%%%%%%%%%%%%%%%%%%%%%%%%%%%%%%%%%%%%%%%%%%%%%%%%%%%%
%%%%%%%%%%%%%%%%%%%%%%%%%%%%%%%%%%%%%%%%%%%%%%%%%%%%%%%%%%%%%%%%%%%%%%%%%%%%%%%%%%%%%%%%%%%%%%%%%%%%%%%%%%%%%%%%
\subsubsection{Construction of fuzzy $\CPn$}\label{secCPn}
%%%%%%%%%%%%%%%%%%%%%%%%%%%%%%%%%%%%%%%%%%%%%%%%%%%%%%%%%%%%%%%%%%%%%%%%%%%%%%%%%%%%%%%%%%%%%%%%%%%%%%%%%%%%%%%%
%%%%%%%%%%%%%%%%%%%%%%%%%%%%%%%%%%%%%%%%%%%%%%%%%%%%%%%%%%%%%%%%%%%%%%%%%%%%%%%%%%%%%%%%%%%%%%%%%%%%%%%%%%%%%%%%

Manifolds with a symplectic structure admit construction of a fuzzy analog. In general, co-adjoint orbit of a compact semisimple Lie group enjoys being a symplectic manifold and such spaces can be quantized in a very natural way, described towards at the end of this section. The Hilbert space which arises in this process is a unitary irreducible representation of this group and this is in a very intimate relationship to quantization of the phase space of classical mechanics to a quantum Hilbert space mentioned in the previous section. We will now describe this procedure in some detail for $\CPn$.

Construction of fuzzy spaces which are not derived from a co-adjoint orbit of a semisimple Lie group, for example higher dimensional spheres, is more involved. One starts from a larger space which is a co-adjoint orbit and the irrelevant factors are projected out \cite{abe1,s4}

In this section, we will present the construction of the fuzzy version of the complex projective spaces, following \cite{cpnnair1,cpnnair2}, some of the original references being \cite{blowcpn,cpnoriginal}. To do this, we will consider $\CPn$ as $SU(n+1)/U(n)$. This follows from the standard definition of $\CPn$ as lines in $\mathbb C^{n+1}$ which go through the origin, i.e. the identification $z\sim \lambda z,z\in\mathbb C ^{n+1}$ for some $\lambda\in\mathbb C$. Since there is a natural action of $g\in SU(n+1)$ on $z$ by $z'=gz$, we get the advertised relation for $SU(n+1)/U(n)$. This means, that the functions on $\CPn$ are those functions on $SU(n+1)$, which are $U(n)$ invariant. This allows for construction of the basis of functions on $\CPn$ in terms of the Wigner $\mathcal D$-functions of $SU(n+1)$. We will do this and then show how these are in one-to-one correspondence with $N\times N$ matrices, which are functions on the fuzzy $\CPn$.
%We will also construct the Hilbert space $\H_N$ using the $\D$-functions.

\podnadpis{Commutative $\CPn$ and the deformation}

Consider the totally symmetric representation of $SU(n+1)$ of rank $l$. The dimension of this representation is
\begin{eqnarray}
	N_l=\frac{(n+l)!}{n!l!}.
	\label{2.5}
\end{eqnarray}

Now consider the following functions on $\CPn$
\begin{eqnarray}
	\Psi^l_m(g)&=&\sqrt {N_l} \D^n_{m,-l}(g),\\
	\D^l_{m,-l}(g)&=&\bra{l,m} \hat g \ket{l,-l},\label{2.7}
\end{eqnarray}
where $g$ is an element of $SU(n+1)$, $\hat g$ is its corresponding operator in this representation, $\ket{k,-k}$ denotes the lowest weight state and $\ket{k,m},m=1,2,\ldots,N_k$ are the states in the representation. The fact that we consider only the lowest weight state $\ket{k,-k}$ ensures, that these functions on $SU(n+1)$ are correctly invariant under $U(n)$. The $\D$-functions can be constructed explicitly using the coherent state representation and the local complex coordinates for $\CPn$ \cite{coherent}. They are completely symmetric holomorphic functions of order $k$, but this will not be needed for what follows.

The basis for functions on regular $\CPn$ is then given by the union of all the functions (\ref{2.7}) for $l=0,1,2,\ldots$. These functions are eigenfunctions of the quadratic Casimir operator with eigenvalues and multiplicities given by
\be\label{11CPnspectrum}
C_2\Psi^l_m(g)=l(l+n)\ ,\ \textrm{dim}(n,l)=\frac{\big((k+n-1)!\big)^2(2l + n)}{\lr{l!}^2\lr{(n -1)!}^2n}\sim\frac{l^{2n-1}}{(n-1)!^2n}
\ee

Now consider space of $N_L\times N_L$ matrices, where $N_L$ is given by (\ref{2.5}) for $k=L$, where $L$ is the cut-off on the number of the modes. These are going to form a basis of functions on the fuzzy $\CPn_F$. We need to show, that as $L\to\infty$ these are in one to one correspondence with the functions on regular $\CPn$, that the matrix product becomes the usual commutative product in this limit and we need to define object like derivatives and integrals.

\podnadpis{Symbols and large $N$-limit of matrices}

We associate an $N_L\times N_L$ matrix $A$ with a function $A(g)$ on the classical $\CPn$ by
\be
A(g)=\sum_{mm'}\D^n_{m,-n}(g)A_{mm'}\D^{*n}_{m',-n}(g).
\label{2.8}
\ee
We call this function a symbol corresponding to $A$ and this object is going to be essential and in the large $N_L$, or large $L$, limit, the matrix $A$ will tend to its symbol $A(g)$. We can also define a product between the functions on $\CPn$ (symbols), which we will call the star product and denote it $\star$, by
\be
A(g)\star B(g)\equiv (AB)(g).
\ee
This means that the star product of the two symbols is given by a symbol of the product of the two matrices. This product is not commutative and deforms the original product on $\CPn$ by contributions that are suppressed by powers of $1/N_L$, see \cite{star1,star3}. The symbol corresponding to the commutator of two matrices becomes the Poisson bracket on $\CPn$ \cite{star2}
\be
([A,B])(g)=\frac{i}{L}\{A,B\}+\O\lr{1/L^2}.
\ee
This reflects the standard procedure of quantum mechanics, where the Poisson brackets of the observables get replaced by commutators of the corresponding operators. The trace of the matrix becomes an integral
\be
\trl{A}=\sum_i A_{ii}=N\int dg\,D^k_{m,-n}A_{mm'}\D^{*n}_{m',-n}(g)=N\int dg\,A(g),
\ee
where $dg$ is the Haar measure on $SU(n+1)$ and we have used the orthogonality property of the Wigner $\D$ functions.

\podnadpis{Matrix-function correspondence}

The last step is to show how the matrices correspond to the functions on commutative $\CPn$. Functions on fuzzy $\CPn_F$ are $N_L\times N_L$ matrices and we can identify the coordinate matrices. These we define as
\be
X_A=-\frac{R}{\sqrt{C_2(L)}} T_A,
\label{2.12}
\ee
where $T_A$ are the generators of $SU(n+1)$ in the symmetric rank $k$ representation and $C_2(k)$ is the value of quadratic Casimir in this representation and $R$ is the radius of the $\CPn$. In the large $L$ limit, the symbol of $X_A$ is $S_{A\,n^2+2n}(g)\equiv 2\trl{g^T t_A g^* t_{n^2+2n}}$. Here, $t_A$'s are the generators of the $SU(n+1)$ that was used to define the $\CPn$ and $t_{n^2+2n}$ is the generator of the $U(1)$ direction in the $U(n)$ subgroup of $SU(n+1)$. Moreover, for any matrix function of the coordinate matrices $F(X_A)$, the corresponding symbol is given given as a function of the coordinate functions $F(S_{A\,n^2+2n}(g))$. This shows, that in the large $L$ limit, the matrices and the functions coincide.

Finally, considering a general matrix $M$ generated by the coordinate matrices (\ref{2.12}), defining the derivative operator as
\be
-i D_A M \equiv [T_A,M]\approx-\frac{i}{L}\frac{n L}{\sqrt{2n(n+1)}}\{S_{A\,n^2+2n},M\},
\ee
one can check that $D_A$'s follow the appropriate $SU(n+1)$ algebra conditions. Also, in explicit realization in terms of complex coordinates on $\CPn$ these reduce to known expressions for derivatives.

\podnadpis{The fuzzy sphere as a special case of $\mathbb C P^1_F$}

The sphere is $\mathbb CP^1$ and the explicit formulae should also illustrate rather abstract treatment of the previous section.

We will deal with representations of $SU(2)$, which are given by the standard angular momentum theory. The spin-$j$ representation is given by the maximal angular momentum $L=j/2$ and $N=2j+1=L+1$. Generators of the group are angular momentum matrices $L_i$ and relation (\ref{2.12}) becomes (\ref{2.2}). The number of spherical harmonics, which form a basis on regular sphere, up to a certain angular momentum is $\sum_{j=0}^L(2j+1)=(L+1)^2=N^2$, which is the number of independent $N\times N$ matrices.

The basis of matrices is given by the symmetrized products of the coordinate matrices $X_i$, i.e. $id,X_i,X_{(i}X_{j)},\ldots$. The same is true for the spherical harmonics, which are formed by the products of $S_{i3}$ with contractions removed, functions with up to $k$ factors of $S_{i3}$ correspond to spherical harmonics of angular momentum up to $k$. There is thus one-to-one correspondence between the matrices and spherical harmonics and as $L\to\infty$, the matrix algebra corresponding to fuzzy sphere becomes the algebra of functions on regular sphere.

\subsubsection{Limits of $\mathbb C P^n_F\to \mathbb R^{2n}$}\label{cpntorn}

After the construction of the fuzzy $\CPn_F$, one is left with several possibilities of taking the large $N$ limit. Let us discuss those in some more detail.

\podnadpis{Commutative $\CPn$}

Most of the previous section was devoted to proving that when one takes the $N_L\to\infty$ limit with a fixed radius $R$, we recover the commutative version of $\CPn$. Matrices become functions, etc. However, let us stress here that this correspondence is geometrical. If we define some structure on top of the geometry, like field theory for example, we are not guaranteed to obtain the commutative counterpart. And as we will see, this is, as a rule, not the case.

\podnadpis{Non-commutative $\Rnn_\theta$}

Here, it is useful to consider the explicit realization of the fuzzy $\CPn_F$ coordinates as
\be
x_i=\frac{R}{\sqrt{\frac{n}{2(n+1)}L^2+\frac{n}{2}L}}L_i \ \ \ , \ \ \ [x_i,x_j]=i\frac{R}{\sqrt{\frac{n}{2(n+1)}L^2+\frac{n}{2}L}}f_{ijk}x_k\ \ \ ,
\ee
where $f_{ijk}$ are the structure constants of $SU(n+1)$, $L_i$ are the generators of $SU(n+1)$ in the corresponding representation and $R$ is the radius of the $\CPn$. If we now scale the radius as $R^2=L\theta\frac{n}{n+1}$, we blow up the $\CPn$ around one point. What is left is a non-commutative $\Rnn_\theta$. We will show this for the case of fuzzy sphere, more general details can be found in \cite{blowcpn}. For $n=1$ we find
\be
[x_1,x_2]=i\frac{2\sqrt{\frac{N\theta}{2}}}{\sqrt{N^2-1}}\sqrt{\frac{N\theta}{2}-x_1-x_2}\to i\theta,
\ee
which means that $x_1,x_2$ are coordinates on the non-commutative $\mathbb R^2_\theta$.

\podnadpis{Commutative $\Rnn$}

If we scale radius with a power of $L$ smaller than $\half$, i.e. $R^2=L^{1-\ep},0>\ep>1$, the commutators of the left-over coordinates vanish and we obtain the commutative $\Rnn$.

\subsubsection{Quantization of Poisson manifolds}

To conclude, let us note that the presented approach was an explicit realization of a more general concept of quantizing a Poisson manifold \cite{stenposi,pois1,pois2}. We will not worry too much about the proper definitions and existence of what we are about to work with and will simply assume that the objects can be defined and behave nicely.

We start with a manifold equipped with an anti-symmetric bracket $\{.,.\}$ satisfying the Jacobi identity. Quantization map is then defined as a map between the algebra of functions on the manifold and a matrix algebra $\A(\mathbb C)$,
\be
\I\ : \ f(x)\to F\in \A.
\ee
Algebra $\A$ is generated by the images of the coordinate functions $X^\mu=\I(x^\mu)$. To be a good quantization, $\I$ has to satisfy
\bear
\I(fg)-\I(f)\I(g)&\to&0\nonumber,\\
\frac{1}{\theta}\big[\I(i\{f,g\})-[\I(f),\I(g)]\big]&\to&0,\nonumber\\
\textrm{as }\theta&\to&0,
\eear
where we have defined a parameter $\theta$ by $\{x_\mu,x_\nu\}=\theta\,\theta_0^{\mu\nu}(x)$. The star product is then defined as a pullback of the matrix product using $\I$,
\be
f\star g\equiv \I^{-1}\big(\I(f)\I(g)\big).
\ee
The integral of a function over the manifold then tends to the trace of the corresponding matrix
\be
\int d\Omega\, f\to \trl{\I(f)},
\ee
where $d\Omega$ is the volume defined by the Poisson structure $\theta\,\theta_0^{\mu\nu}$. Finally, when we consider the manifold to be embedded in some $\mathbb R^d$ by coordinate functions $x^a$, images of these define the coordinate matrices $X^a\equiv \I(x^a)$. These matrices then follow the same embedding conditions as the original coordinate functions did.

When the algebra $\A$ is finite dimensional, we refer to the resulting non-commutative space as a fuzzy space.

Finally, note, that (\ref{2.8}) is a particular choice of the quantization map $\I$.

%% file: 1_2fields.tex
\subsubsection{Formulation of the scalar field theory}

As in the case of the regular space, the scalar field on a fuzzy space is a power series in the coordinate functions and thus an element of the algebra discussed in the previous section. It is itself an ${N\times N}$ matrix, which can be expressed as
\be
M=\sum_{l,m} c_{l,m}T_{l,m} \ \ \ , \ \ \ l=0,1,\ldots,L \ \ \ , \ \ \ m=1,2,\ldots,\dim(n,l)\ \ \ ,
\label{expansion}
\ee
where $T_{l,m}$ are the polarization tensors, the eigenfunctions of the laplacian and the analogue of the spherical harmonic functions
\bal
\sum_{i=1}^{(n+1)^2-1}[L_i,[L_i,T_{l,m}]]=&l(l+n)T_{l,m}\ ,\nonumber\\
\trl{T_{l,m} T_{l',m'}}=&\delta_{ll'}\delta_{mm'}\ ,\nonumber\\
\sum_m\left( T_{l,m}T_{l',m'}\right)_{ij}=&\frac{\dim(n,l)}{N}\delta_{ll'}\delta_{ij}\ ,
\end{align}
where $L_i$ the $SU(n+1)$ generators in the $N$ dimensional representation. The field theory is defined by the action. In what follows we work with the Euclidean signature.

The field theory on the fuzzy space is then defined by ''starring" all the products in the commutative action. The free field action is then
\be
S_0[\phi]=\int dx \lr{\half(\partial_i \phi)\star(\partial_i \phi)+\half r \phi\star\phi}.
\ee
The symbol $r$ for the square of the mass is not standard in the field theory literature, but is very common in the matrix context and we will use it throughout the text.

By construction, this action has the desired commutative limit. As we have seen in the previous section, in the case of fuzzy spaces, when we represent the field $\phi$ with a $N\times N$ matrix $M$, the integral becomes a trace and $\partial_i$ becomes a commutator with $L_i$ and the star product is just the regular matrix product. Therefore, this action becomes
\bear\label{freeaction}
S_0&=&\frac{V_n}{N}\trsl{-\half\frac{1}{R^2}[L_i,M][L_i,M]+\half r M^2}=\nonumber\\&=&\frac{V_n}{N}\slr{\half \frac{1}{R^2}\trl{M[L_i,[L_i,M]]}+\half r \trl{ M^2}}\ ,
\eear
with $V_n$ the volume of $\CPn$.

We will denote the kinetic part of the action by $\K$ and for the standard kinetic term we get
\be\label{12standkint}
\K M=[L_i,[L_i,M]]\ .
\ee
Later, we will work with more general kinetic terms $\K$. We will assume that $\K$ vanishes for the identity mode $l=0$ and that the polarization tensors are eigenfunctions of $\K$ with
\be \K T_{l,m}=K(l)T_{l,m}\ .\ee
Clearly for the standard kinetic term $K(l)=l(l+n)$.

We will rescale the fields to absorb the $1/N$ and the volume factors. Using the expansion (\ref{expansion}) the free field action becomes
\be
S_0=\sum_{l,m}\half\big(l(l+n)+r\big)(c_{l,m})^2
\ee
The correlator of the two components of the field is
\be
\contraction{}{c}{_A^l }{c}
c_{l,m} c_{l,m'}=\frac{1}{l(l+n)+r}\delta_{ll'}\delta_{mm'}
\ee
which is an expression analogous to the usual propagator $(p^2-m^2)^{-1}$. Using this, one can compute the free field correlation functions. The full interacting field action is then given as $S=S_0+S_I$, with
\be
S_I[M]=\sum g_k \lr{\frac{N}{V_n}}^{1-\frac{k}{2}}\trl{M^k}=\sum \tilde g_k\trl{M^k}\ .
\ee

The field theory is defined by the functional correlations
\be
\left<F\right>=\frac{\int dM\ e^{-S} F[M]}{\int dM\ e^{-S}}\ ,
\label{functional}
\ee
or by the generating function for the correlators
\be
Z(J)=\frac{1}{\int dM\ e^{-S}}\int dM\ e^{-S+Tr(JM)}\ ,
\ee
or in any other usual way. One then derives the Feynman rules and Feynman diagrams \cite{fuzzy2}.

\subsubsection{The UV/IR mixing}

The key property of the noncommutative field theories for our purposes will be the UV/IR mixing \cite{uvir2}. It arises due to the nonlocality of the theory and introduces an interplay of the long and short distant processes. There is no longer a clear separation of scales and the theory becomes nonrenormalizable. Most surprisingly, this phenomenon persists in the commutative limit and the naive ''starred'' theory has a very different limit than the theory we did start with.

Let us present a short discussion of the appearance of the UV/IR mixing in the case of the noncommutative euclidean space $\mathbb R_\theta^{2n}$ \cite{noncom2}. We will get to the case of the fuzzy spaces shortly.

The action of the quartic scalar field on $\mathbb R_\theta^{2n}$ is given by
\be
S(\phi)=\int d^{2n}x \lr{\half (\partial \phi)^2+\half r \phi^2+g \phi\star\phi\star\phi\star\phi}\ ,
\ee
where we have taken into account the fact, that for the quadratic expressions $\int dx\,f\star g=\int dx\,fg$. The propagators of the theory thus do not change and the vertex contribution acquires a phase
\be
V(p_1,\ldots,p_4)=\prod_{a<b} e^{-\frac{i}{2}\theta^{ij} (p_a)_i(p_b)_j}\ ,
\ee
which is invariant only under cyclic permutations of the momenta. This leads to the planar self-energy diagram contribution
\be
\frac{1}{3}\int \frac{d^{2n}k}{(2\pi)^{2n}}\frac{1}{k^2+r}
\ee
and the nonplanar self-energy diagram contribution
\be
\frac{1}{6}\int \frac{d^{2n}k}{(2\pi)^{2n}}\frac{e^{i\theta^{ij} (p)_i(k)_j }}{k^2+r}\ .
\ee
The planar diagram is the same as for the commutative theory and needs to be regularized by introduction of a cutoff $\Lambda$. This introduces an effective regulator for the nonplanar diagram $\Lambda_{eff}=1/(\Lambda^{-2}-\theta^{ij} (p)_i(p)_j)$. This is finite in the large $\Lambda$ limit and the nonplanar diagram is thus regulated by the noncommutativity. But then, the divergence reapers in the $p\to 0$ limit. Moreover, if we try to use such formulas for construction of the effective action, the corrected propagator becomes a complicated nonlocal expression which cannot be considered as a mass renormalization.

As we have seen in the section \ref{cpntorn}, the noncommutative euclidean space can be obtained as a particular limit of the fuzzy $\CPFn$. It is not surprising that the UV/IR mixing will exhibit itself also here \cite{uvir1,uvirkuk}. There are no divergences, as all the integrals are finite, but the UV/IR mixing leads to a finite difference between the planar and the non-planar diagrams of the theory. In the case of the fuzzy sphere, this translates into a the finite, non-vanishing difference between the planar and the non-planar contributions to the two point functions at one loop, which is equal to
\be
\frac{2}{N^2-1}\sum_{l=0}^{N-1}\frac{l(l+1)(2l+1)}{l(l+1)+r}\ .
\ee
At the level of large $N$ effective action, this introduces a non-local, momentum dependent contribution, which cannot be canceled by a counter term and thus is referred to as a non-commutative anomaly \cite{uvir1}. When treating the sphere as an approximation of the flat space and taking the large $R$ limit as in the section \ref{cpntorn}, this results into an infrared divergence in the non-planar contribution to the above one loop divergence.

In order to remove this mixing and reproduce the expected commutative limit, we need to redefine the naive action (\ref{freeaction}). One possible procedure is to define a normal ordering of the interaction term, which results into the cancellation of the extra non-commutative contribution \cite{uvir}. This leads into the following modification of the action for fuzzy sphere theory
\be
S[M]=\frac{V_n}{N}\trsl{\half \frac{1}{R^2}MC_2\mathcal Z(C_2)M+\half M\lr{t-\frac{g}{2}\mathcal R(t)}M+g M^4},
\label{2.30}
\ee
where $C_2M=[L_i,[L_i,M]]$ is the quadratic Casimir of $SU(2)$. Function $\mathcal Z(C_2)$ is a power series in $C_2$ starting with $1+\kappa Q(C_2)$ and $\kappa$ is related to $r$ and $g$. Both $\mathcal Z$ and $\mathcal R$ are related to the renormalization of the wave function and the mass of the theory.

For the case of $\R^{2n}_\theta$, a different approach was suggested in \cite{gw}. Here, the UV/IR mixing is treated as an effect of the asymmetry between the high energy and the low energy regimes of the non-commutative theory and the models is modified to restore this symmetry. This is achieved by introduction of a harmonic oscillator-like term into the action, which modifies the dispersion relation.

%% file: 1_3phases.tex
As was mentioned in the previous section \ref{sec12}, the commutative limit of the fuzzy scalar field theory is much more complicated than one would expect. The UV/IR mixing of the non-commutative theory on the plane is related to the non-commutative anomaly on the fuzzy sphere. This, on the other hand was argued to be the source of a distinct phase in the phase diagram of the theory, which is classically not present and which survives the commutative limit \cite{phaseover,noncomphase1}.

For review of the phase structure of fuzzy field theories, see \cite{fuzzy3,num14panero1,medina}.

\subsubsection{Description of the phases of the theory in flat space}

Scalar $\phi^4$ theory on commutative $\mathbb R^2$ allows for two phases \cite{comr2}. A symmetric phase, where the field configurations oscillate around the symmetric vacuum $\phi=0$ and a phase which spontaneously breaks the $\phi\to-\phi$ symmetry of the theory. In this phase, the field oscillates around one of the minima of the potential and the phase transition between these two is of the second order. Later, using the lattice techniques, the critical line of the theory has been identified \cite{comr2num}. These two phases are referred to as a disorder phase and non-uniform order phase.

The phase structure of the non-commutative theories \cite{noncomphase1,noncomphase2,noncomphase3} predicted existence of a third, striped phase or the uniform order phase. In this phase, the field is non-vanishing in both of the minima of the potential, if those are sufficiently deep. Existence of this third phase is a nonlocal, non-commutative effect and the phase does not disappear in the commutative limit of the theory. This is the result of the UV/IR mixing and reflects the fact that the naive non-commutative theory does not reproduce the expected commutative counterpart. If we consider modifications of the action which remove the UV/IR mixing and reproduce the expected commutative limit, it is natural to expect that such modification would remove this phase.

Since non-commutative theory on the plane can be obtained from the theory on the fuzzy sphere, a third phase of the theory is expected also there.

\subsubsection{Numerical simulations for the fuzzy sphere}

There are several numerical results for different fuzzy and noncommutative space \cite{rsfnum,num14panero2,fuzzydiscnum}. The simulations investigate the commutative limit of the noncommutative theory. In this work, we will concentrate on the case of the fuzzy sphere and the quartic scalar field.

In \cite{nummartin}, the three phases of the scalar field theory on the fuzzy sphere, disordered, uniform ordered and non-uniform ordered, were identified. The explicit formula for the critical lines between the uniform/disordered phase and the non-uniform/disordered phase was obtained, but the non-uniform/uniform phase transition was not accessible.

The phase transition between the disorder and non-uniform order phase has been identified and carefully studied in \cite{paneronum}. The split of the eigenvalue density was observed and it was shown that a gap around the zero eigenvalue develops for a critical value of the coupling. The coordinates of this critical point have been found for a range of values of $g$ and $r$.

\begin{figure}[t]                 
\begin{center}
\includegraphics[width=0.8\textwidth]{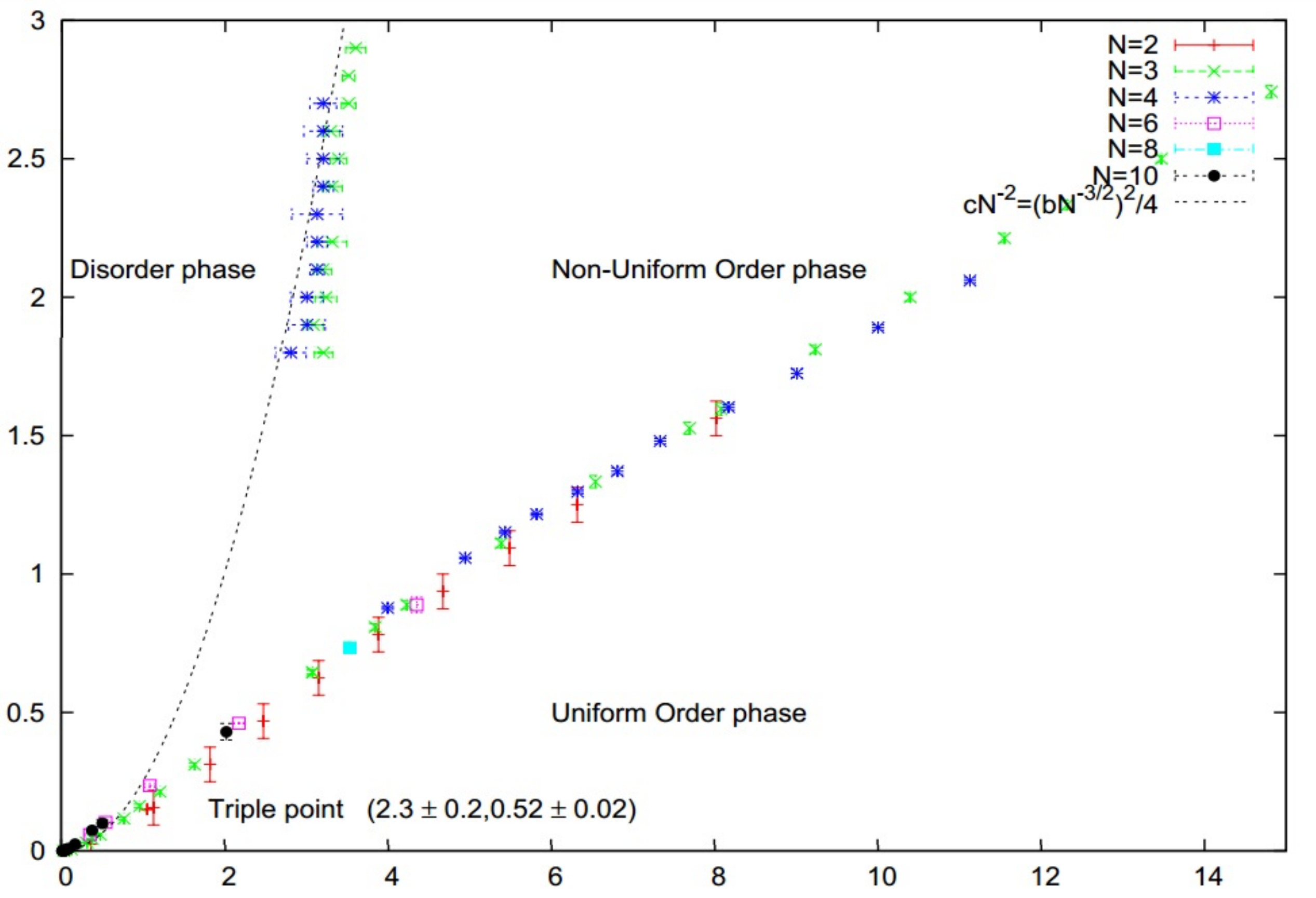}
\end{center}
\vspace*{-0.5cm}
\caption{(Color online) The phase diagram for the fuzzy sphere obtained numerically in \cite{denjoenum2}, presented here with a kind permission of the authors. The horizontal axis is related to $-r$ and the vertical axis is related to the coupling $g$.}\label{plotnum}  
\end{figure}

In \cite{denjoenum1,denjoenum2} the whole phase diagram of the theory was obtained. The result is shown in the figure \ref{plotnum}.
All three phases of the theory are identified, with the corresponding boundaries of co-existence. This work was also able to pin down the triple point, with coordinates, in our notation,
\be
g_c=0.13\pm0.005\ ,\ r_c=-2.3\pm0.2\ ,\label{cnum2}
\ee
A more recent work \cite{num14} used a different approached and obtained a diagram with the same qualitative features and identified the location of the triple point to be
\be
g_c=0.145\pm0.025\ ,\ r_c=-2.49\pm0.07\ .\label{cnum3}
\ee
We thus conclude, that the numerical simulations of the quartic scalar field theory on the fuzzy sphere predict three phases in the phase diagram even in the commutative limit and the location of the triple point in the interval
\be\label{13numericalptriple}
g_c\in(0.12,0.17)\ .
\ee

%% file: 2_0intro.tex
This sections reviews matrix models, with the most important notions and techniques. We explain properties of ensembles of random matrices and investigate the limit of a large matrix size. We then show computational technique of the saddle point approximation and how it can be used to calculate quantities of interest in this limit. In the second part of this section we study more complicated matrix ensembles characterized by a higher power of a trace of the matrix in the probability distribution and show how to use the saddle point approximation there. We elaborate on several examples to make the ideas as clear and understandable as possible.

A number of wonderful reviews of matrix model techniques is available. A classic, yet a mathematically minded \cite{metha}, several other reviews \cite{matrix1,matrix11,rndmath,matrixnum} or very readable original papers \cite{brezin,itzub}. Some of the techniques are nicely reviewed in \cite{matrix2,matrix3,string}.

%% file: 2_1matrixmodels.tex
We start with the description of the single trace Hermitian matrix models and the treatment of the limit of large matrix size.

\subsubsection{Ensembles}

As mentioned in the introduction, random matrix theory is given by the ensemble of matrices $M$ with the integration measure $dM$ and the probability measure $\exp{-S(M)}$. The expected value of a function $f(M)$ of the random matrix is then computed as
\begin{eqnarray}
	\avg{f}=\frac{1}{Z}\int dM e^{-S(M)}f(M)\ .\label{3.1}
\end{eqnarray}
Here, the normalization $1/Z$ is such that $\avg{1}=1$. 

The ensemble usually has some symmetry and it is assumed that the integration measure $dM$, as well as the weight $S(M)$ and any reasonable function $f(M)$ are symmetric also.

The choice of the matrix ensemble is then dictated by the physical or mathematical setup. Most usually, matrix $M$ is hermitian, real symmetric or quaternionic self-dual, with $SU(N)$, $SO(N)$ and $Sp(2N)$, or a sub-group, being the symmetry group. Matrix $M$ can be also directly an unitary, orthogonal or symplectic matrix. In some mathematical applications, such as statistics or number theory, more complicated matrices, which need not to be square, arise, see e.g. \cite{matrixnum}.

The most general choice of the probability measure is then dictated by the choice of the ensemble and the symmetry group. To be more precise, let $M$ be a square $N\times N$ matrix with symmetry group $G$, with the action $M\to gMg^{-1}$ with $g\in G$. We then write the measure as $e^{-N^2 S(M)}$, the factor of $N^2$ makes it explicit that the measure is of the same order as $dM$ and thus contributes. The action $S(M)$ is then finite in the limit of large $N$. Then, the most general invariant measure is given by
\begin{eqnarray}\label{21orequival}
	S(M)=\frac{1}{N}\sum_{n=0}^N g_n \trl{M^n}\ ,
\end{eqnarray}
where the factor of $1/N$ ensures proper large $N$ limit, as the sum of $N$ eigenvalues is of the order $N$. Any other invariant function can be re-expressed in terms of the first $N$ traces. The constant and the linear terms are usually not considered, since they can be absorbed into redefinition of $M$. Also, the $M^2$ term is usually considered separately and we write
\begin{eqnarray}
	S(M)=\half r \frac{1}{N}\trl{M^2}+\sum_{n=3}^N g_n\frac{1}{N} \trl{M^n}\ .\label{3.3}
\end{eqnarray}
This is because of the fact that the $n\geq3$ terms introduce ``interactions'' into the theory, which means that this renders also some of the higher order correlation functions nontrivial. Theory with only $\trl{M^2}$ term can be solved exactly and the extra terms can be considered as a perturbation.

We will treat the measure $dM$ explicitly only for the case of $N\times N$ Hermitian matrices. In the general case, upon the diagonalization of the matrix, the measure will become
\begin{eqnarray}
	dM=J(\lambda)d\Lambda dU\ ,
\end{eqnarray}
where $d\Lambda$ is the measure on the space of eigenvalues $\lambda$ of $M$, $dU$ is the measure of the symmetry group $G$ and $J(\lambda)$ is Jacobian corresponding to the change of variables from $M_{ij}$ to $\lambda$ and $U$. In the case of Hermitian matrices, $G=SU(N)$ and the integration measure is given by
\begin{eqnarray}
	dM=\prod_{i=1}^n M_{ii}\prod_{i<j} d\textrm{Re}M_{ij}\,d\textrm{Im}M_{ij}\ .
\end{eqnarray}
We can diagonalize the matrix
\begin{eqnarray}
	M=U\Lambda U^\dagger\ ,
\end{eqnarray}
where $\Lambda=\textrm{diag}\,\lr{\lambda_1,\ldots,\lambda_N}$ is the diagonal matrix of the eigenvalues of $M$. The integration measure then becomes
\begin{eqnarray}
	dM=\lr{\prod_{i<j}(\lambda_i-\lambda_j)^2}\lr{\prod_{i=1}^N d\lambda_i} dU\ .\label{3.7}
\end{eqnarray}
There are number of ways how to compute the Jacobian, which turned out to be the square of the Vandermonde determinant in this case. Thanks to the invariance of the measure, $J$ depends only on $\lambda$'s, and we can compute it in the vicinity of $U=1$. Here
\begin{eqnarray}
	dM=dU \Lambda+d\Lambda +\Lambda dU^\dagger=d\Lambda+\Lambda dU-dU\Lambda\ ,
\end{eqnarray}
where we have used $dU^\dagger=-dU$. This means that
\begin{eqnarray}
	dM_{ij}=d\lambda_i\delta_{ij}+(\lambda_i-\lambda_j)dU_{ij}\ ,
\end{eqnarray}
the change of variables is diagonal and the Jacobian is just the product of the factors $(\lambda_i-\lambda_j)$.

\subsubsection{Planar limit as the leading order in the limit of large matrices}

One is usually interested in the behavior of the results in the limit of a very large matrix. For the case of $N\times N$ Hermitian matrices this means $N\to\infty$ limit. There are many reasons for this, being mathematical, physical and also practical.

From the mathematical point of view, the quantities like eigenvalue distribution become continuous in the large matrix limit. Also in this limit, certain properties of eigenvalue distribution become independent of the exact probability distribution and depend only on the symmetry group of the ensemble. This notion is called universality and allows to study certain properties on the simplest ensembles \cite{metha}.

When random matrices describe a physical system, large matrix limit is the limit of large number of constituents. When studying the spectra of large nuclei, this means large number of levels. In condensed matter this means large number of electrons or other particles of interest. When one uses the random matrix to describe a lattice or to discretize some surface, large $N$ limit is the continuum limit. In all these cases the limit is well justified by the physics of the problem.

When one considers field theory on a fuzzy space, large $N$ limit is the limit of commutative theory and if the fuzzy structure was introduced to regulate the theory, this removes the regulator. Also, if there is noncommutativity present in the nature, it is very small and this justifies the limit of large $N$. The subleading contributions then represent the non-commutative correction.

And lastly, the large $N$ limit provides a considerable simplification to the calculations. As we will see, the diagrams involved in computation of the averages give contributions with different powers of $N$ and this power is related to the topology of the diagram. The leading order is given by the planar diagrams and this simplifies greatly the resulting combinatorial problem. The subleading corrections can then be computed systematically.

To compute expectation values of invariant functions $f(M)$, one has to compute correlators of the form $\avg{\trl{M^k}[W(M)]^l}$. To compute these, we use the Wick's theorem. One has to sum over all possible pairing of $M$'s in the expectation value, weighted by a propagator $\contraction{}{M}{{}_{ij}}{M}M_{ij}M_{kl}$ for each pairing. Since the contractions are done with the Gaussian, or free, measure, the propagator is given by the inverse of the quadratic part of (\ref{3.3}) and is equal to
\begin{eqnarray}
	\contraction{}{M}{{}_{ij}}{M}M_{ij}M_{kl}=\frac{1}{Nr}\delta_{il}\delta_{jk}\ .
\end{eqnarray}

\begin{figure}[tb]              
\begin{center}
\vspace{-0.3cm}
\includegraphics[width=0.3\textwidth,clip]{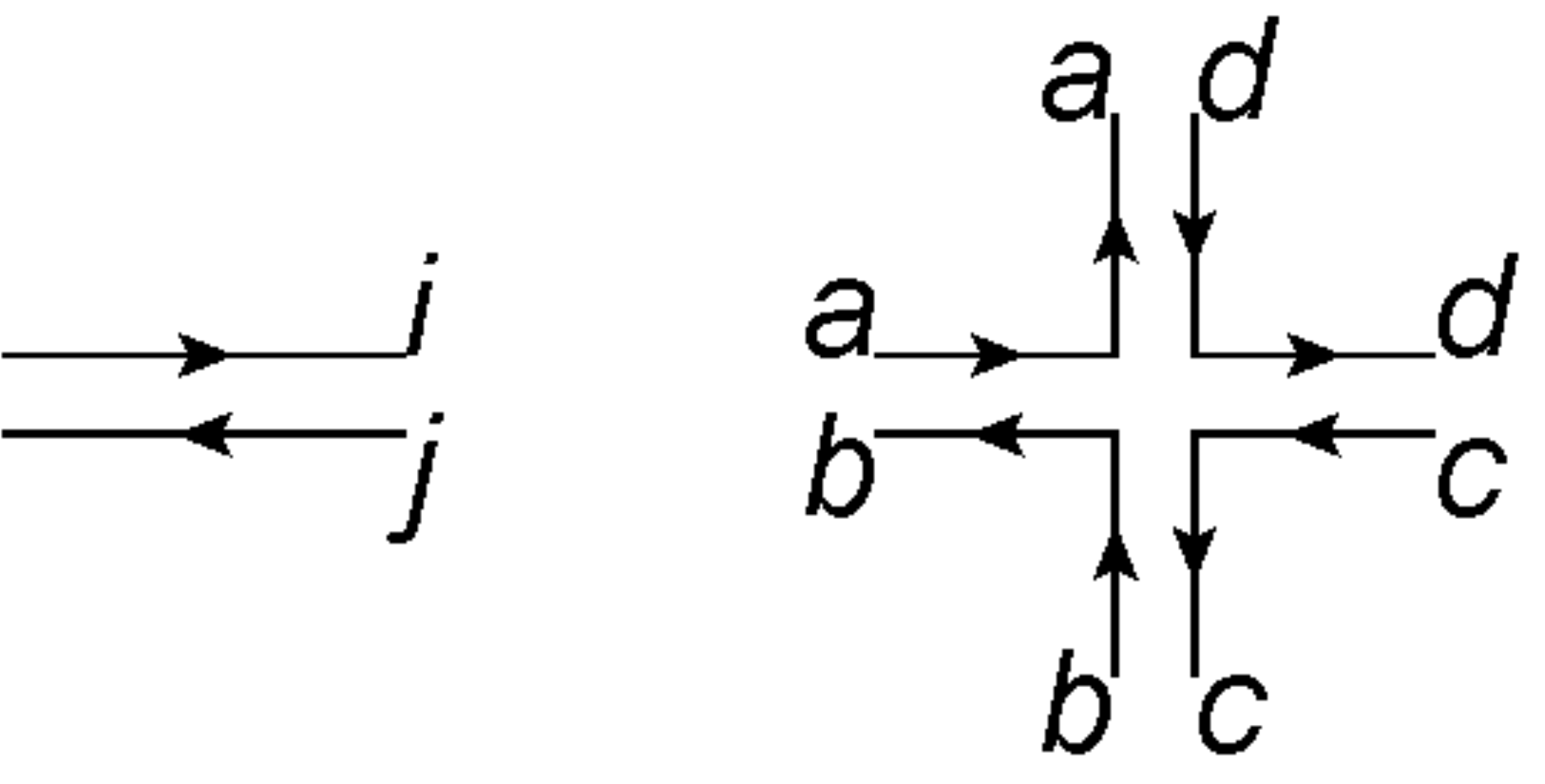}  
\end{center}
\vspace{-0.7cm}
\caption{Figure shows the double-line notation of a matrix element $M_{ij}$ and the double line notation of a quartic vertex $M_{ab}M_{bc}M_{cd}M_{da}$.}\label{notation}
\vspace{-0.4cm}
\end{figure}
To compute the higher correlators, we use method of fat graphs due to 't Hooft \cite{thooft}. We will indicate the matrix $M$ in the diagrams by a double line, which represent the double index structure of $M_{ij}$. Since the matrix is in the adjoint representation of $SU(N)$, indexes $i$ and $j$ can be viewed as in the fundamental and anti-fundamental representation and the lines and the arrows reflect this structure. This is illustrated in the figure \ref{notation}. Vertices are then given by a star of $n$ double lined legs. Let us illustrate this at the computation of the first order correction to the two point function $\avg{M_{ij}M_{kl}}$ of the theory with quartic interaction $g\trl{M^4}$.

\begin{figure}[tb]              
\begin{center}
a)\includegraphics[width=0.3\textwidth,clip]{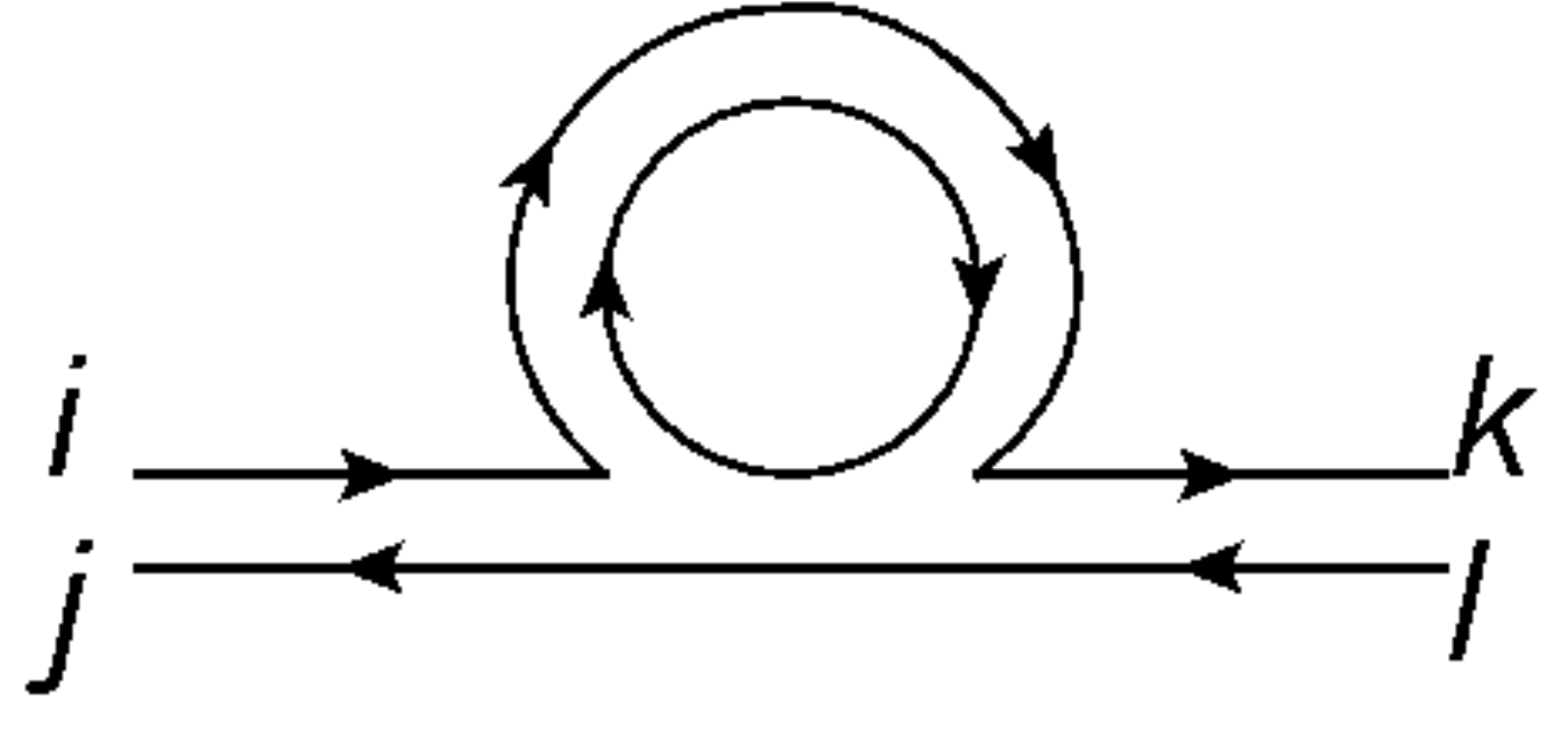}\ \ 
b)\includegraphics[width=0.3\textwidth,clip]{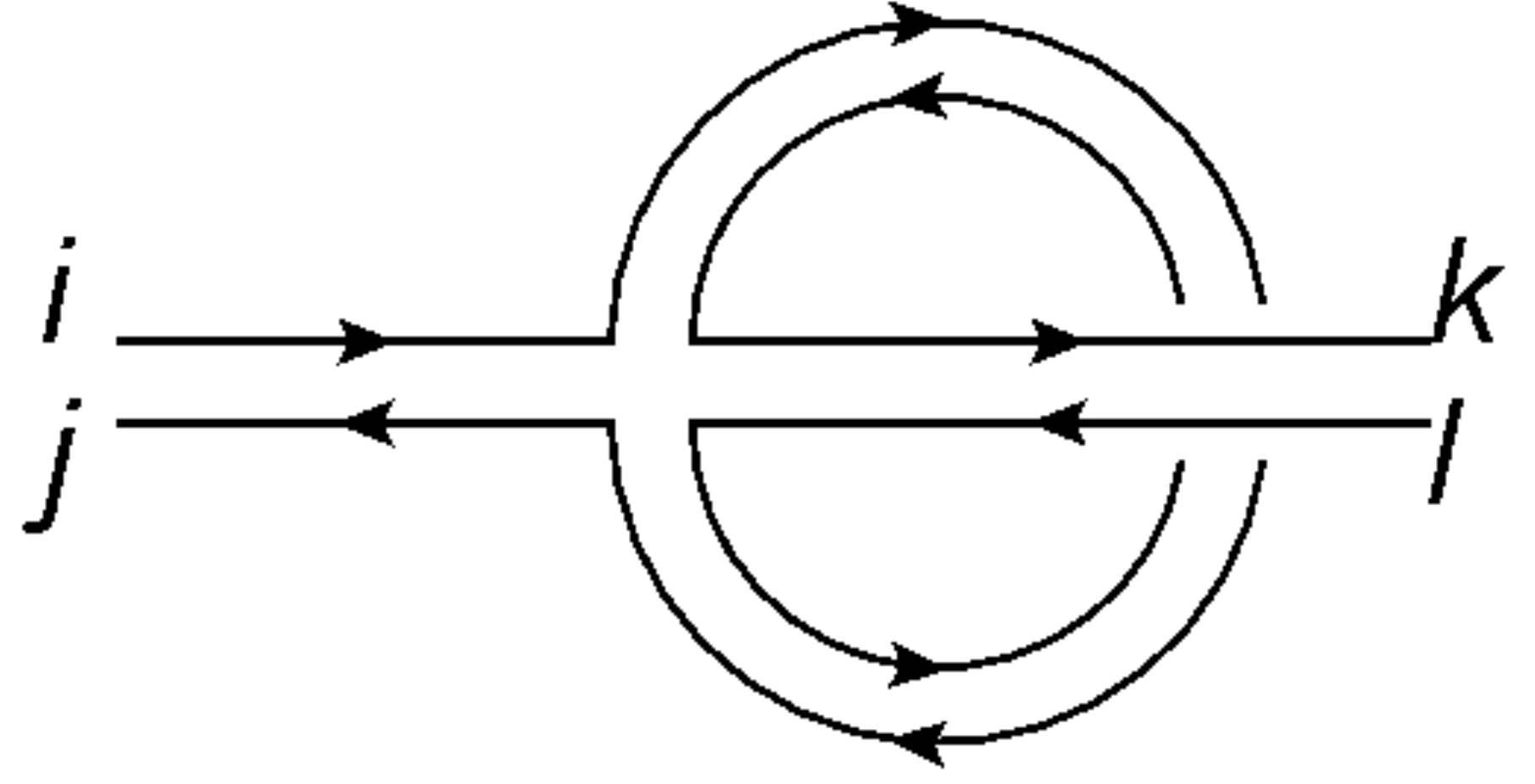}
\end{center}
%\vspace{-20mm}
\vspace{-0.5cm}
\caption{a) shows the planar diagram contribution to the $g^1$ order of the two point function $\avg{M_{ij}M_{kl}\trl{M^4}}$, b) the nonplanar contribution.}\label{fourpoint} 
\end{figure}

The relevant diagrams are shown in the figure \ref{fourpoint}. There are two possible kinds of contractions, one planar \ref{fourpoint}a and one non-planar \ref{fourpoint}b. They give respective contributions
\begin{eqnarray}
	\textrm{planar}=Ng\frac{1}{Nr}\delta_{ia}\delta_{jb}\frac{1}{Nr}\delta_{bl}\delta_{ik}\frac{1}{Nr}\delta_{dd}\delta_{ic}=\frac{1}{N}g\lr{\frac{1}{r}}^3\delta_{ik}\delta_{jl}\label{3.11}
\end{eqnarray}
and
\begin{eqnarray}
	\textrm{non-planar}&=&Ng\frac{1}{Nr}\delta_{ia}\delta_{jb}\frac{1}{Nr}\delta_{ab}\delta_{cd}\frac{1}{Nr}\delta_{dk}\delta_{cl}\nonumber\\&=&\frac{1}{N^2}g\lr{\frac{1}{r}}^3\delta_{ij}\delta_{kl}\ .\label{3.12}
\end{eqnarray}
Following the summation of indexes on delta functions we observe that the overall factor of $N$ depends on the number of closed lines in our fat graph. Each line produces, after the summation of all but one of the indexes, factor of $\delta_{ii}$, which then gives a factor of $N$ when summed over the last index.

Now comes a crucial observation for a general diagram. Let it have $E$ propagators, or edges, $V_n$ vertices with $n$ legs and $h$ closed loops. The factor for this diagram is then
\begin{eqnarray}
	\lr{\frac{1}{Nr}}^E N^h \prod_{n=3}^N (-Ng_n)^{V_n}\ ,\label{3.13}
\end{eqnarray}
where $V=\sum V_n$ is the total number of vertices. If we now consider the diagram as a Riemann surface of genus $g$, we have the topological relation
\begin{eqnarray}
	2-2g=h-E+V.
\end{eqnarray}
So (\ref{3.13}) can be rewritten as
\begin{eqnarray}
	N^{2}\Bigg[\lr{\frac{1}{r}}^E \prod (-g_n)^{V_n}\Bigg]\Bigg[\frac{1}{N^2}\Bigg]^g\ .
\end{eqnarray}
We see that the first factor is given by the type of the diagram. The second factor is given by the topology of the diagram and diagrams of the same type, with higher genus $g$ are all suppressed by a factor of $1/N^{2g}$. This also means that the leading order contribution is given by the $g=0$, i.e. planar diagrams.

To connect the two-point functions (\ref{3.11}) and (\ref{3.12}) to this result, we need to realize that the two diagrams cannot be realized as Riemann surfaces, since they are not an average of an invariant function of the form $\trl{M^k}$. In order to be able to do so, we need to contract the two external matrices with $\delta_{ik}\delta_{jl}$, i.e. to close the loops, and consider them as a new, $2$-point vertex, which brings an extra factor of $N$. Then, we see that the planar diagram really has $N^2$ dependence, and the non-planar $N^0$.

%% file: 2_21sadlepoint.tex
This section introduces our main tool to analyze the matrix models, the saddle point approximation.

\subsubsection{General aspects of the saddle point method}

In this section, we will show how to obtain the eigenvalue distribution of the random matrix without explicit computation of any expectation values. We will later show that doing that and counting the diagrams leads to the same result as computed here.

We absorb the Jacobian in (\ref{3.7}) into the action and obtain a theory governed by the measure
\begin{eqnarray}
	N^2 S_{V}(\lambda)=N^2\slr{\half r\frac{1}{N}\sum_{i=1}^N \lambda_i^2+W(\lambda)-\frac{2}{N^2}\sum_{i<j} \log|\lambda_i-\lambda_j|}\ ,\label{3.16}
\end{eqnarray}
where we have denoted the $n\geq3$ part of the measure as 
\be
W(\lambda)=\sum_k g_k\frac{1}{N}\sum_{i=1}^N \lambda_i^k\ .
\ee
In some literature the notation effective in (\ref{3.16}) is standard, but since we will encounter different kind of effective quantities two more times in this text, we shall not use this notation. Since we will need the notation for this object only rarely, we will denote it $S_V$ and reserve the term effective action for something else.

We will assume, that the matrix and the parameters are scaled in such way, that the terms are of order $1$ in the large $N$ limit. The key feature will be the fact, that the sums in the expression are of order $N$ in the large $N$ limit. The Vandermonde term is thus already of order $1$ and any scaling of the matrix ads only a constant term to it, which can be disregarded.

We further introduce scaled quantities
\be
r=\tilde r N^{\theta_r}\ , \ g_k=\tilde g_k N^{\theta_k}\ ,\ M=\tilde M N^{\theta_x}\ .
\ee
The mass terms behaves like $N^{1+\theta_r+2\theta_x}$ and the $k$-th term of the potential part like $N^{1+\theta_k+k \theta_x}$. And we want both these to scale as $N^2$, so we obtain conditions
\be\label{22generalscaling} 1=\theta_r+2\theta_x \ , \ 1=\theta_k+k \theta_x\ .\ee
These do not fix the scaling uniquely and for the simplest choices we get
\bse
\bal
\theta_r=0\ ,&\ \theta_x=\frac{1}{2}\ ,\ \theta_k=1-\frac{k}{2}\\
\theta_r=1\ ,&\ \theta_x=0\ ,\ \theta_k=1\ .
\end{align}
\ese
We will not denote the scaled quantities by tilde and write expressions like (\ref{3.16}), but we will keep in the back of our mind that such form is achievable for example by using the scaling (\ref{22generalscaling}).

Therefore as $N\to\infty$, the integral in (\ref{3.1}) will be dominated by the saddle-point configuration of the eigenvalues $\lambda^E_i$, which extremizes the action (\ref{3.16}), or in other words has the largest probability. The average of an invariant function $f(M)$ is then given, in the large $N$ limit, by
\be
\avg{f}=\sum_{i=1}^Nf\lr{\lambda^E_i}\ .
\ee
We vary the action with respect to $\lambda_i$ to obtain
\begin{eqnarray}
	r\lambda^E_i+W'(\lambda^E_i)=\frac{2}{N}\sum_{i\neq j}\frac{1}{\lambda^E_i-\lambda^E_j}\ ,
	\label{3.17}
\end{eqnarray}
for $i=1,2,\ldots,N$. This form of the equation will be very useful later, but often we will work with the eigenvalue distribution. For the solutions of the equation $\lambda^E_i$, it is formally defined as
\begin{eqnarray}
	\rho(\lambda)=\frac{1}{N}\sum_{i=1}^N\delta(\lambda-\lambda^E_i)\ .
\end{eqnarray}
This function becomes continuous in the large $N$ limit. For a function of the eigenvalues $f(\lambda)$ we then have in the large $N$ limit
\begin{eqnarray}
	\sum_{i=1}^Nf(\lambda_i)\to N\int_\C d\lambda\,\rho(\lambda)f(\lambda)\ ,\label{3.21}
\end{eqnarray}
where the integral is over the support of the distribution, which is a bounded interval or a bounded union of intervals, thanks to the requirement on the scaling of the action $S$. Using this property, we change (\ref{3.17}) to
\begin{eqnarray}
	r\lambda+W'(\lambda)=2P\int d\lambda'\frac{\rho(\lambda')}{\lambda-\lambda'}\ .
	\label{3.20}
\end{eqnarray}
Here, $P\int$ denotes the principal value of the integral. We introduce a function, called the planar resolvent,
\begin{eqnarray}
	\omega_0(z)=\frac{1}{N}\sum_i\frac{1}{z-\lambda^E_i}=\int d\lambda\frac{\rho(\lambda)}{z-\lambda}\ .\label{3.23}
\end{eqnarray}
It is very important to note, that for large $|z|$, we have $\omega_0\to1/z$ thanks to the normalization of $\rho$.

Computing a square of the resolvent
\begin{align}
\omega_0(z)^2=&\frac{1}{N^2}\sum_{i,j}\lr{\frac{1}{z-\lambda_i^E}}\lr{\frac{1}{z-\lambda_j^E}}=\no
=&-\frac{1}{N}\omega_0'(z)+\frac{2}{N^2}\sum_{i,j}\lr{\frac{1}{\lambda_i^E-\lambda_i^E}}{\frac{1}{z-\lambda_j^E}}\ .
\end{align}
We neglect the second term, as it is subdominant in the large-$N$ limit. We then rewrite this equation using the saddle point condition (\ref{3.17}) as
\be
\omega_0(z)^2=\lr{r\lambda^E_i+W'(\lambda^E_i)}\omega_0(z)+P(z)\ ,
\ee
where
\be
P(z)=\frac{1}{N}\sum_i\frac{rz+W'(z)-r\lambda^E_i-W'(\lambda^E_i)}{z-\lambda_i^E}\ .
\ee
This is a quadratic equation for the resolvent, which can be solved as
\be\label{22solresol}
\omega_0(z)=\half\slr{rz+W'(z)-\sqrt{\slr{r\lambda^E_i+W'(\lambda^E_i)}^2-4P(z)}}\ .
\ee
The polynomial $P$ is not know yet, but it is much simpler to determine than $\omega_0$ directly from (\ref{3.20}).

Also note, that the resolvent is not well defined on the support of the eigenvalue distribution, which is clear from both the definition (\ref{3.23}) and the solution (\ref{22solresol}).

In terms of the resolvent function the eigenvalue distribution can be computed using the discontinuity equation
\begin{eqnarray}
	\rho(\lambda)=-\frac{1}{2\pi i}\slr{\omega_0(\lambda+i\ep)-\omega_0(\lambda-i\ep)}\ ,
	\label{3.22}
\end{eqnarray}
and equation (\ref{3.20}) becomes an equation for the resolvent
\begin{eqnarray}
	\omega_0(z+i\ep)-\omega_0(z-i\ep)=-rz-W'(z)\ .\label{3.25}
\end{eqnarray}

\subsubsection{One cut and multiple cut assumptions}

To proceed further, one has to make an assumption about the topology of the support of the distribution $\C$. Namely on the number of the disjoint intervals on the real axis which form this support, which are going to be referred to as cuts.

\podnadpis{One cut assumption}

If $\C$ is given by the interval $[b,a]$, the equation (\ref{3.25}) is solved by
\begin{eqnarray}
	\omega_0(z)=\half \oint_\C \frac{dz'}{2\pi i}\frac{r z'+W'(z')}{z-z'}\sqrt{\frac{(z-a)(z-b)}{(z'-a)(z'-b)}}\ .
	\label{3.24}
\end{eqnarray}
To see this, we first realize that from (\ref{22solresol}) we must have
\be
\lr{r\lambda^E_i+W'(\lambda^E_i)}^2-4P(z)=M(z)^2(z-a)(z-b)
\ee
for some polynomial $M(z)$, given by the condition
\be\label{22polyM}
M(z)=\textrm{Pol}\,\frac{r\lambda^E_i+W'(\lambda^E_i)}{\sqrt{(z-a)(z-b)}}\ ,
\ee
where $\textrm{Pol}\,$ denotes the polynomial part and which is the result of dividing (\ref{22solresol}) and taking the large-$N$ limit. It can be deduced also from
\begin{eqnarray}
	M(z)=\oint_0 \frac{dz'}{2\pi i}\frac{r/z'+W'(1/z')}{1-z z'}\frac{1}{\sqrt{(1-az')(1-bz')}}\ ,\label{3.29}
\end{eqnarray}
with contour around $z=0$. Once $M$ is known, the ends of the cut are again given by the asymptotics $\omega_0(z)\to1/z$ and
\bse\label{3.27}
\begin{align}
	\oint_\C\frac{dz'}{2\pi i}\frac{r z'+W'(z')}{\sqrt{(z'-a)(z'-b)}}=&0\ ,\label{3.27a}\\
	\oint_\C\frac{dz'}{2\pi i}\frac{z'\slr{r z'+W'(z')}}{\sqrt{(z'-a)(z'-b)}}=&2\ .\label{3.27b}
\end{align}
\ese
Once $M(z)$ is known, these are equivalent to requiring
\be\label{22polyres}
\textrm{Pol}\,\omega_0(z)=\textrm{Pol}\,\half\slr{rz+W'(z)-M(z)\sqrt{(z-a)(z-b)}}=0\ .
\ee
Finally, the distribution is given by
\be\label{22polyrho}
\rho(\lambda)=\frac{1}{2\pi i}M(\lambda)\sqrt{(\lambda-a)(\lambda-b)}=\frac{1}{2\pi}M(\lambda)\sqrt{(a-\lambda)(\lambda-b)}\ .
\ee
We have assumed that $M(z)>0$. If this is not the case, we are clearly running into trouble. The one cut assumption is not valid and we need to go further.

Knowing the distribution, one can now compute various expectation values of the theory. For example the normalized traces of powers of $M$ are given by
\begin{eqnarray}
	\frac{1}{N}\avg{\trl{M^k}}=\int_\C d\lambda\,\lambda^k\rho(\lambda).\label{3.30}
\end{eqnarray}
Similarly, the planar free energy is given by
\begin{eqnarray}\label{22freeenergy}
	\F_0=-\frac{1}{N^2}\log \lr{\int dM\,e^{-N^2 S(M)}}=-\frac{1}{N^2}\log e^{-N^2S_{V}[\rho(\lambda)]}=S_{V}[\rho(\lambda)]\ .
\end{eqnarray}
Note that
\bal
S_{V}[\rho(\lambda)]=&\half r\int d\lambda\ \lambda^2\rho(\lambda)+\int d\lambda\ W(\lambda) \rho(\lambda)-\no&-2\int d\lambda\,d\lambda'\ \rho(\lambda)\rho(\lambda') \log|\lambda-\lambda'|\ .
\end{align}
This formula can however be simplified a little into a form more suitable for numerical integration we will do later. For this purpose, we introduce a Lagrange multiplier $\xi$ into the action fixing the normalization of the eigenvalue distribution
\bal
S_V[\rho(\lambda)]=&\half r\int d\lambda\ \lambda^2\rho(\lambda)+\int d\lambda\ W(\lambda) \rho(\lambda)-\no&-2\int d\lambda\,d\lambda'\ \rho(\lambda)\rho(\lambda') \log|\lambda-\lambda'|+\xi\lr{\int d\lambda \rho(\lambda)-1}\ .
\end{align}
Varying this equation with respect to the density we obtain\footnote{The variation of this equation with respect to $\lambda$ would yield the saddle point equation (\ref{3.20}).}
\be
\half r\lambda^2+W(\lambda)-2\int d\lambda' \rho(\lambda')\log|\lambda-\lambda'|+\xi=0\ ,
\ee
which has to hold for any value of $\lambda$. The choice is free, we will mostly chose the larges eigenvalue such that we do not have to think too much about the absolute value in the logarithm but other choices are possible. Using this, the free energy can be simplified to
\be
\F_0=\half\slr{\int d\lambda\lr{\half r \lambda^2 +W(\lambda)}\rho(\lambda)-\xi}\ .
\ee
Often in the literature the free energy is defined without the free part
\be
	\F_0=-\frac{1}{N^2}\slr{\log \lr{\int dM\,e^{-N^2 \lr{\half r \lambda^2+W(\lambda)}}}
	-\log \lr{\int dM\,e^{-N^2 \lr{\half r \lambda^2}}}}\ .
\ee
This is equivalent to removing the vacuum bubbles. We will not do it here, as the free energy is going to be a tool for determining the most probable solution at given values of parameters. This answer is unaffected by this removal so we choose the simplest possible definition of $\F$.

We should also mention that the resolvent is connected to the moments of the distribution by
\begin{eqnarray}
	z\omega(z)=\frac{1}{N}\sum_{k=0}^\infty \frac{\avg{\trl{M^k}}}{z^k}=\sum_{k=0}^\infty \frac{c_k}{z^k}.\label{3.33}
\end{eqnarray}
This is indeed the defining relation for the full resolvent with contributions from diagrams of any topology. The planar part is then given by the planar diagrams of $\avg{\trl{M^k}}$, which together with (\ref{3.30}) gives our original definition of planar resolvent (\ref{3.23}).

\podnadpis{Two and more cuts}

The situation gets more complicated when the potential has more than one minimum. If the minima are deep enough, the gas of particles can split into two or more disjoint parts, each located in one minimum. In the language of the eigenvalue density, the one cut assumption is no longer valid and we need to assume a more complicated support of the distribution.

We illustrate the approach on the case of two minima of the potential. We will therefore assume that the support $\C$ is given by the union of two intervals $[a,b]$ and $[c,d]$. Equation (\ref{22solresol}) the becomes
\begin{eqnarray}
	\omega_0(z)=\half\slr{r z+W'(z)-M(z)\sqrt{(z-a)(z-b)(z-c)(z-d)}}
	\label{3.40}
\end{eqnarray}
and the endpoints of the intervals are given by
\bse
\begin{align}
	\oint_\C\frac{dz'}{2\pi i}\frac{r z'+W'(z')}{\sqrt{(z'-a)(z'-b)}}=&0\ ,\\
	\oint_\C\frac{dz'}{2\pi i}\frac{z'\slr{r z'+W'(z')}}{\sqrt{(z'-a)(z'-b)}}=&0\ ,\\
	\oint_\C\frac{dz'}{2\pi i}\frac{(z')^2\slr{r z'+W'(z')}}{\sqrt{(z'-a)(z'-b)}}=&2\ ,
\end{align}
\ese
or by
\be
\textrm{Pol}\,\omega_0(z)=\textrm{Pol}\,\half\slr{r z+W'(z)-M(z)\sqrt{(z-a)(z-b)(z-c)(z-d)}}=0\ .
\ee
Recall that these are given by the $1/z$ asymptotics of $\omega_0(z)$. There are too few conditions to determine the endpoints! In general for $s$-cut case, there are $s+1$ conditions but $2s$ unknowns to be determined. Here, we have to make some extra assumptions.

The rescue is the free energy (\ref{22freeenergy}). Note that the solution with the the lowest free energy has the largest probability and will dominate the large-$N$ limit.

One can show that assuming the free energy of the system to be minimal gives exactly the extra $s-1$ conditions that are needed. One introduces the filling fraction $x_s$ of the eigenvalues in the $s$-th cut as
\begin{eqnarray}
	x_s=\int_{\C_s} d\lambda\,\rho(\lambda)\ ,\ \sum_s x_s=1,
\end{eqnarray}
and the variation of the free energy as a function of $x_s$ should vanish. This condition has also a very intuitive physical meaning. Recall the problem as a $N$ particle gas of eigenvalues and let’s get back to the two cut case. These now sit in two wells. If we try to move one eigenvalue from one well to the other, this should cost us no energy in the equilibrium case. If we could gain something, the fluctuations of the eigenvalues would eventually make this change spontaneously. If we lose energy this process would happen the other way and at the end of the day, we would reach balance given by the no-work-done condition.

The force on the $i$-the eigenvalue is
\be
f(\lambda)=-r\lambda_i+W'(\lambda_i)+\frac{2}{N}\sum_{i\neq j}\frac{1}{\lambda_i-\lambda_j}=-r\lambda_i+W'(\lambda_i)+2\omega_0(\lambda_i)\ .
\ee
If we now fix the eigenvalues and try to move one eigenvalue from one cut to the next one, the work required is
\be
\int_c^b d\lambda\,f(\lambda)=-\int_c^b d\lambda\,M(\lambda)\sqrt{(z-a)(z-b)(z-c)(z-d)}\ .
\ee
The no-work-done condition then gives
\begin{eqnarray}\label{twocutminimal}
	\int_b^c M(z)\sqrt{(z-a)(z-b)(z-c)(z-d)}=0.
\end{eqnarray}
For a multiple cut solution, similar condition is straightforwardly derived and has to hold between any two neighboring cuts.

%% file: 2_22wigner.tex
\subsubsection{The Wigner semicircle distribution}

The simplest example is clearly the no potential case $W(\lambda)=0$ and the probability measure given by
\be
S(M)=\half r \trl{M^2}\ .
\ee
The first the condition (\ref{3.27a}) or (\ref{22polyres}) becomes
\be
\frac{r}{4}\lr{a+b}=0\ \Rightarrow\ b=-a\ ,
\ee
which is the expected symmetric distribution for the symmetric potential. The second condition (\ref{3.27}) or (\ref{22polyres}) then yields
\be
\frac{ra^2}{4}=1
\ee
and 
\be\label{22radwigner}
a=2/\sqrt r\ .
\ee
Equation (\ref{3.29}) or (\ref{22polyM}) then yields $M(z)=r$ and we finally obtain
\begin{eqnarray}
	\omega_0(z)=\half\lr{r z-r\sqrt{z^2-\frac{4}{r}}},
\end{eqnarray}
and using the discontinuity equation (\ref{3.22}) or the expression (\ref{22polyrho}), we obtain the celebrated Wigner semicircle law
\begin{eqnarray}\label{22semicirc}
	\rho(\lambda)=\frac{r}{2\pi}\sqrt{\frac{4}{r}-\lambda^2}\ , \ \lambda^2<\frac{4}{r}\ .
\end{eqnarray}
Note that this distribution is normalized to $1$, which is the result of the assumed scaling. To find out this scaling explicitly, we can either look at the general formulas (\ref{22generalscaling}) or do something a little different.

We first solve the model without any rescaling. This normalizes $\rho(\lambda)$ to $N$, which is the number of eigenvalues, rather than to $1$. This yields an eigenvalue distribution
\begin{eqnarray}
	\rho(\lambda)=\frac{r}{2\pi}\sqrt{\frac{4 N}{r}-\lambda^2}\ , \ \lambda^2<\frac{4N}{r}\ .
\end{eqnarray}
Again, introducing rescaled quantities
\be\label{22scalingwig}
r=\tilde r N^{\theta_r}\ ,\ M=\tilde M N^{\theta_x}\ ,
\ee
this becomes
\begin{eqnarray}
	\frac{\rho(\lambda)}{N}=N^{2\theta_x+\theta_r-1}\frac{\tilde r}{2\pi}\sqrt{\frac{4}{\tilde r}N^{1-\theta_r-2\theta_x}-\tilde\lambda^2}\ ,\ \tilde\lambda^2<\frac{4}{r}N^{1-\theta_r-2\theta_x}\ .
\end{eqnarray}
So we see we obtain the same condition
\be 
1-\theta_r-2\theta_x=0\ .\label{22scalecond}
\ee
for both the radius and the distribution to finite in the large $N$. And this is clearly equivalent to the general formula (\ref{22generalscaling}).

Let us note, that without the scaling, the eigenvalues would spread to the infinity. This means that the Vandermonde repulsion would win over the potential and to include the effect of the potential, we need to enhance it by a corresponding power of $N$.

\begin{figure}[!t]
       \begin{center}
      \includegraphics[width=0.4\textwidth]{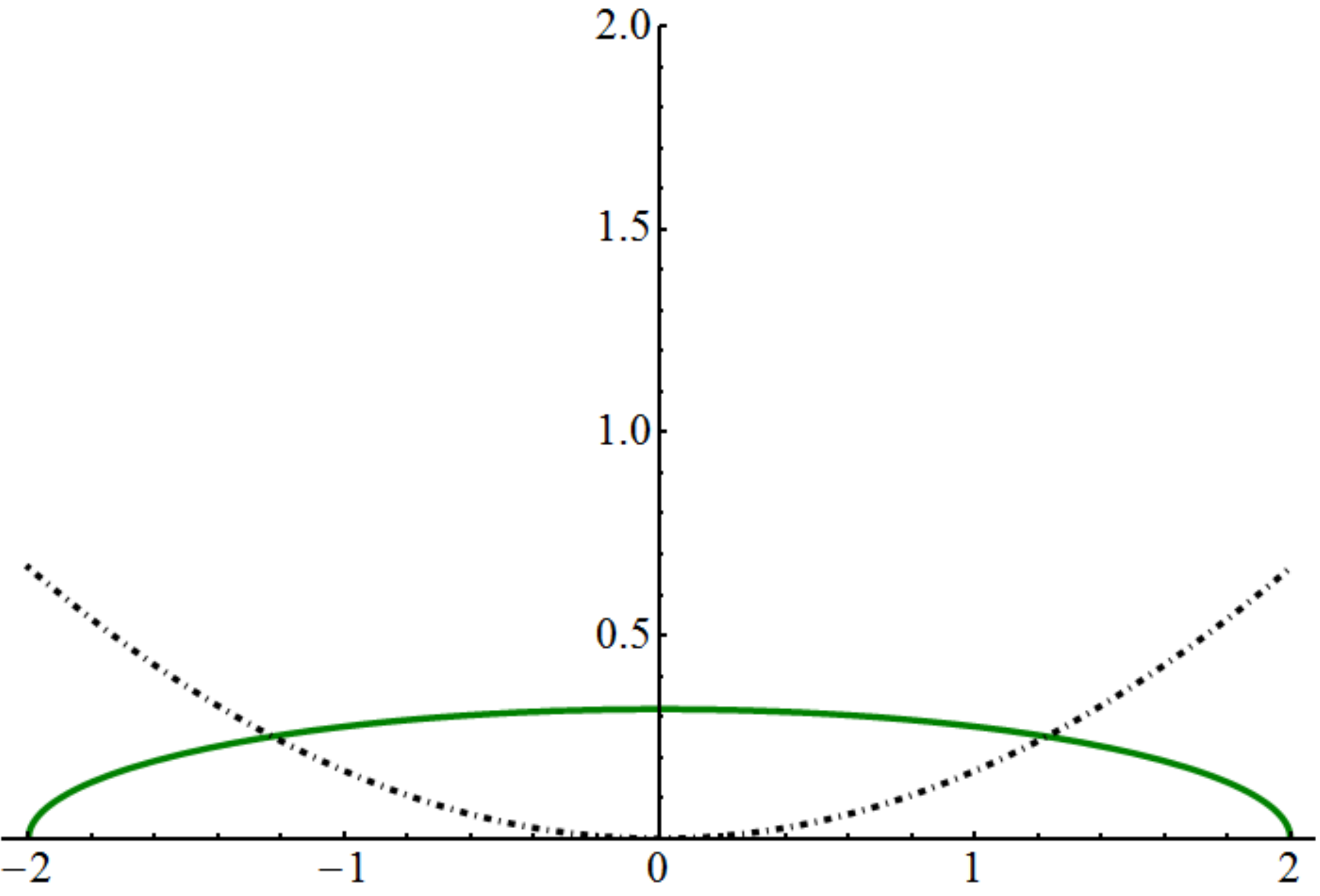}
			\includegraphics[width=0.4\textwidth]{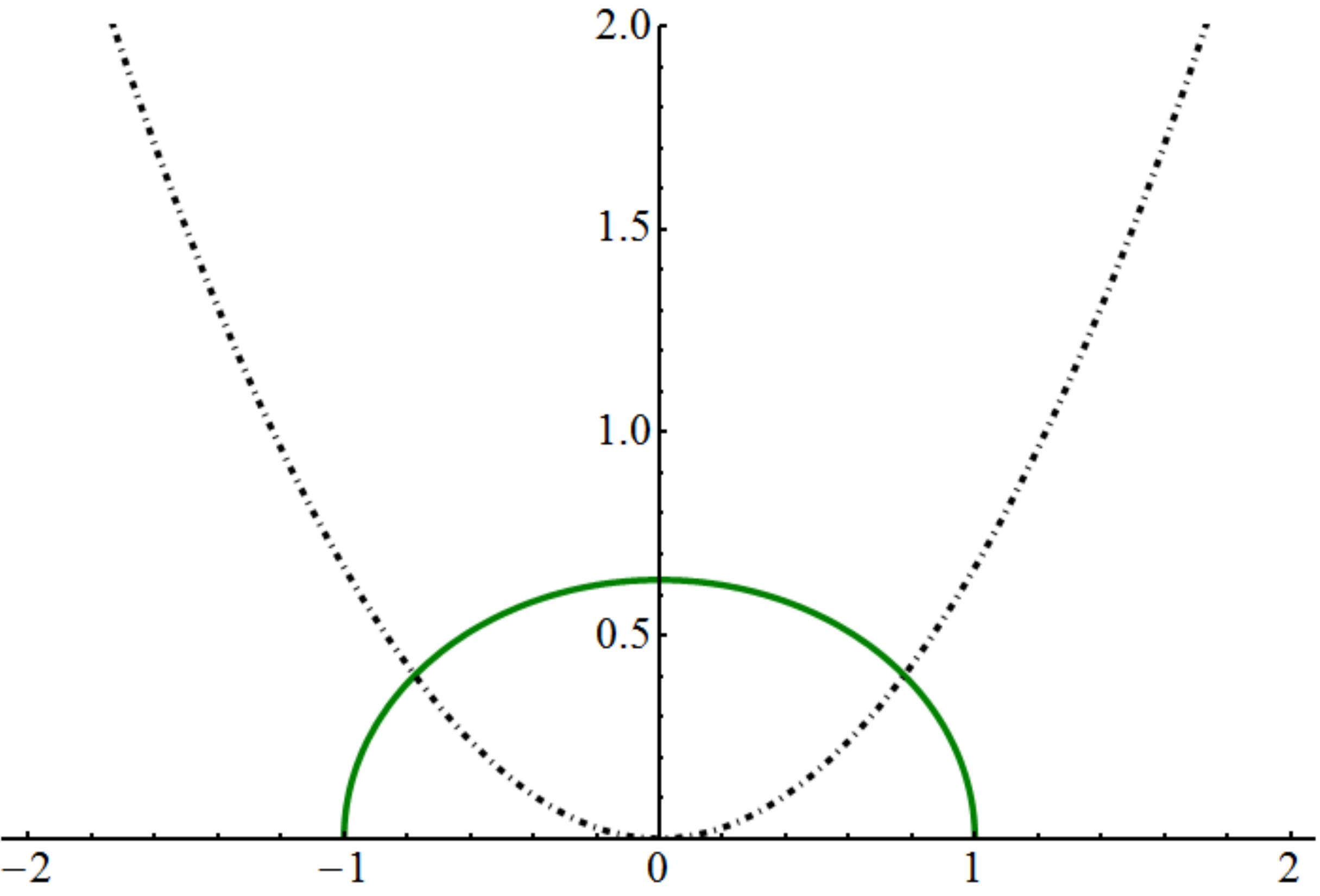}\\
			\includegraphics[width=0.4\textwidth]{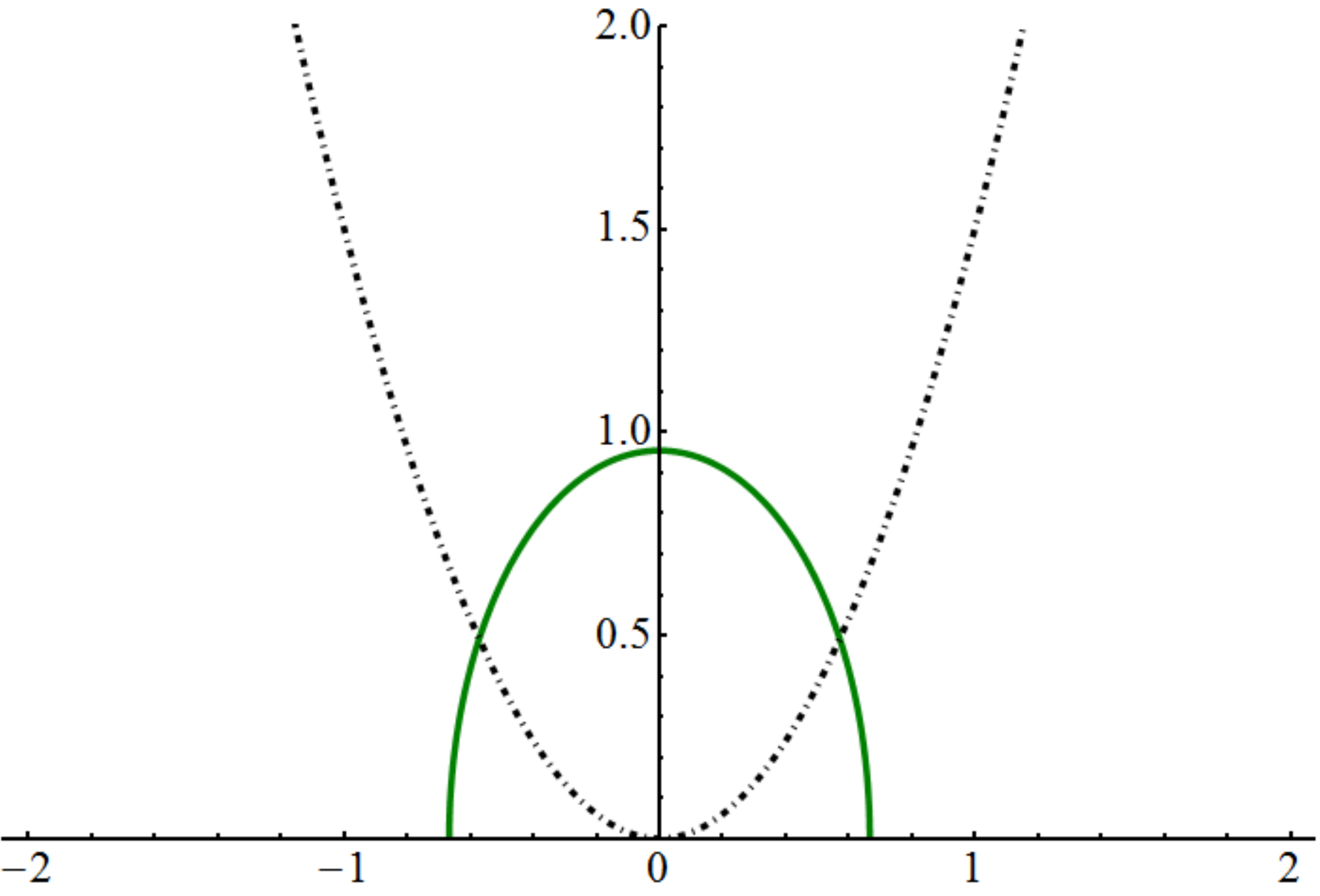}
			\includegraphics[width=0.4\textwidth]{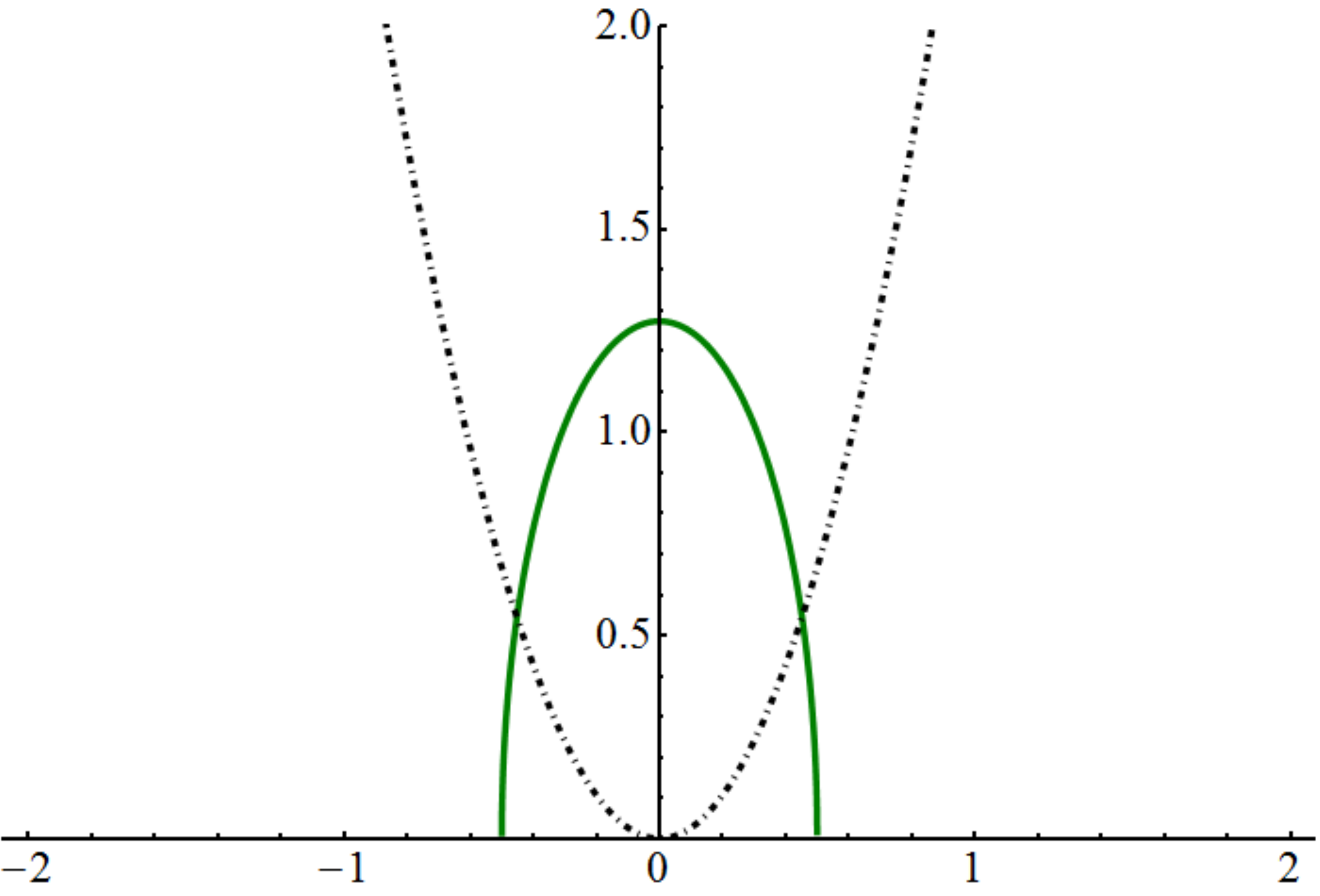}\\
			\includegraphics[width=0.4\textwidth]{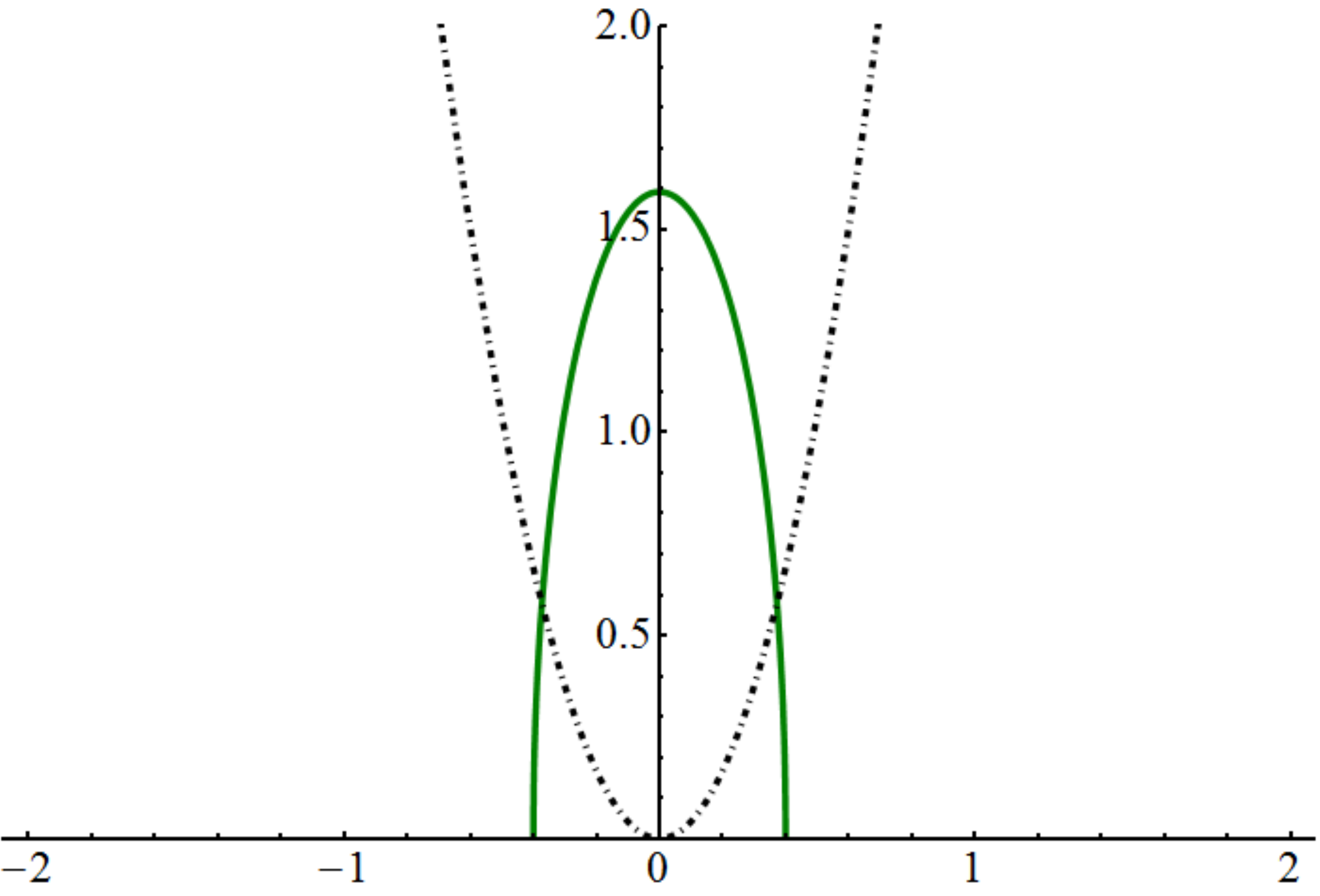}
			\includegraphics[width=0.4\textwidth]{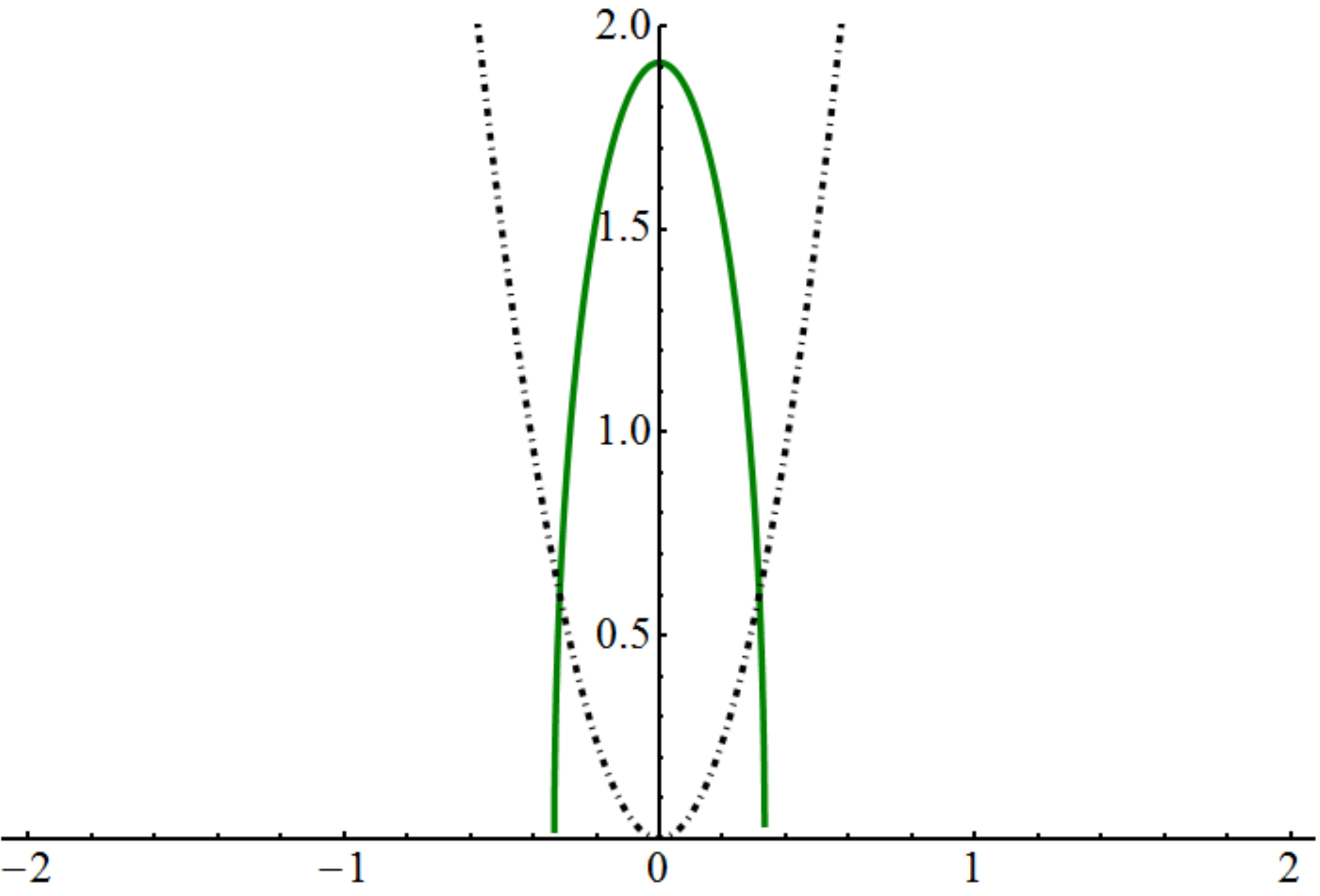}
	\caption{(Color online)  The semicircle distribution (\ref{22semicirc}) in solid green, together with the corresponding potential, in dot-dashed, for vales of $r=\{1,4,9,16,25,36\}$, left to right, top to bottom. The horizontal axis represents the eigenvalue, the vertical either the potential or the distribution. We have introduced a green color for the one cut solution.}
       \label{fig22wigner}
	\end{center}
	\end{figure}

On a little different front, with such scaling, the second moment of the distribution is then given by
\be
c_2=\frac{r a^4}{16}=\frac{1}{r}
\ee
and reflects the fact that the distribution is getting narrower as we increase the $r$. This is on the other hand very intuitive in the gas analogy, where the steeper potential confines the eigenvalues in a smaller region. The figure \ref{fig22wigner} illustrates the distribution for several values of $r$.

%% file: 2_23quartic.tex
\subsubsection{The quartic potential}

We now introduce a simple interaction potential $W(\lambda)=g \lambda^4$, which corresponds to the term $g\trl{M^4}$ in the measure and we have
\be\label{22actionquartic}
S(M)=\half r \trl{M^2}+g\trl{M^4}\ .
\ee

\podnadpis{Positive $r$}

Again, the condition (\ref{3.27a}) or (\ref{22polyres}) yields 
\be\label{22dificult}
\frac{1}{8} (a + b) (5 a^2 g - 2 a b g + 5 b^2 g + 2 r)=0\ ,
\ee
which has a clear solution $a=-b=\sqrt{\delta}$. In terms of this the second condition gives
\begin{eqnarray}\label{22quartcond1}
	\frac{3}{4} \delta^2 g + \frac{1}{4}\delta r=1\ .
\end{eqnarray}
The radius of the distribution is then given by
\begin{eqnarray}\label{22quarticdelta}
	\delta=\frac{1}{6g}\lr{\sqrt{r^2+48g}-r}.
\end{eqnarray}
Equation (\ref{3.29}) or (\ref{22polyM}) then yields 
\be
M(z)=4gz^2+2g\delta+r
\ee
and we finally obtain
\begin{eqnarray}
	\omega_0(z)=\half\slr{4 g z^3+r z-(4gz^2+2g\delta+r)\sqrt{z^2-\delta}},
\end{eqnarray}
and
\begin{eqnarray}
	\rho(\lambda)=\frac{1}{2\pi}\lr{r+2g \delta+4g\lambda^2}\sqrt{\delta-\lambda^2}\ , \ \lambda^2<\delta.
	\label{3.38}
\end{eqnarray}
The free energy of this distribution is given by
\be
\F=\frac{-r^2 \delta^2 + 40 r \delta}{384} - \half\log\lr{\frac{\delta}{4}} + \frac{3}{8}\ .
\ee
These results were first obtained in \cite{brezin}. Again, this distribution is normalized to $1$. If we used the normalization to $N$, the result for the square of the radius would be
\begin{eqnarray}
	\delta=\frac{1}{6g}\lr{\sqrt{r^2+48gN}-r}
\end{eqnarray}
and it is left as an exercise, that this leads to the same scaling as the general formulas (\ref{22generalscaling}).

The moments of the distribution can then be deducted from the $1/z$ expansion (\ref{3.33}) or by explicit integration and are
\begin{subequations}\label{2momentsphi4}
\begin{align}
c_2=&\frac{1}{4}\delta^3 g + \frac{1}{16}\delta^2 r=\frac{\delta}{4}+\frac{\delta^3 g}{16}\ ,\label{2momentsphi4a}\\
c_4=&\frac{9}{64} \delta^4 g +\frac{1}{32}\delta^3 r=\frac{\delta^2}{8}+\frac{3 \delta^4 g}{64}\ .
\end{align}
\end{subequations}

For positive $r$, this is the end of the story. 

\podnadpis{Negative $r$}

However if we allow for negative $r$, things change. Potential has then two minima and we expect a two cut solution to emerge. However not immediately for any value of $r$, since the potential price eigenvalues have to pay for being close to the origin has to be large enough to overcome the repulsion of eigenvalues. In other words, the peak that the potential develops at the origin has to he high enough to split the eigenvalues apart.

There are two ways to treat this situation. First, we can look at the one cut distribution (\ref{3.38}) and see, that for some value of $r$, the distribution becomes negative. This indicates that something is going wrong and we have to start over. Clearly, the distribution becomes negative at ${x=0}$, so the condition becomes ${M(0)=0}$ and
\be\label{22agreeent}
r+\frac{1}{3}\lr{\sqrt{r^2+48g}-r}=0\ \Rightarrow\ r=-4\sqrt g\ .
\ee
This tells us, where the behavior of the solution should change, but does not tell us what the new solution is. 

\podnadpis{Two cut solution}

The second approach is to directly look for this solution. To find it, we look for a symmetric two cut solution with the support
\be
\C=\lr{-\sqrt{D+\delta},-\sqrt{D-\delta}}\cup\lr{\sqrt{D-\delta},\sqrt{D+\delta}}\ .
\ee
The conditions (\ref{3.27}),(\ref{22polyres}) become
\bse\label{33cond2cut}
\bal
4Dg+r=&0\ ,\\
\delta^2=&\frac{1}{g}
\end{align}
\ese
and are trivial to solve. For the solution to be well defined, we must have ${D-\delta>0}$, which yields
\be
r<-4\sqrt{g}\ ,
\ee
in agreement with (\ref{22agreeent}). The polynomial ${M(z)}$ becomes simply
\be
M(z)=4g|z|
\ee
and eigenvalue distribution is then
\begin{eqnarray}\label{22twocut}
	\rho(\lambda)=\frac{2 g |\lambda|}{\pi}\sqrt{\lr{\delta^2-(D-\lambda^2)^2}}\ ,
\end{eqnarray}
which has the following free energy
\be 
\F=-\frac{r^2}{16g} + \frac{1}{4}\log g + 3/8
\ee
and the following second and fourth moment
\bse\label{22moments2cut}
\bal
c_2=&D \delta^2 g=D\ ,\label{22moments2cuta}\\
c_4=&D^2 \delta^2 g + \frac{1}{4}\delta^4 g=D+\frac{1}{4g}\ .
\end{align}
\ese
It is interesting to note, that at the transition point
\be\label{22basictrln}
r=-4\sqrt g\ ,
\ee the distributions (\ref{3.38}) and (\ref{22twocut}) coincide, as do the moments (\ref{2momentsphi4}) and (\ref{22moments2cut}).

	\begin{figure}[!tb]
       \begin{center}
    \vspace*{-0.3cm}
      \includegraphics[width=0.4\textwidth]{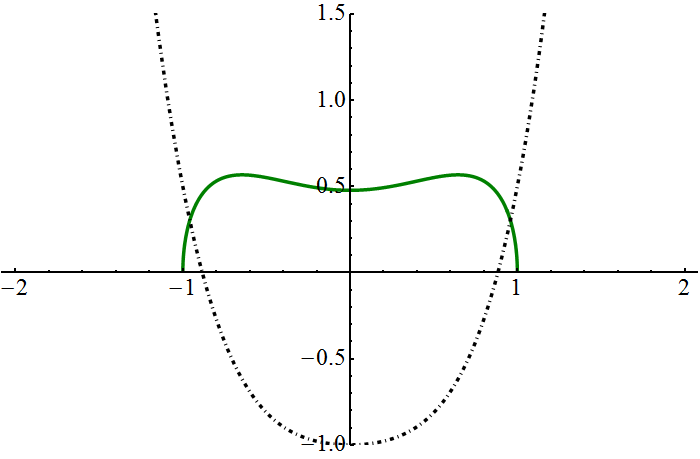}
			\includegraphics[width=0.4\textwidth]{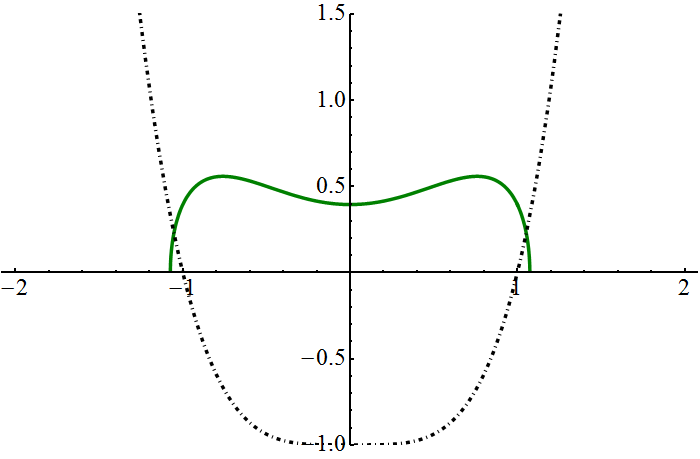}\\
			\includegraphics[width=0.4\textwidth]{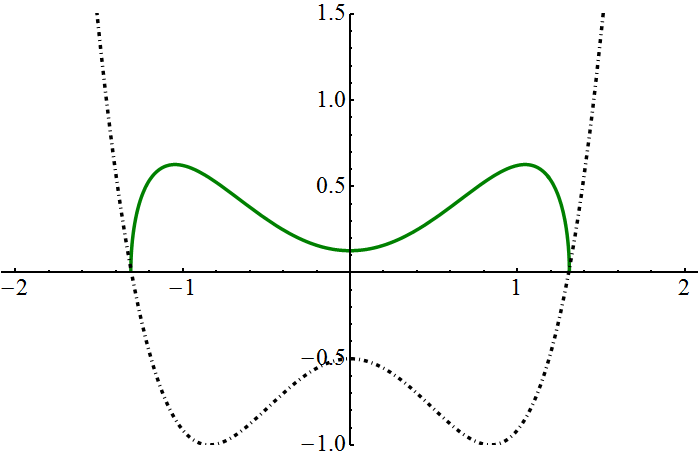}
			\includegraphics[width=0.4\textwidth]{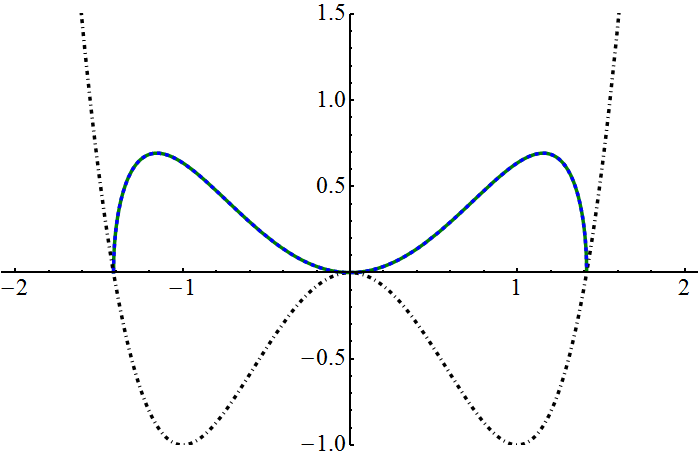}\\
			\includegraphics[width=0.4\textwidth]{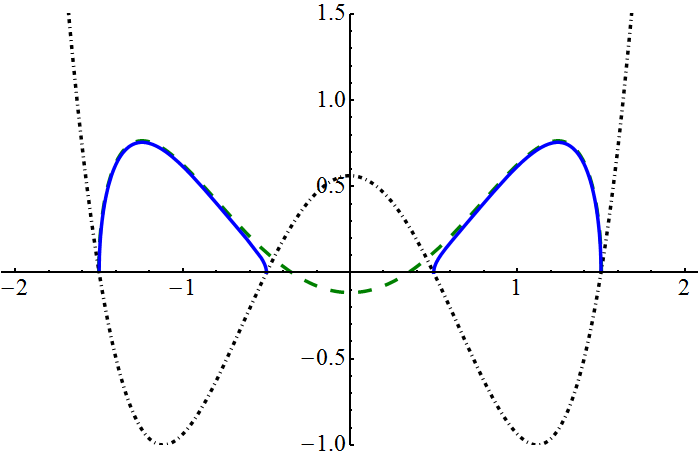}
			\includegraphics[width=0.4\textwidth]{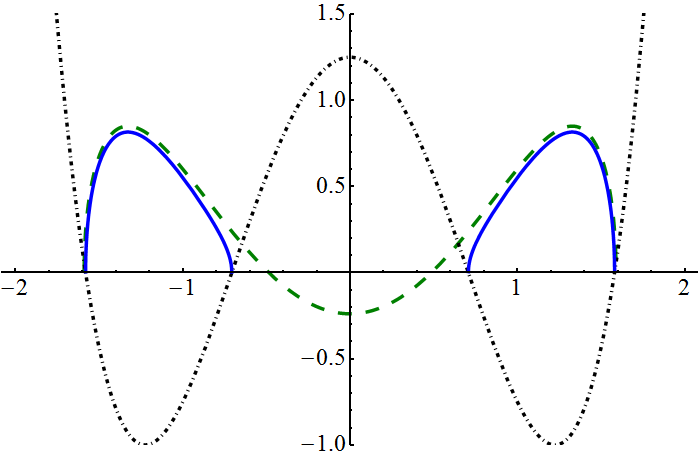}\\
			\includegraphics[width=0.4\textwidth]{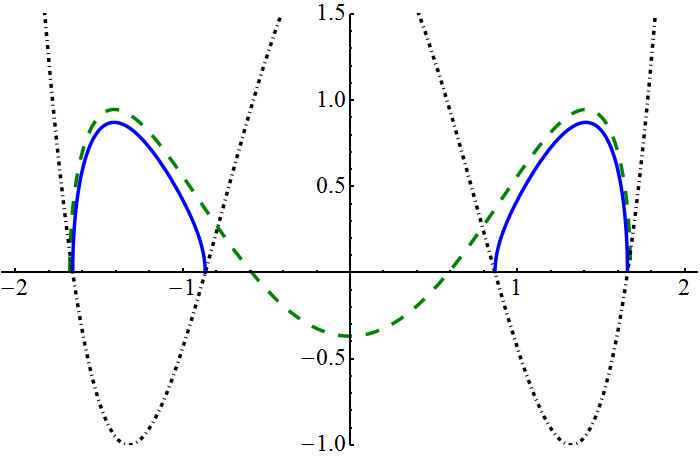}
			\includegraphics[width=0.4\textwidth]{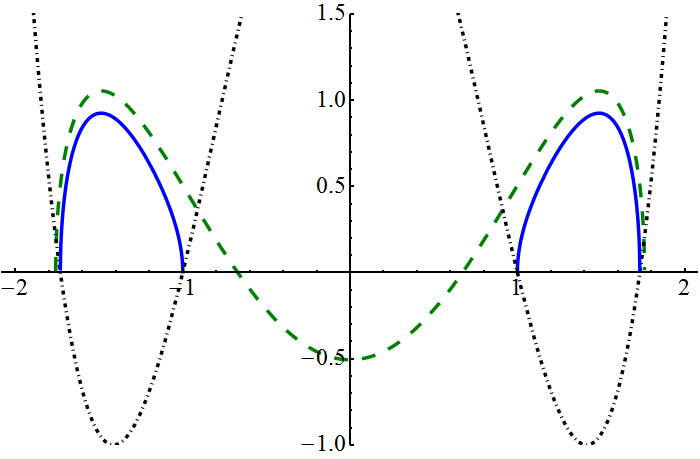}
    \vspace*{-0.3cm}
	\caption{(Color online) The comparison of the one cut and the two cut solutions to the quartic model (\ref{22actionquartic}). The plots are for the value $g=1$ and from right to left, top to bottom $r=\{1,0,-4/\sqrt{2},-4,-5$, $-6$, $-7$, $-8\}$. We can see that beyond $r=-4\sqrt{g}$ the one cut solution becomes negative and that at this point the one cut and the two cut solutions coincide. We have introduced a blue color for the two cut solution. Note that the potential, the dot-dashed line, has been shifted so that the minimum is at the bottom of the picture and not lower. The horizontal axis represents the eigenvalue, the vertical either the potential or the distribution. The solid line is the realized solution, the dashed line is the formal one cut solution for $r<-4\sqrt{g}$.}
    \vspace*{-1.0cm}
       \label{fig22quarticdistr}
	\end{center}
	\end{figure}

The figure \ref{fig22quarticdistr} shows the eigenvalue distribution for $g=1$ and several different values of $r$, together with the potential. We can observe the interval being split by the peak.

\podnadpis{Asymmetric one solution}

However, as it turns out this is still not the end of the story. It is difficult to see from the equation (\ref{22dificult}), but the model does have a different one cut solution \cite{shimishimi}. 

This can be seen in the particle gas analogy. The particles repel each other so generally if we place all of them in one of the wells they spread. Potential confines the particles and it is not hard to imagine a situation, where the wells of the potential are steep enough to win over the repulsion of the eigenvalues before they start to leak over the barrier.

If we rewrite the boundaries of the interval a little, namely
\be\label{22generalonecut}
\C=\lr{D-\sqrt\delta,D+\sqrt\delta}\ ,
\ee
the condition (\ref{22polyres}) becomes
\be
D\lr{2 D^2 g + 3 \delta g + \half D r}=0\ ,
\ee
which has apart from the symmetric solution $D=0$ an asymmetric solution
\bse\label{22asymonecutconditions}
\be
2 D^2 g + 3 \delta g + \half D r=0\ .
\ee
This is supplemented by the second condition form (\ref{22polyres})
\be
3 D^2 \delta g + \frac{3}{4} \delta^2 g + \frac{1}{4}\delta r=1\ .
\ee
\ese
The polynomial $M(z)$ (\ref{22polyM}) is 
\be\label{23Masym1cut}
M(z)=\ 4 D^2 g + 2 \delta g + r + 4 D g z + 4 g z^2\ ,
\ee
and the solution of the equations (\ref{22asymonecutconditions}) is given by
\be
\delta=\frac{-r-\sqrt{-60g+r^2}}{15 g}\ ,\ D=\pm\sqrt{\frac{-3r+2\sqrt{-60g+r^2}}{20 g}}\ 
\ee
and there is plenty to be noticed about it. First, there are two solutions with the same $\delta$ and opposite $D$, which says that the distribution can live in either well of the potential. Second, it exist only for negative $r$, as $\delta$ has to be positive. And since both $D$ and $\delta$ have to be real, we obtain ${r<-2\sqrt{15 g}}$. Finally, for $|r|$ very large, we get
\be\label{22asymbound}
\delta=0\ , \ D=\pm\half\sqrt{\frac{-r}{g}}\ ,
\ee
which is easily seen to be the location of the minima of the potential. And as expected, in the limit ${r\to-\infty}$, as the walls of the potential well become very steep, the eigenvalues become localized at its bottom.

And finally, the formula for the distribution itself is
\bal\label{22quarasym}
\rho(\lambda)=&\frac{1}{2\pi}\lr{4 D^2 g+ 4 D g x + 2 \delta g + r  + 4 g x^2}\sqrt{\lr{D+\sqrt{\delta}-x}\lr{x-D+\sqrt{\delta}}}=\no=&\frac{1}{2\pi}\lr{4 D^2 g+ 4 D g x + 2 \delta g + r  + 4 g x^2}\sqrt{\delta-\lr{x-D}^2}
\end{align}
and the free energy of this solution is given by the rather unappealing formula
\bal
\F=&\frac{\delta}{8}\Bigg[\frac{9}{16}g^2 \delta^3+\delta^2\lr{\frac{5}{8}gr+\frac{27}{2}g^2D^2}
+\delta\lr{33g^2D^2+\frac{15}{2}g r D^2+\frac{r^2}{8}+\frac{15}{g}}\no
&\hspace*{-0.4cm}+15 g^2 D^6+7gr D^4+\lr{12g+\half r^2}D^2+3 r\Bigg]+\frac{1}{4}r D^2+\half g D^4-\half \log\lr{\frac{\delta}{4}}-1
\end{align}

\begin{figure}[!tb]
       \begin{center}
      \includegraphics[width=0.4\textwidth]{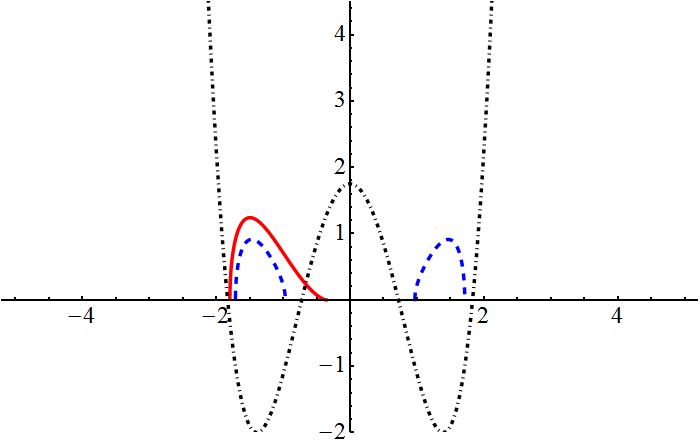}
			\includegraphics[width=0.4\textwidth]{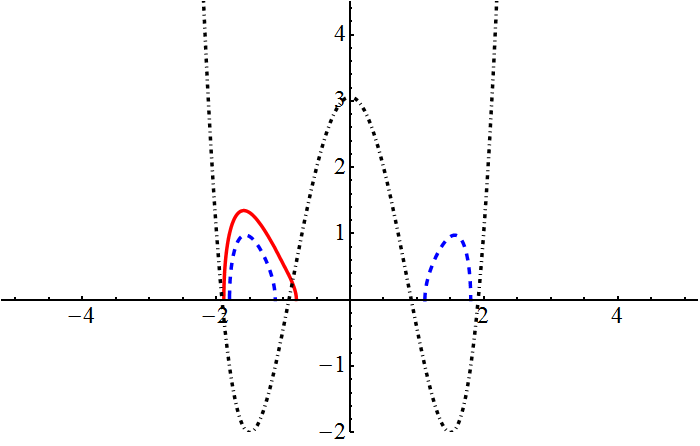}\\
			\includegraphics[width=0.4\textwidth]{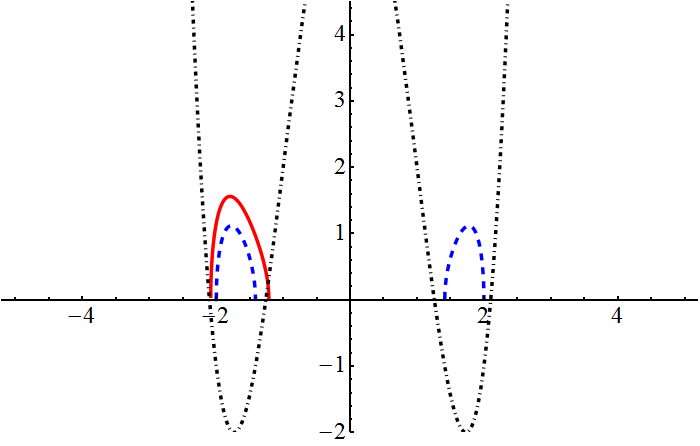}
			\includegraphics[width=0.4\textwidth]{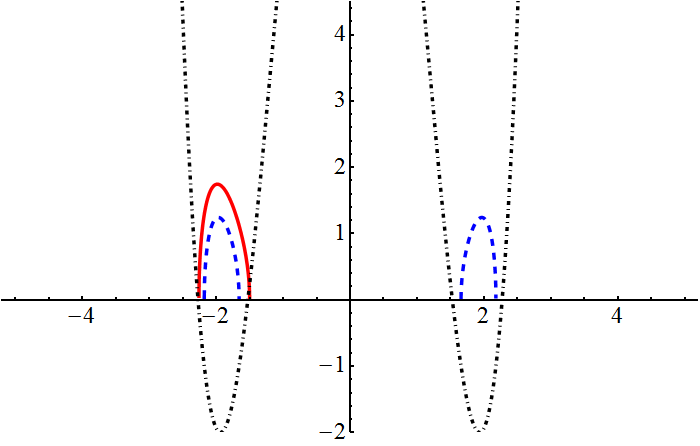}\\
			\includegraphics[width=0.4\textwidth]{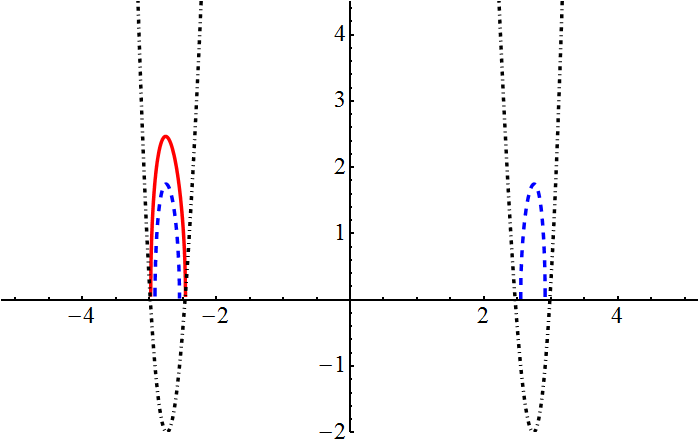}
			\includegraphics[width=0.4\textwidth]{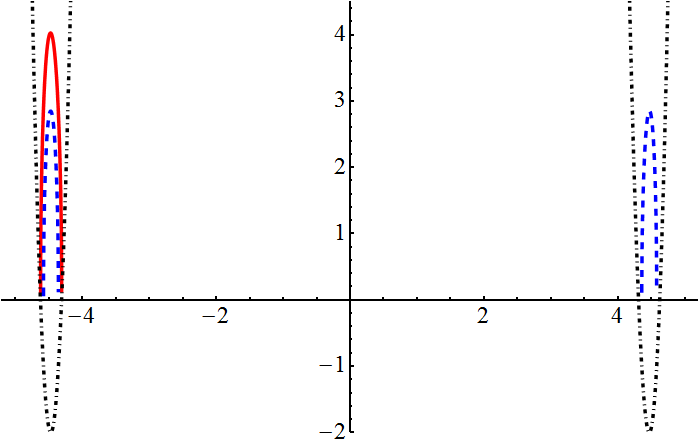}
	\caption{(Color online) Plot of the asymmetric one cut solution (\ref{22quarasym}) to the model (\ref{22actionquartic}) for the value $g=1$ and for left to right, top to bottom $r=\{-2\sqrt{15},-9,-12,-15,-30,-80\}$, together with (a part of) the potential, shown in dot-dashed. Note that the potential has been shifted so that the minimum is at the bottom of the picture and not lower. The horizontal axis represents the eigenvalue, tho vertical either the potential or the distribution. We have introduced the red color for the asymmetric one cut solution. The dashed line is the two cut solution, which is energetically preferred.}
       \label{fig22quartiasymcdistr}
	\end{center}
	\end{figure}

In the figure \ref{fig22quartiasymcdistr}, the asymmetric one cut solution is plotted, together with the two cut solution, for the value of $g=1$.

It is interesting to compute the asymmetric eigenvalue density just when it starts to be possible. We plug $r=-2\sqrt{15 g}$ into (\ref{22asymbound}) and obtain for the edges of the interval
\be
\pm \lr{\frac{1}{60g}}^{1/4}\ \textrm{and}\ \pm5\lr{\frac{1}{60g}}^{1/4}\ .
\ee
Note that the interval does not start at $0$, but at some depth into the well. This is easily understood in the particle gas analogy. The particle at the edge of the interval is being repelled by the rest of the particles. So there has to be a force acting on it in the opposite direction and thus it cannot sit at the top of the peak at $x=0$.

The last question to answer is, which of the two solutions that exist in the region $r<-2\sqrt{15 g}$ is eventually realized. This depends how we look at the model. We can either view the large $N$ limit as a process. Then $1/N^2$ plays the role of a temperature and the large $N$ limit is the limit of freezing the particles in the well. However if we assume that this process takes some "time", in the sense that the fluctuations of the eigenvalues can take one state from another sufficiently many times, the solution with the lowest free energy (\ref{22freeenergy}) is realized. Or in other words the more probable solution.

If we on the other hand assume that the system is in the ${N=\infty}$ state to begin with, there are no thermal fluctuations and the eigenvalues cannot get from one well into another even if it lowers the free energy. This is a view we are not going to take! First of all because we usually do have some limiting process in mind\footnote{In our case for example the process of taking the noncommutativity parameter to zero.} and because it just leaves room for too many solutions, as for example any two cut solution, once it exists, would be stable.

So we are to determine which of the two solutions, the two cut (\ref{22twocut}) and the asymmetric one cut (\ref{22quarasym}) has lower free energy in the region where both exist. Without summoning any explicit formulas for the free energy, it is easy to solve the dispute. In the two cut case, eigenvalues are further away from each other, so the energy of the interaction lowers. They are also in region of lower potential, so the potential energy lowers. The free energy is thus lower in the two cut case and the two cut solution is the preferred one everywhere.

%DOPLNIT SOLUTION FOR NEGATIVE G

\begin{figure}[tb]
       \begin{center}
       \includegraphics[width=0.7\textwidth]{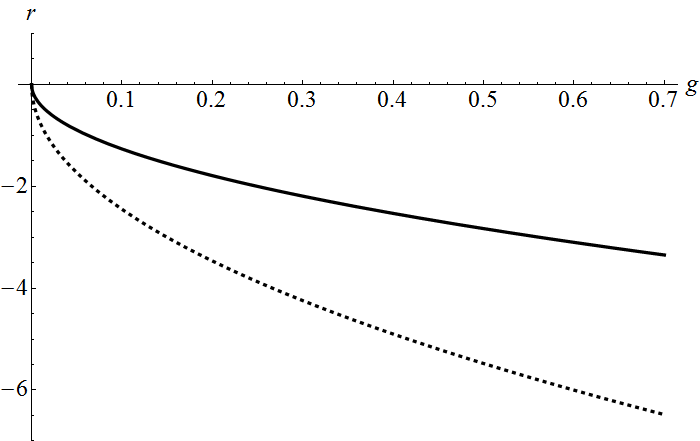}
	\caption{The phase diagram of the quartic model (\ref{22actionquartic}). Above the solid line, the eigenvalue distribution is in the one cut phase. Under the solid line, the eigenvalue distribution is in the two cut case. Under the dashed line, an asymmetric one cut phase exists, but is not realized as it has higher free energy.}
       \label{fig22quartiphasediag}
	\end{center}
	\end{figure}

\begin{figure}[tb]
       \begin{center}
       \includegraphics[width=0.8\textwidth]{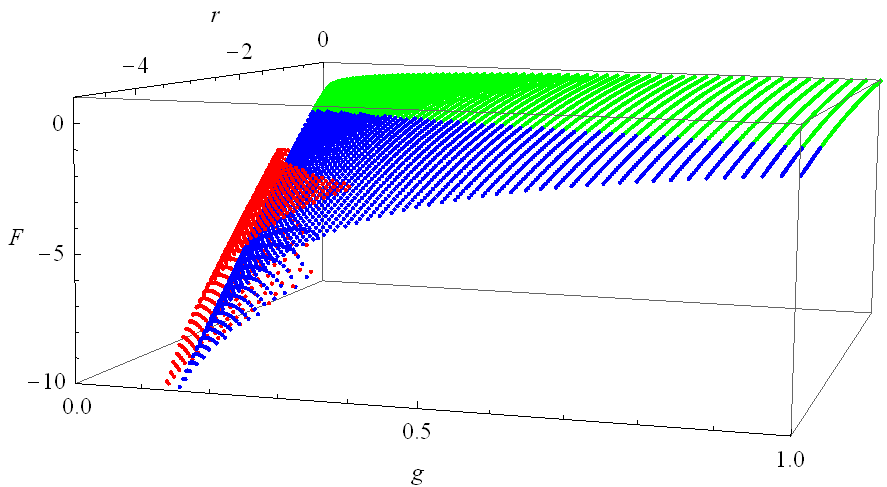}
	\caption{(Color online) The free energy diagram of the quartic model (\ref{22actionquartic}). The green region denotes the one cut solution, the blue region the two cut solution and the red region the asymmetric one cut solution. We can see that the asymmetric solution has higher free energy everywhere it exists.}
       \label{2_23quarticfreeenergy}
	\end{center}
	\end{figure}

So we got our first phase diagram! It is shown in the figure \ref{fig22quartiphasediag} and describes the phase structure of the quartic model (\ref{22actionquartic}). The figure \ref{2_23quarticfreeenergy} shows the free energies of the two three solutions, clearly indicating that the asymmetric one cut solution has always higher free energy.

We got only symmetric solutions as the energetically preferred solutions in this model. We can use the notion of freezing the solution to obtain a preferred asymmetric solution in the case of the symmetric potential (\ref{22actionquartic}).

We first introduce a symmetry breaking term of the form $\trl{M}$ or $\trl{M^3}$ which lowers one of the wells of the potential. If the term is strong enough, the asymmetric solution living in this well will become the preferred solution since its free energy will be the lowest. After we take the large-$N$ limit, we remove the symmetry breaking. This brings the two wells back to the same level again. However the eigenvalues are now frozen in one well and cannot get to the other one, obtaining an asymmetric solution.

We will not employ this strategy though. The matrix model we will need to study as a description of the fuzzy field theory will not allow to remove the symmetry breaking term, so we will have to study the asymmetric models completely. But before that, let us summarize what we have learned in this section.

%% file: 2_24lesons.tex
\subsubsection{Lessons learned}

We have seen that the particle picture for the eigenvalues is very useful. There are two forces on the particles, one from the outside potential which pushes the eigenvalues towards the minima and one from the interaction among eigenvalues which pushes them away from each other. These two forces compete, until the equilibrium is reached, where the net force on each particle vanishes.

If the potential has more than one minimum, there are sometimes more equilibria available, simply because the wells allow for more complicated stable situations. We then assume that the particles can find the configuration of the lowest free energy in the process of lowering temperature, which is equivalent to taking the large-$N$ limit.

And we have seen that if the potential is even, the resulting distribution is symmetric. To get an asymmetric solution, we need to add odd terms into the potential, which we are going to do in the next section.

%% file: 2_3asymquartic.tex
The most general quartic action for the matrix $M$ is
\be
S(M)=a \trl{M}+\half r \trl{M^2}+h\trl{M^3}+g\trl{M^4}\ .
\ee
With a shift and rescaling of the matrix and dropping an irrelevant constant term, we can bring this to the form
\be\label{23asymqrt}
S(M)=\trl{M}+\half r \trl{M^2}+g\trl{M^4}\ .
\ee
Such ensemble can be treated by the saddle point approximation in the same way we treated the symmetric case in the section \ref{sec22}.

\podnadpis{One cut solutions}

First, let us consider the one cut solution of this problem. We will discuss the two cut solution afterward. We have already encounter an ansatz for the general one cut support (\ref{22generalonecut})
\be
\C=(D-\sqrt\delta,D+\sqrt\delta)\ .
\ee
(\ref{22polyres}) now leads to equations
\begin{subequations}\label{23kuk}
\begin{align}
\half + 2 D^3 g + 3 D \delta g + \half D r=&0\ ,\label{23kuk1}\\
3 D^2 \delta g + \frac{3}{4} \delta^2 g + \frac{1}{4}\delta r=&1\ ,\label{23kuk2}
\end{align}
\end{subequations}
which are slightly more complicated than (\ref{22asymonecutconditions}). We invite the reader to show, that in the free case these equations give rise a distribution, which is the original semicircle shifted by $1/r$ to the left, which is exactly what we would expect.

The polynomial $M(z)$ (\ref{22polyM}) is the same as (\ref{23Masym1cut}), since it depends only on the form of the cut
\be\label{23onecutasymM}
M(z)=\ 4 D^2 g + 2 \delta g + r + 4 D g z + 4 g z^2\ .
\ee

For a general coupling, because of to the extra term $1/2$, the equations are not solvable anymore, as solving (\ref{23kuk1}) for $\delta$ turns (\ref{23kuk2}) into the equation of sixth order in $D$. The moments of the distribution are given in terms of the solution of these equations by
\begin{subequations}\label{2momentseffectiveasymmetric}
\begin{align}
c_1=&3 D^3 \delta g + \frac{3}{2} D \delta^2 g + \frac{1}{4}D \delta r\ ,\\
c_2=&3 D^4 \delta g + 3 D^2 \delta^2 g + \frac{1}{4}\delta^3 g +  \frac{1}{4} D^2 \delta r + \frac{1}{16}\delta^2 r\ ,\\
c_3=&3 D^5 \delta g + \frac{21}{4} D^3 \delta^2 g + \frac{9}{8} D \delta^3 g + 
 \frac{1}{4} D^3 \delta r + \frac{3}{16} D \delta^2 r\ ,\\
c_4=&3 D^6 \delta g + \frac{33}{4} D^4 \delta^2 g + \frac{27}{8} D^2 \delta^3 g + \frac{9}{64}
  \delta^4 g + \frac{1}{4} D^4 \delta r + \frac{3}{8} D^2 \delta^2 r + \frac{1}{32}\delta^3 r\ .
\end{align}
\end{subequations}

Every one cut solution of this model is an asymmetric one. However, we are going to distinguish two kinds of one cut solutions, which are rather different. We are going to them almost-symmetric solution and the asymmetric solution.

As suggested by the name, the almost-symmetric solution is not going to be too different from the symmetric solution (\ref{3.38}). We will define it by the condition that the extremum of the polynomial $M(z)$ is within the supporting interval. This means that the eigenvalue distribution will have two peaks, even though not of the same height.

The asymmetric solution will have this extremum outside of the supporting interval. This means that the eigenvalue distribution will have only one peak and all the eigenvalues will be located in one well only.

The extremum of $M$ given by (\ref{23onecutasymM}) occurs at $z=-D/2$, so in terms of the solution of (\ref{23kuk}), the solution is asymmetric if
\be
\frac{3}{2}D+\sqrt{\delta}<0\ .
\ee
To find the phase transition, we need to look for the moment when the asymmetric solution does not exists, i.e. when the distribution becomes negative. Since the extremum of the potential is outside the interval, this will happen at the edge of the interval and we get the condition for the boundary of existence of the asymmetric solution
\be\label{23asymphasecondition}
M(D+\sqrt\delta)=12 D^2 g + 12 D \sqrt{\delta} g + 6 \delta g + r=0\ .
\ee
For the almost-symmetric solution, the distribution becomes negative at the minimum of the potential, so the boundary is given by
\be
M(-D/2)=3 D^2 g + 2 \delta g + r=0\ .
\ee
These two equations, together with the conditions (\ref{23kuk}) determine the two boundary lines of the regions, where the two solutions exist. They cannot be solved analytically, but after some algebra $D$ and $\delta$ can be removed for the almost symmetric case to give a condition just in terms of $r$ and $g$
%\bse
\bal\label{23asymconditions}
45 \sqrt g -  \sqrt 15 \slr{4 r + \sqrt{-80 g + 9 r^2}}\sqrt{-2 r +    \sqrt{-80 g + 9 r^2}}=0\ ,
\end{align}
%\ese
This condition can now be solved either numerically or perturbatively.

\begin{figure}[!tb]
       \begin{center}
       \includegraphics[width=0.8\textwidth]{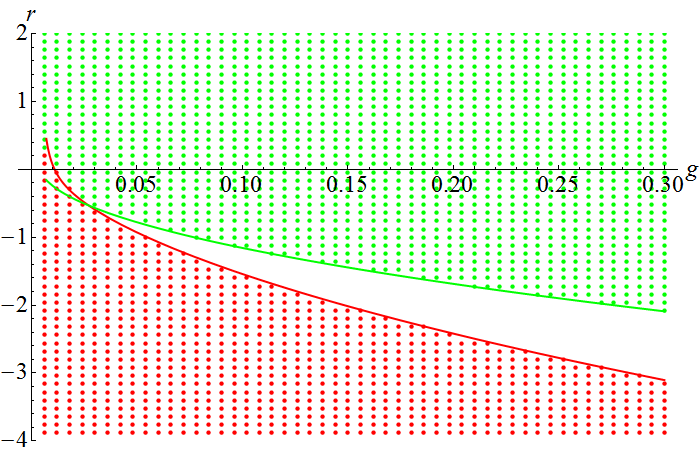}
	\caption{(Color online) The numerical phase diagram of the asymmetric quartic model (\ref{23asymqrt}). The red region denotes the asymmetric one cut solution and the red line the perturbative boundary line (\ref{23bound2}). The green region denotes the almost symmetric one cut solution and the green line the perturbative boundary line (\ref{bound1}). Note the difference in the $g$ axis.}
       \label{fig22asymquartiphasediag}
	\end{center}
	\end{figure}

When doing the perturbative calculation for either case, we introduce a factor of $\ep$ in front of the linear term, see where the $\ep$ propagates into (\ref{23asymconditions}) and look for the solution as a power series in $\ep$. The result is
\be\label{bound1}
r(g)=\frac{3}{32} + \frac{2343}{8388608 g} + \frac{123}{32768 \sqrt g} - 4 \sqrt g
\ee
for the almost symmetric solution and
\bal\label{23bound2}
r(g)=&- 2 \sqrt{15} \sqrt g+ \frac{15^{1/4} g^{1/4}}{\sqrt{2}} +\frac{1}{12}+ \frac{199}{5184 \sqrt{2} 15^{1/4} g^{1/4}}+ \frac{733}{62208 \sqrt{15} \sqrt g}+\no
&+ \frac{49807}{ 5971968 \sqrt{2} 15^{3/4} g^{3/4}}+\frac{ 2431}{11337408 g}+ \frac{244091717}{ 1393140695040 \sqrt{2} 15^{1/4} g^{5/4}}
\end{align}
for the asymmetric solution. These two lines are shown in the numerical phase diagram of the theory in the figure \ref{fig22asymquartiphasediag}. Note that they intersect and beyond this intersection there is no true distinction between the two solutions. We will see this better once we plot the solutions a little later. In the empty region, only a two cut solution exists.

\begin{figure}[!tb]
       \begin{center}
       \includegraphics[width=0.6\textwidth]{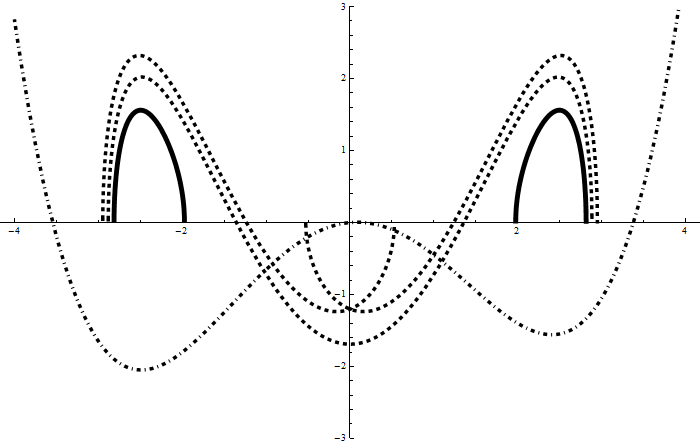}
	\caption{All the real solutions to equations (\ref{23kuk}) obtained numerically at ${r=-12,g=1/2}$. The dot-dashed line is the potential, the dashed lines are the negative solutions and are not well defined. The solid solutions are well defined but the solution in the right well has higher free energy.}
       \label{fig23problem}
	\end{center}
	\end{figure}

To obtain this phase diagram of the one cut part of the problem numerically, we have chosen particular values of $r$ and $g$, solve the equations (\ref{23kuk}) with these particular values. From the numerical point of view, it is rather challenging to navigate one's way around a slew of solutions. For example figure \ref{fig23problem} shows all the numerical solution one obtains for the values ${r=-12,g=1/2}$. Once drawn, it is clear which solution is the correct one. It is however useful to be able to determine the correct solution just from the values of $D$ and $\delta$. Once all the imaginary results and negative $\delta$'s are thrown away, it is the solution with $D$ closest to the minimum of the potential.

We have exhausted the discussion of the one cut solution and will proceed to the description of the two cut solution. It has to be the solution of the model in the blank region in the figure \ref{fig22asymquartiphasediag}, but as we will see it exists also outside this area.

\podnadpis{Two cut solution}

The ansatz for a general two cut solution is
\be
\C=(D_1-\delta_1,D_1+\delta_1)\cup(D_2-\delta_2,D_2+\delta_2)\ .
\ee
The three conditions we get from (\ref{22polyres}), recall that it is the condition ${\omega(z)\sim 1/z}$ for large $z$, are then
\begin{subequations}\label{23conditions}
\begin{align}
2 D_1^2 g + 2 D_1 D_2 g + 2 D_2^2 g + \delta_1^2 g + \delta_2^2 g +\half r&=0\ ,\\
\half - 2 D_1^2 D_2 g - 2 D_1 D_2^2 g + 2 D_1 \delta_1^2 g + 
 2 D_2 \delta_2^2 g&=0\ ,\\
2 D_1^2 \delta_1^2 g - D_1 D_2 \delta_1^2 g - D_2^2 \delta_1^2 g + \frac{1}{4}
 \delta_1^4 g- D_1^2 \delta_2^2 g-&\no - D_1 D_2 \delta_2^2 g + 
 2 D_2^2 \delta_2^2 g - \half \delta_1^2 \delta_2^2 g + \frac{1}{4}\delta_2^4 g&=1\ .
\end{align}
\end{subequations}
The polynomial $M(z)$ is given by
\be
M(z)=4\lr{D_1+D_2+g}
\ee
and the distribution becomes
\be\label{23asymdistr}
	\rho(\lambda)=\frac{2|D_1+D_2+g|}{\pi}\sqrt{\lr{\delta_2^2-(D_1-\lambda)^2}\lr{\delta_2^2-(D_2-\lambda)^2}}\ ,
\ee
The moments of the distribution are given by
\begin{subequations}\label{2asymconditions}
\begin{align}\label{2asym2cutmoments}
c_1=&
2 D_1^3 \delta_1^2 g - D_1^2 D_2 \delta_1^2 g - D_1 D_2^2 \delta_1^2 g + D_1 \delta_1^4 g - D_1^2 D_2 \delta_2^2 g\no&
 - D_1 D_2^2 \delta_2^2 g 
+ 2 D_2^3 \delta_2^2 g - D_1 \delta_1^2 \delta_2^2 g -  D_2 \delta_1^2 \delta_2^2 g + D_2 \delta_2^4 g\ ,\\
c_2=&
2 D_1^4 \delta_1^2 g 
- D_1^3 D_2 \delta_1^2 g 
- D_1^2 D_2^2 \delta_1^2 g 
+  \frac{9}{4} D_1^2 \delta_1^4 g 
- \frac{1}{4} D_1 D_2 \delta_1^4 g \no&
-  \frac{1}{4} D_2^2 \delta_1^4 g 
+ \frac{1}{8}\delta_1^6 g) 
- D_1^2 D_2^2 \delta_2^2 g 
-  D_1 D_2^3 \delta_2^2 g 
+ 2 D_2^4 \delta_2^2 g \no&
 -D_1^2 \delta_1^2 \delta_2^2 g 
- \frac{3}{2} D_1 D_2 \delta_1^2 \delta_2^2 g 
-  D_2^2 \delta_1^2 \delta_2^2 g 
- \frac{1}{8} \delta_1^4 \delta_2^2 g \no&
-  \frac{1}{4} D_1^2 \delta_2^4 g 
- \frac{1}{4} D_1 D_2 \delta_2^4 g 
+  \frac{9}{4} D_2^2 \delta_2^4 g 
- \frac{1}{8} \delta_1^2 \delta_2^4 g 
+ \frac{1}{8}\delta_2^6 g)\ ,\\
c_3=&2 D_1^5 \delta_1^2 g 
- D_1^4 D_2 \delta_1^2 g 
- D_1^3 D_2^2 \delta_1^2 g 
+  4 D_1^3 \delta_1^4 g 
- \frac{3}{4} D_1^2 D_2 \delta_1^4 g \no&
-  \frac{3}{4} D_1 D_2^2 \delta_1^4 g 
+ \frac{3}{4} D_1 \delta_1^6 g 
- D_1^2 D_2^3 \delta_2^2 g 
- D_1 D_2^4 \delta_2^2 g 
+ 2 D_2^5 \delta_2^2 g \no&
-  D_1^3 \delta_1^2 \delta_2^2 g 
- \frac{3}{2} D_1^2 D_2 \delta_1^2 \delta_2^2 g 
-  \frac{3}{2} D_1 D_2^2 \delta_1^2 \delta_2^2 g 
- D_2^3 \delta_1^2 \delta_2^2 g 
- \frac{1}{2} D_1 \delta_1^4 \delta_2^2 g \no&
- \frac{1}{4} D_2 \delta_1^4 \delta_2^2 g 
-  \frac{3}{4} D_1^2 D_2 \delta_2^4 g 
- \frac{3}{4} D_1 D_2^2 \delta_2^4 g 
+ 4 D_2^3 \delta_2^4 g 
- \frac{1}{4} D_1 \delta_1^2 \delta_2^4 g \no&
-  \frac{1}{2} D_2 \delta_1^2 \delta_2^4 g 
+ \frac{3}{4} D_2 \delta_2^6 g\ ,\\
c_4=&
2 D_1^6 \delta_1^2 g 
- D_1^5 D_2 \delta_1^2 g 
- D_1^4 D_2^2 \delta_1^2 g 
+  \frac{25}{4} D_1^4 \delta_1^4 g 
- \frac{3}{2} D_1^3 D_2 \delta_1^4 g\no& 
-  \frac{3}{2} D_1^2 D_2^2 \delta_1^4 g 
+ \frac{5}{2} D_1^2 \delta_1^6 g 
- \frac{1}{8} D_1 D_2 \delta_1^6 g 
- \frac{1}{8} D_2^2 \delta_1^6 g 
+ \frac{5}{64} \delta_1^8 g \no&
-  D_1^2 D_2^4 \delta_2^2 g 
- D_1 D_2^5 \delta_2^2 g 
+ 2 D_2^6 \delta_2^2 g 
-  D_1^4 \delta_1^2 \delta_2^2 g 
- \frac{3}{2} D_1^3 D_2 \delta_1^2 \delta_2^2 g \no&
- \frac{3}{2} D_1^2 D_2^2 \delta_1^2 \delta_2^2 g 
-  \frac{3}{2} D_1 D_2^3 \delta_1^2 \delta_2^2 g 
- D_2^4 \delta_1^2 \delta_2^2 g 
-  \frac{9}{8} D_1^2 \delta_1^4 \delta_2^2 g 
- \frac{7}{8} D_1 D_2 \delta_1^4 \delta_2^2 g \no&
-\frac{1}{4} D_2^2 \delta_1^4 \delta_2^2 g 
- \frac{1}{16} \delta_1^6 \delta_2^2 g 
-  \frac{3}{2} D_1^2 D_2^2 \delta_2^4 g 
- \frac{3}{2} D_1 D_2^3 \delta_2^4 g 
+  \frac{25}{4} D_2^4 \delta_2^4 g \no&
- \frac{1}{4} D_1^2 \delta_1^2 \delta_2^4 g 
- \frac{7}{8} D_1 D_2 \delta_1^2 \delta_2^4 g 
- \frac{9}{8} D_2^2 \delta_1^2 \delta_2^4 g
 -  \frac{1}{32} \delta_1^4 \delta_2^4 g 
- \frac{1}{8} D_1^2 \delta_2^6 g \no&
-  \frac{1}{8} D_1 D_2 \delta_2^6 g 
+ \frac{5}{2} D_2^2 \delta_2^6 g
- \frac{1}{16} \delta_1^2 \delta_2^6 g 
+  \frac{5}{64} \delta_2^8 g\ .
\end{align}
Finally, the condition for the minimal free energy (\ref{twocutminimal}) for the asymmetric two cut solution (\ref{23asymdistr}) becomes
\begin{align}\label{ap1condition}
0=&\Big(-4 D_1^5+12 D_1^4 D_2-8 D_1^3 D_2^2-8 D_1^2 D_2^3+12 D_1 D_2^4-4 D_2^5-10 D_1^3 \delta_1^2\no&
+22 D_1^2 D_2 \delta_1^2-14 D_1 D_2^2 \delta_1^2+2 D_2^3 \delta_1^2+14 D_1 \delta_1^4+2 D_2 \delta_1^4-8 D_1^3 \delta_1 \delta_2 \no&
+8 D_1^2 D_2 \delta_1 \delta_2+8 D_1 D_2^2 \delta_1 \delta_2
-8 D_2^3 \delta_1 \delta_2-28 D_1 \delta_1^3 \delta_2-4 D_2 \delta_1^3 \delta_2+2 D_1^3 \delta_2^2\no&
-14 D_1^2 D_2 \delta_2^2+22 D_1 D_2^2 \delta_2^2-10 D_2^3 \delta_2^2+16 D_1 \delta_1^2 \delta_2^2+16 D_2 \delta_1^2 \delta_2^2-4 D_1 \delta_1 \delta_2^3\no&
-28 D_2 \delta_1 \delta_2^3
+2 D_1 \delta_2^4+14 D_2 \delta_2^4\Big) E\left[\frac{(D_1-D_2)^2-(\delta_1+\delta_2)^2}{(D_1-D_2)^2-(\delta_1-\delta_2)^2}\right]+\no&
+\Big(6 D_1^5-12 D_1^3 D_2^2+6 D_1 D_2^4+6 D_1^4 \delta_1-12 D_1^2 D_2^2 \delta_1+6 D_2^4 \delta_1+29 D_1^3 \delta_1^2\no&
-12 D_1^2 D_2 \delta_1^2+15 D_1 D_2^2 \delta_1^2+16 D_2^3 \delta_1^2+29 D_1^2 \delta_1^3+20 D_1 D_2 \delta_1^3-D_2^2 \delta_1^3\no&
+D_1 \delta_1^4-4 D_2 \delta_1^4+\delta_1^5+8 D_1^3 \delta_1 \delta_2-8 D_1^2 D_2 \delta_1 \delta_2-8 D_1 D_2^2 \delta_1 \delta_2+8 D_2^3 \delta_1 \delta_2\no&
+28 D_1 \delta_1^3 \delta_2 +4 D_2 \delta_1^3 \delta_2+15 D_1^3 \delta_2^2+36 D_1^2 D_2 \delta_2^2-3 D_1 D_2^2 \delta_2^2-D_1^2 \delta_1 \delta_2^2\no&
+20 D_1 D_2 \delta_1 \delta_2^2+29 D_2^2 \delta_1 \delta_2^2-30 D_1 \delta_1^2 \delta_2^2-28 D_2 \delta_1^2 \delta_2^2-2 \delta_1^3 \delta_2^2\no&
+4 D_1 \delta_1 \delta_2^3+28 D_2 \delta_1 \delta_2^3-3 D_1 \delta_2^4
+\delta_1 \delta_2^4\Big) K\left[\frac{(D_1-D_2)^2-(\delta_1+\delta_2)^2}{(D_1-D_2)^2-(\delta_1-\delta_2)^2}\right]+\no&
+\Big(-12 D_1^4 \delta_1+24 D_1^2 D_2^2 \delta_1-12 D_2^4 \delta_1-58 D_1^2 \delta_1^3-40 D_1 D_2 \delta_1^3\no&
+2 D_2^2 \delta_1^3-2 \delta_1^5+2 D_1^2 \delta_1 \delta_2^2-40 D_1 D_2 \delta_1 \delta_2^2-58 D_2^2 \delta_1 \delta_2^2\no&
+4 \delta_1^3 \delta_2^2-2 \delta_1 \delta_2^4\Big) \Pi\left[\frac{D_1-D_2-\delta_1-\delta_2}{D_1-D_2+\delta_1-\delta_2},\frac{(D_1-D_2)^2-(\delta_1+\delta_2)^2}{(D_1-D_2)^2-(\delta_1-\delta_2)^2}\right]
\end{align}
\end{subequations}
where $K$ is the complete elliptic integral of the first kind, $E$ is the complete elliptic integral of the second kind and $\Pi$ is the complete elliptic integral of the third kind.

The mission is now straightforward. Solve equations (\ref{2asymconditions}) and obtain the parameters $D_{1,2},\delta_{1,2}$. However these equations are not only beyond any chance to solve analytically, but also their numerical treatment is extremely complicated. Available computer programs cannot be used, since the equations are very sensitive to the starting point used as the first trial in the numerical computation and there is very little we can do systematically here.

We have thus programed our own algorithm to deal with the equations (\ref{2asymconditions}) described in the appendix \ref{ap2}. It is almost surely far from being optimal both in precision and speed, but it gives results sufficiently fast and precise for our purposes.

	\begin{figure}[!tb]
       \begin{center}
      \includegraphics[width=0.8\textwidth]{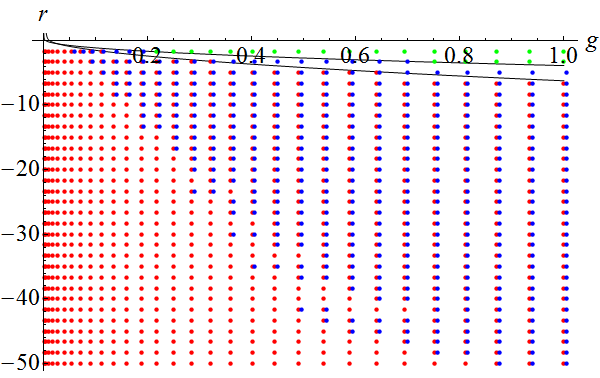}
	\caption{(Color online) The phase diagram of the asymmetric model (\ref{23asymqrt}) including the two cut phase. The color code is as before, green for the almost symmetric one cut, red for the asymmetric one cut and blue for the two cut solution. Note that the dots of the two cut solution have been shifted slightly to the right to the right not to overlap with the red dots. See the text for details.}
       \label{fig23asymdiagram}
	\end{center}
	\end{figure}

Using this algorithm, we obtain the phase diagram shown in the figure \ref{fig23asymdiagram}. The first thing to notice is no overlap of the region of the symmetric one cut and the two cut solutions, i.e. green and the blue regions, as is expected. And there is an overlap between the asymmetric one cut and the two cut, i.e. the blue and the red regions, as expected. In this region, both the asymmetric one cut and the two cut solutions exists. And we identify a region, where only an asymmetric one cut solution exists. We see, that our algorithm did miss several points along the boundary of this area. The numerics becomes very sensitive and it is extremely challenging to pin down the boundary precisely. Fortunately, we will not need the exact location of the boundary for our purposes.

\begin{figure}[!t]
       \begin{center}
      \vspace*{-0.3cm}
      \includegraphics[width=0.35\textwidth]{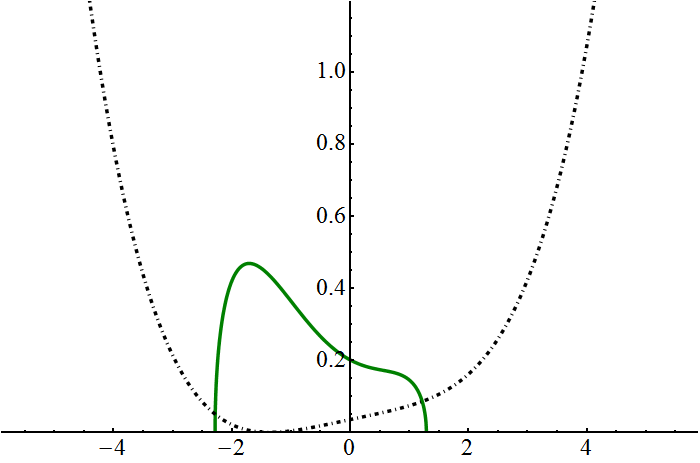}
			\includegraphics[width=0.35\textwidth]{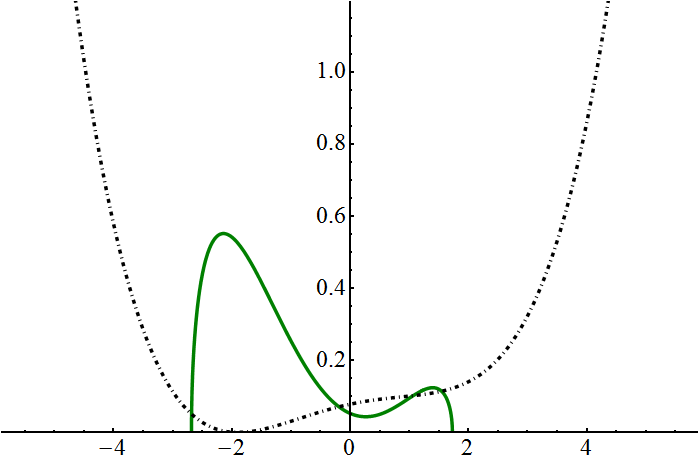}\\
			\includegraphics[width=0.35\textwidth]{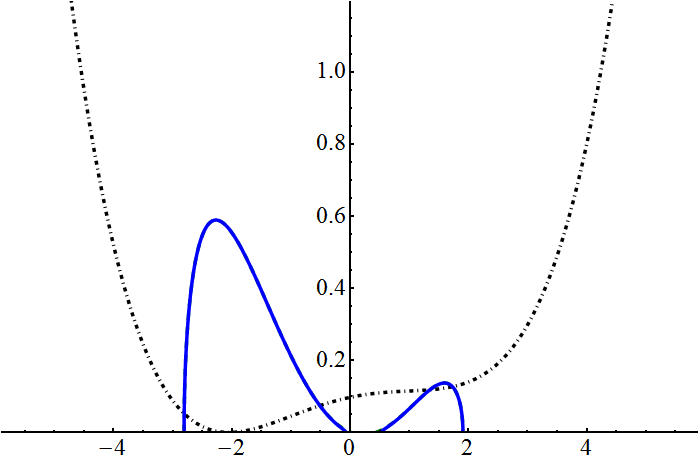}
			\includegraphics[width=0.35\textwidth]{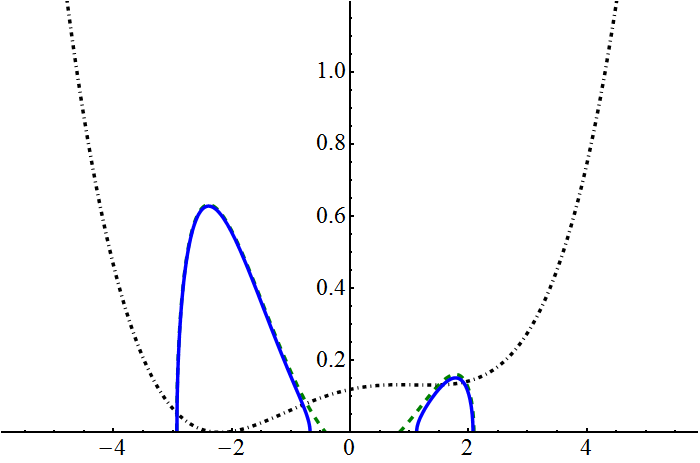}\\
			\includegraphics[width=0.35\textwidth]{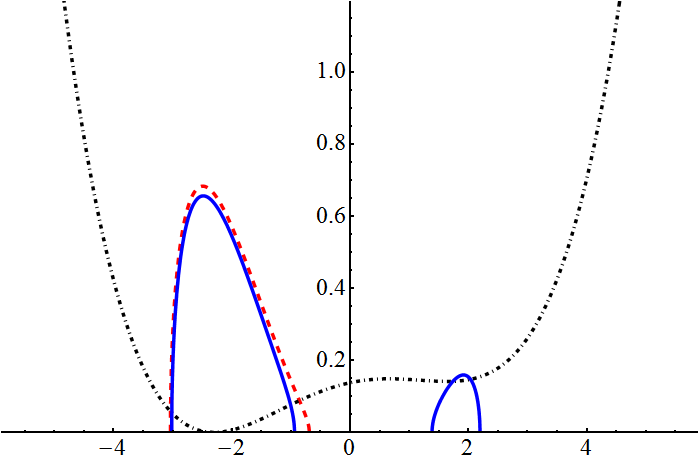}
			\includegraphics[width=0.35\textwidth]{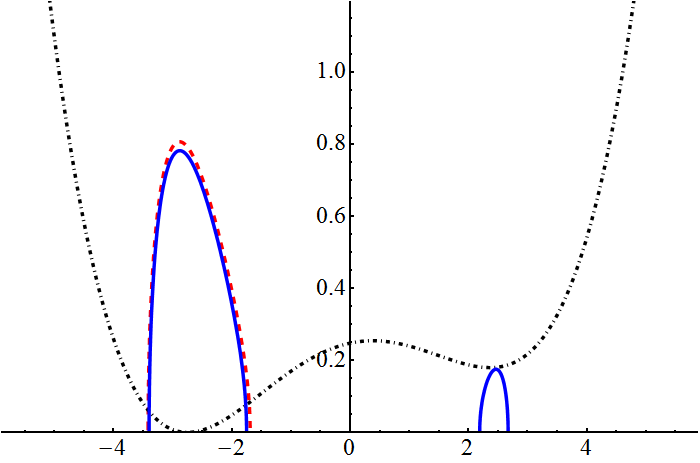}\\
			\includegraphics[width=0.35\textwidth]{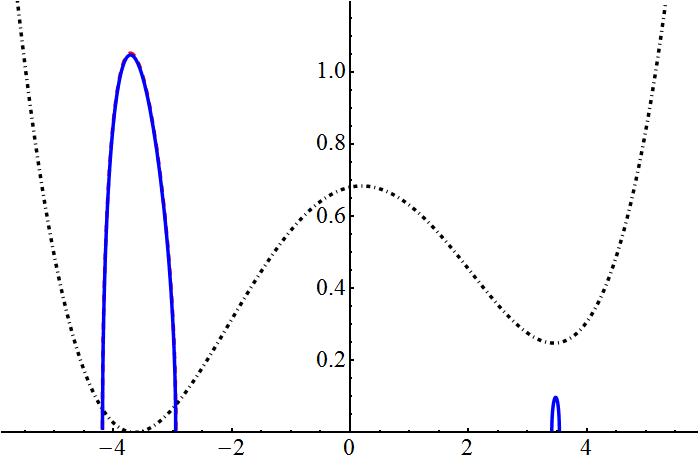}
			\includegraphics[width=0.35\textwidth]{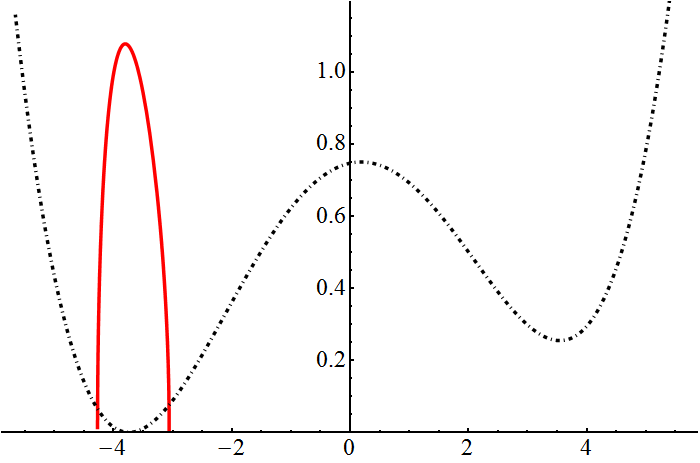}
      \vspace*{-0.1cm}
	\caption{(Color online) Plots of the eigenvalue distribution of the asymmetric quartic model (\ref{23asymqrt}). The plots are for the value $g=0.1$ and from right to left, top to bottom $r=\{0,-0.91,-1.19$, $-1.47$, $-1.68$, $-2.66,$ $-5.04,-5.32\}$.
In dashed are shown the solutions which either do not exist or have higher free energy. We can see that beyond some value of $r$, the asymmetric one cut solution becomes the proffered one, thanks to the large enough difference between the two minima. Note that the potential has been shifted vertically and rescaled so that it fits into the figure, so the differences between the extrema were not preserved. But it has been not shifted horizontally, so the position of the extrema of the potential have been preserved.}
       \label{fig22asymquarticdistr}
      \vspace*{-0.9cm}
	\end{center}
	\end{figure}

For a better idea what happens there, we give a series of plots of the distributions in the figure \ref{fig22asymquarticdistr}.

As we lower the value of $r$, the two wells of the potential become too deep for a solution extending over both of the wells to exists and we obtain a two cut solution. As we go lower with $r$, the walls of the left well become steep enough to confine all the eigenvalues in the left well, which has a lower potential.\footnote{There might be a solution living also in the right well, but it always has a higher free energy.} However, such situation has a large free energy. But as we go even further lower with the value of $r$, the difference between the right and the left well becomes so large, that the energy loss  for the eigenvalue when moving into from the right to the left well is high enough to pay for the energy gain in the interaction energy. The two cut solution with the lowest free energy is the asymmetric one cut solution living in the left well.

	\begin{figure}[!tb]       \begin{center}
			\includegraphics[width=0.8\textwidth]{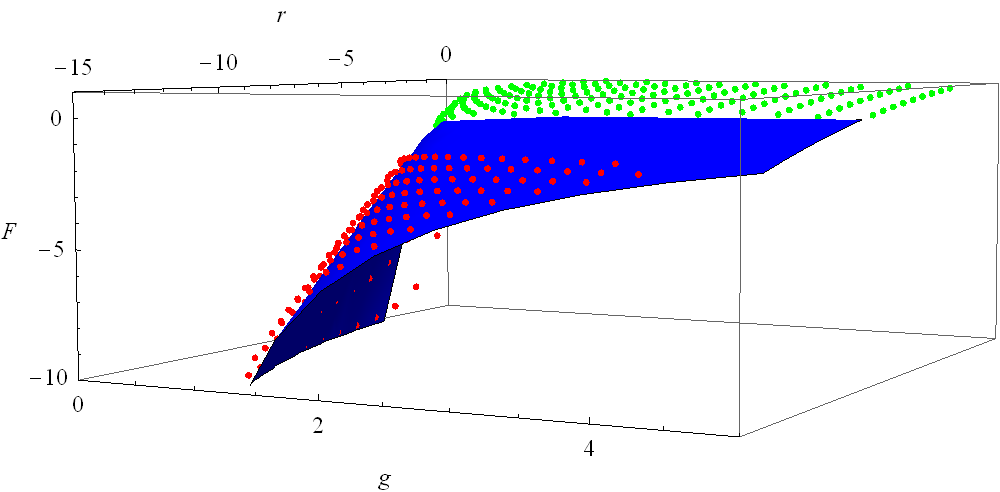}
	\caption{(Color online) The free energy diagram of the asymmetric model (\ref{23asymqrt}). The free energy of the two cut solution is shown as a filled area to emphasize its location in the diagram. See the text for details.}
       \label{fig23asymdiagramfree}
	\end{center}
   \vspace*{-0.4cm}
	\end{figure}

This is shown in the free energy diagram in the figure \ref{fig23asymdiagramfree}. We can see how the free energy of the two cut solution interpolates between the energies of the two one cut solutions.

Let us note, that to our best knowledge the asymmetric quartic model has not been analyzed in this detail in the literature.

%% file: 2_4multitrace.tex
Up to now, we have considered only matrix models with simple polynomial action (\ref{3.16}). We would now like to generalize our treatment for more complicated models. Recalling the formula for the moments of the distribution (\ref{3.30}) we see that (\ref{21orequival}) is simply
\be S(M)=\sum g_n c_n\ ,\ee
a linear expression in moments. A natural way to generalize such action is to include nonlinear terms into the action and consider a general action
\be S(M)=f(c_1,c_2,\ldots)+\half r \trl{M^2}+g \trl{M^4}\ ,\label{24genmlti}\ee
where we have considered an addition of a new term to the original quartic action. We will refer to this kind of models as multitrace models, opposing to the single trace models (\ref{21orequival}). Such models arise when the matrix $M$ is coupled to some external matrix $A$ via some interaction term $\trl{MA}$ or $\trl{MAMA}$ \cite{newcritical}. Such terms are not invariant under $M\to UMU^\dagger$ and the angular integral will not be trivial anymore. We will see an explicit example of this for the case of the scalar field on the fuzzy sphere in the sections \ref{sec31} and \ref{sec32}.

The integration will introduce a multitrace term of the form (\ref{24genmlti}), with the function $f$ given by the form of the interaction and the eigenvalues of the matrix $A$ \cite{itzub}. We will not consider any general cases here and after a swift description of the saddle point approximation we will work our way through several examples slowly building apparatus towards the matrix models which describe the scalar field theory on the fuzzy spaces.

\subsubsection{General aspects of multitrace matrix models}

To derive the saddle point equation for the action (\ref{24genmlti}), we need to realize that the variation of
\be
c_m^n=\lr{\sum_j \lambda_j^m}^n
\ee
with respect to the $i$-th eigenvalue is
\be
d(c_m^n)=n\lr{\sum_j \lambda_j^m}^{n-1}\pd{}{\lambda_i}\lr{\sum_j \lambda_j^m}d\lambda_i=nc_m^{n-1}m\lambda_i^{m-1}d\lambda_i
\ee
and we get
\be\label{24_derofmoment}
\pd{c_m^n}{\lambda_i}=nc_m^{n-1}m\lambda_i^{m-1}\ .
\ee
If we have more than one moment coupled to each other, we straightforwardly get
\be
\pd{c_{m_1}c_{m_2}}{\lambda_i}=c_{m_1}m_2\lambda_i^{m_2}+c_{m_2}m_1\lambda_i^{m_1}
\ee
and similarly for more moments coupled in a more complicated way. This way we obtain the saddle point equation of a general form
\be \sum_n \pd{f}{c_n} n \lambda_i^{n-1}+r \lambda_i+4 g \lambda_i=\frac{2}{N}\sum_{i\neq j}\frac{1}{\lambda_i-\lambda_j}\ .\ee

This means, that multitrace terms will introduce terms into saddle point equation, which include the moments of the distribution multiplied by $\lambda_i$. If we view the saddle point equation as the equilibrium condition for the gas of particles, these terms generate a force on the $i$-th eigenvalue which depends on the position of other eigenvalues. This means that they introduce further selfinteraction among the eigenvalues. However this is not a standard pair interaction between two particles. Multitrace terms rather couple the eigenvalue to an overall distribution of all the eigenvalues.

The approach we will use is based on rewriting the saddle point equation as an equation of an effective matrix model, which has only single traces in the potential. This is the second time we have encountered the word effective. There will be one more setting where we will use the word effective. To distinguish this effectivnes from the future one, we will always refer to this as an effective single trace model, action of the effective single trace model, moments of the effective single trace model, etc.

The form of the saddle point equation will thus be
\be \sum_n g_{n,eff}\lambda_i^{n-1}=\frac{2}{N}\sum_{i\neq j}\frac{1}{\lambda_i-\lambda_j}\ .\ee
This way, appearance of $c_m^n$ will be seen as a correction to the coupling constant $g_m\to g_{m,eff}$. We then compute the eigenvalue distribution of this effective single trace model as if these parameters were constant. After that the moments of the effective distribution will be functions of the effective couplings and thus of the moments themselves. Equations for the moments like (\ref{2momentsphi4}) for example will thus turn into a set of selfconsistency conditions. Simply because
\be c_n=\int d\lambda\,\lambda^n\rho(\lambda,g_{n,eff})=\int d\lambda\,\lambda^n\rho(\lambda,c_1,c_2,\ldots)\ \ee
the eigenvalue distribution to be used to compute the moments is a function of moments themselves. The solution will be valid however only in the large $N$ limit, as what we effectively do is we rewrite
\be \trl{M^n}^m=\avg{\trl{M^{n}}^{m-1}}\trl{M^n}\ \ee
which uses the factorization property valid only in the large $N$ limit.

Technically, there is really not much more to it. It is good to keep in mind the picture, where the eigenvalues create an outside force among themselves that acts on them. With that in mind, we shall proceed to some, gradually more complicated, examples.

We will first study matrix models which have the standard quartic potential in the action accompanied by a multitrace part that contains even moments of the distribution
\be
S(M)=f(c_2,c_4,\ldots)+\half r\trl{M^2}+g\trl{M^4}\ .
\ee
We thus expect the resulting distributions to be symmetric. Later, we will get to some simple examples of models involving also asymmetric moments in the multitrace.

\subsubsection{Second moment multitrace models}

Let us start with a very simple model which we will use to illustrate the method outlined in the previous part. As the name of the section suggests, we will deal with models involving only the second moment in the multitrace.

\podnadpis{The most simple model}

Let the multitrace part be given just by the square of the second moment
\be\label{24simplecorrection}
S(M)=c_2^2+\half r\trl{M^2}+g\trl{M^4}\ .
\ee
According to (\ref{24_derofmoment}), the saddle point equation for this model is given by
\be
2c_2\lambda_i+r\lambda_i+4 g \lambda_i^2=\frac{2}{N}\sum_{i\neq j}\frac{1}{\lambda_i-\lambda_j}
\ee
and can be rewritten as
\be
\slr{r+2c_2}\lambda_i+4 g \lambda_i^2=\frac{2}{N}\sum_{i\neq j}\frac{1}{\lambda_i-\lambda_j}\ .
\ee
If we now denote 
\be \reff=r+2c_2\ ,\label{24verysimplereff}\ee
this corresponds to the well-known and by now very familiar quartic model. The one cut solution (\ref{3.38}), together with the expression for the second moment (\ref{2momentsphi4a}) leads to a rather unappealing unsolvable equation. However, at the phase transition, things simplify. First of all at the phase transition
\be\label{24firstsimple}
\reff=r+2c_2=-4\sqrt{g}
\ee
and as we have seen at the phase transition the second moment of the distribution is independent of $\reff$ and is equal to $1/\sqrt g$. A little surprisingly we obtain a phase transition of the line
\be\label{24firsttrafoline}
r=-4\sqrt g-\frac{2}{\sqrt g}\ .
\ee
We are tempted to conclude that for any point in the parameter space under this line, we obtain a two cut solution but this is something not to be taken for granted. We will investigate this further.

The conditions for the two cut solution (\ref{33cond2cut}) are much simpler. ${\reff=r+2c_2}$ with (\ref{22moments2cuta}) yields
\be
c_2=\frac{-r}{2+4g}\ .
\ee
This is a well-defined positive number everywhere under (\ref{24firsttrafoline}) and thus the solution exists in the whole region. To understand better what is happening, let us investigate how the phase diagram of the effective single trace model and the multitrace model are connected.

The strategy is as follows. We take a point in the phase diagram of the effective single trace model ${(\reff,\geff)}$ for some particular values of the parameters. The moments of the effective distribution as well as its free energy are then known. We then invert the relation (\ref{24verysimplereff}) to obtain
\be
r=\reff-2c_2\ .
\ee
This way, we obtain a point ${(r,g)}$ in the phase diagram of the original model. We then repeat this procedure for different values of the effective parameters, obtaining a deformation of the phase diagram of the effective single trace model.

	\begin{figure}
       \begin{center}
       \includegraphics[width=0.8\textwidth]{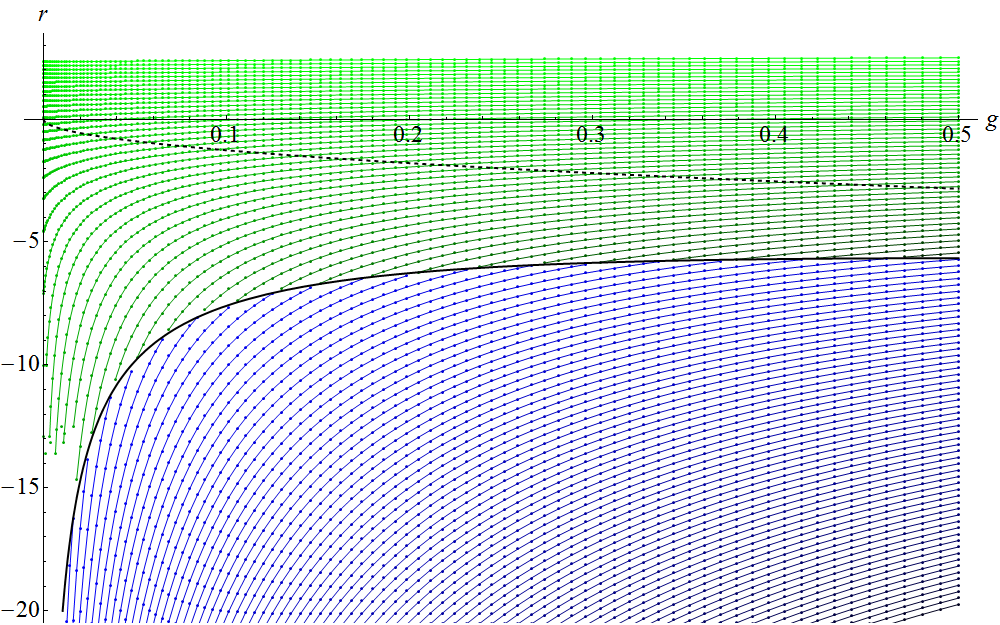}
	\caption{(Color online) The numerical phase diagram of the simple multitrace model (\ref{24simplecorrection}). The green region denotes the one cut solution, the blue region denotes the two cut solution. The black line is the analytic transition line (\ref{24firsttrafoline}), the dashed line is the transition line of the single trace model. The green and blue lines connect the points of constant $\reff$ to stress the deformation of the original phase diagram.}
       \label{fig24symplemultiphase}
	\end{center}
	\end{figure}

The resulting phase diagram is shown in the figure \ref{fig24symplemultiphase}, together with the transition line (\ref{24firsttrafoline}). The lines of constant $\reff$ are also shown to make the deformation more clear.

The most striking feature of this diagram is the region near the negative $r$ axis, where a one cut solution exists even for negative $r$ and small $g$. This is very different from the original model with no multitraces.

This is due to the attractive nature of the interaction the extra term $c_2^2$ introduces. The fact that the force is indeed attractive can be seen either from the effective description where it corresponds to parabolic potential or directly in the original description (\ref{24simplecorrection}) where we see, that bigger $c_2$ costs energy. Eigenvalues will therefore not like to have a large second moment and will be attracted toward the origin.

In the free theory with negative $r$, the eigenvalues would normally spread out to infinity. But now, while they spread the second moment of their distribution rises and at some point energy lost by going further down the potential and away from each other is outgained by the rise of the energy due to larger second moment and the distribution stabilizes.
	\begin{figure}[!t]
       \begin{center}
      \includegraphics[width=0.4\textwidth]{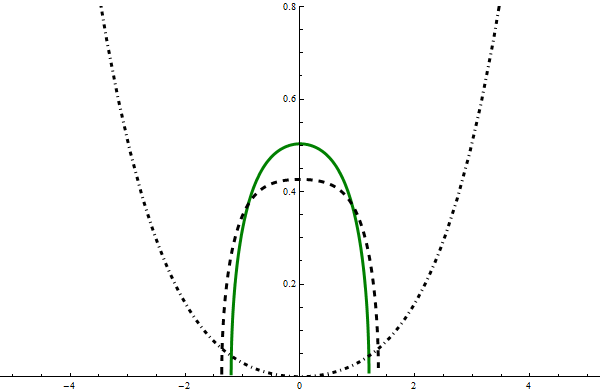}
			\includegraphics[width=0.4\textwidth]{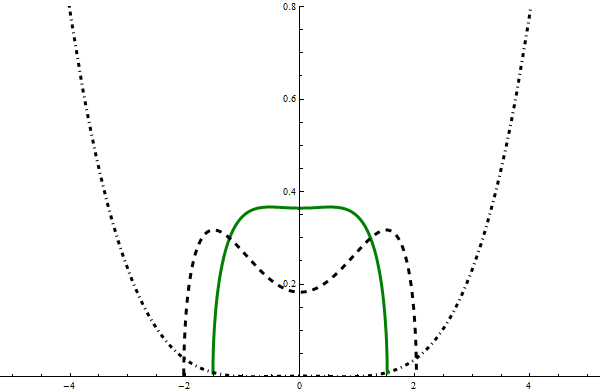}\\
			\includegraphics[width=0.4\textwidth]{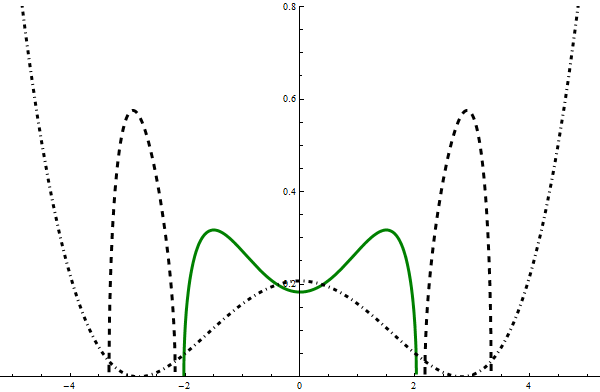}
			\includegraphics[width=0.4\textwidth]{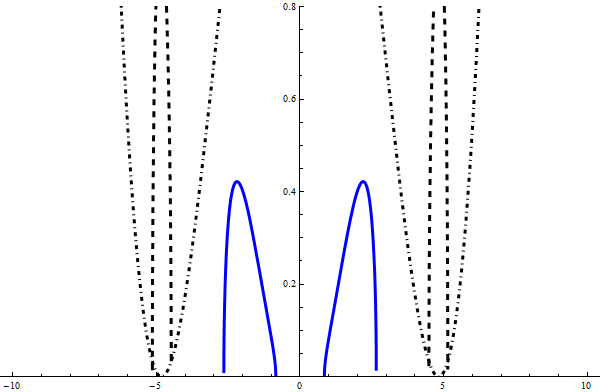}\\
			\includegraphics[width=0.4\textwidth]{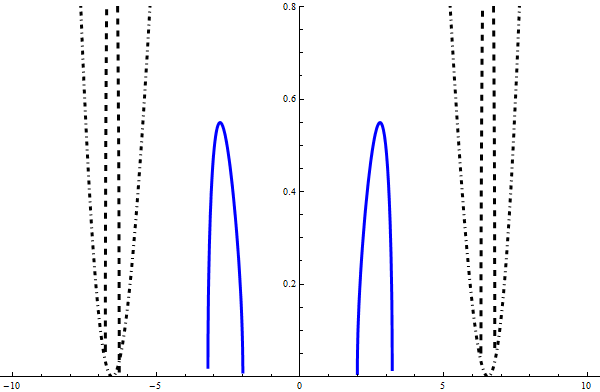}
			\includegraphics[width=0.4\textwidth]{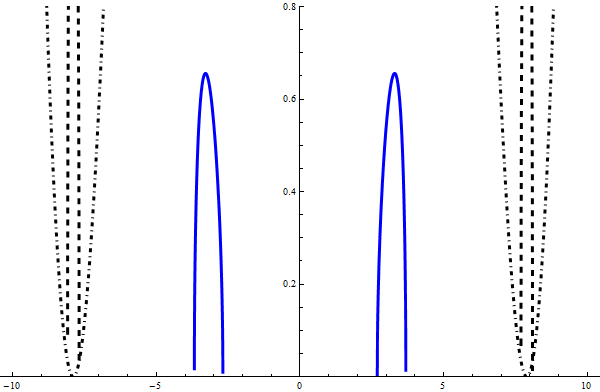}
	\caption{(Color online) The eigenvalue distributions for the simplest multitrace model (\ref{24simplecorrection}) for the value $g=0.1$. The values of $r$ are from left to right, bottom to top 
	\[r=\{1.594,-0.252,-3.152,-9.3,-17.1,-24.9\} \]
	The potential is shown and the eigenvalue distribution for the model with no multitrace term is shown in dashed.} 
       \label{figsimplemulti1}
	\end{center}
	\end{figure}

This is illustrated in the plots of the eigenvalue distribution in the figure \ref{figsimplemulti1}. Note that the potential has been shifted vertically and rescaled so that it fits into the figure, so the differences between the extrema were not preserved. But it has been not shifted horizontally, so the position of the extrema of the potential have been preserved.

The natural question is whether a solution could be stabilized in the case of negative coupling $g$? The answer is no, as we see in the phase diagram. The interaction is not strong enough to overcome a strong outward force of the negative quartic potential. But this also gives an answer how to do that. Introduce $c_4^2$ terms into the action. From the point of view of the transformation of the diagram this will lead to a deformation in the $g$ direction. From the point of eigenvalues this will introduce interaction strong enough to overcome the mentioned strong force.

\podnadpis{The second most simple model}

But before we do that, let us complete the study of quadratic multitrace models with the study of the model
\be
S(M)=-c_2^2+\half r\trl{M^2}+g\trl{M^4}\ .\label{actionnextsimple}
\ee
At first, this model does look similar to (\ref{24simplecorrection}), but after the previous discussion we know better. The force introduced pushes the eigenvalues outward, as having a high second moment lowers the energy. The algebra is similar to the previous case and we obtain the phase transition line
\be
r=-4\sqrt{g}+\frac{2}{\sqrt g}\label{nextsimpletrafo}
\ee
and for the second moment of the two cut solution
\be
c_2=\frac{-r}{4g-2}\ .\label{nexsimplec2}
\ee
This formula already suggests some strange behavior. For $r<0$, this expression is positive only for $g>1/2$. The system does not have a two cut solution for $r<0,g<1/2$! But it has a reasonable two cut solution even for $r>0$ if $g<1/2$!

This can again be understood in the particle gas picture. The multitrace term introduces a repulsive force among the eigenvalues. They like to have a large second moment and as long as the potential well is not too steep (which means $g$ small enough) they will start to split. Only when they get far enough from the origin, the potential becomes steep enough to provide enough force to tame the repulsion. And as we will see for small $r$ and $g$, not even that.

	\begin{figure}
       \begin{center}
       \includegraphics[width=0.8\textwidth]{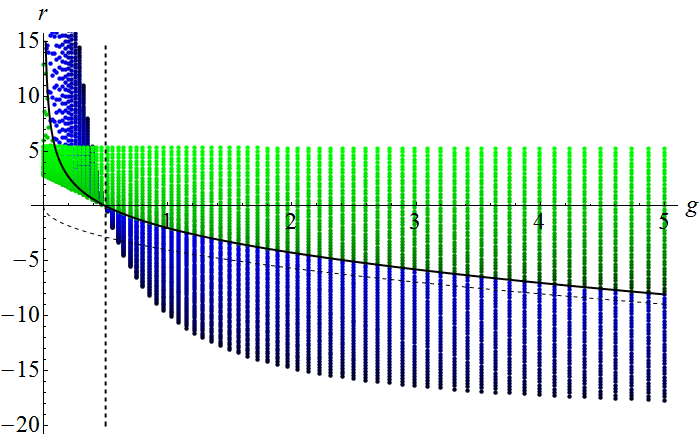}
	\caption{(Color online) The numerical phase diagram of the multitrace model (\ref{actionnextsimple}). The green region denotes the one cut solution, the blue region denotes the two cut solution. The black line is the analytic transition line (\ref{nextsimpletrafo}), the dashed line is the transition line of the single trace model and the horizontal line denotes the $g=1/2$.}
       \label{fig24phasenextsimple}
	\end{center}
	\end{figure}

And the phase diagram in the figure \ref{fig24phasenextsimple} confirms this. Moreover we see that there is a small region in the $r>0,g>0$ part of the diagram where no solution exists. And it is not clear whether there is a two cut solution in the whole area $g<1/2$ above and $g>1/2$ below the line (\ref{nextsimpletrafo}). To resolve this issue, we consider the transformation of the line $\reff=-x\sqrt g$ for $x>4$. This line is transformed into the line
\be r(g)=x\lr{-\sqrt g+\frac{1}{2\sqrt g}}\ .\ee
Therefore, the whole region $r>0,g<1/2$ is covered by some $x$ as is the whole region $r<0,g>1/2$.

\begin{figure}
       \begin{center}
       \includegraphics[width=0.48\textwidth]{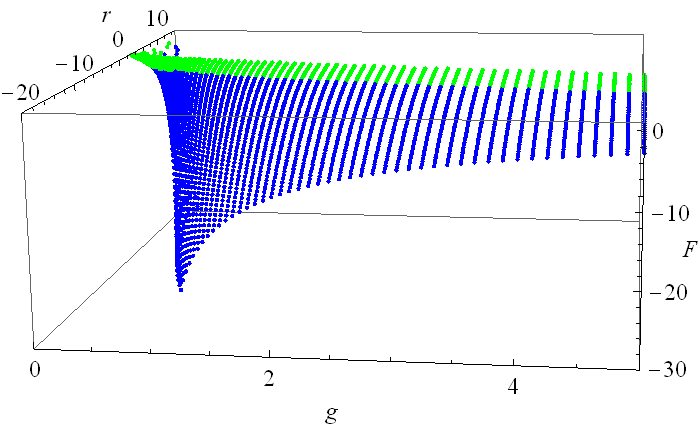}
			\includegraphics[width=0.48\textwidth]{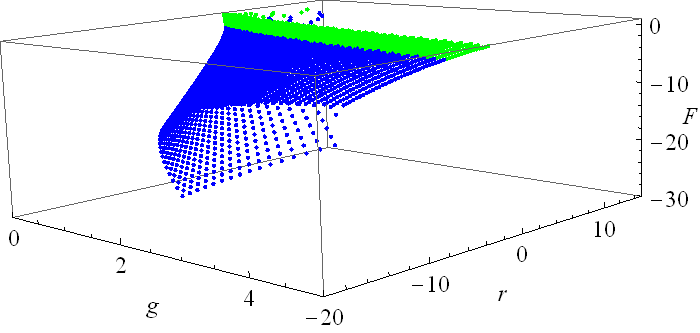}
	\caption{(Color online) The numerical free energy diagram of the multitrace model (\ref{actionnextsimple}). Again, the green region denotes the one cut solution, the blue region denotes the two cut solution. See the text for detailed discussion.}
       \label{fig24phasenextsimplefree}
	\end{center}
    \vspace*{-0.4cm}
	\end{figure}

Another question is what happens in the region of overlap between the one cut and the two cut solutions. The answer is in the free energy diagram in the figure \ref{fig24phasenextsimplefree}.

The free energy diagram shows what has happened. The part close to the negative $r$ axis has been folded upwards under the part close to the positive $r$ axis. Since the free energy of this part is lower, the preferred solution is this ''folded'' one. So apart from the usual matrix phase transition at the line (\ref{nextsimpletrafo}), the model exhibits a new, ''multitrace'' phase transition at the line $g=1/2,r>0$, where the usual one cut solution changes into the ''folded'' two cut solution. This phase transition is of the first order, as the free energy is discontinuous.

The last unanswered question is the region where no solution is available close to the origin. To find its boundary, we again consider the transformation of lines $\reff=-x\sqrt g$, but now for $x<4$. After some algebra we find out that such lines transform to
\be
r(g)=-x\sqrt{g}+\frac{72 x+x^3+48 \sqrt{48+x^2}+x^2 \sqrt{48+x^2}}{432 \sqrt{g}}\ .
\ee
The boundary is then given by the envelope of these curves for all values of $x<4$. To find this envelope, we consider two curves for $x$ and $x+dx$ and find their intersection, which gives one point of the enveloping curve. We solve this equation for $g$ and plug into the formula for $r$. At the end of the day, we find the boundary of existence of the one cut solution do be

\bal
r(g)=\frac{1}{108 \sqrt{2} g^{3/2}}\Bigg(
\begin{array}{l}
\left(\pm1\pm24 g\mp180 g^2\right) \sqrt{-12+\frac{1}{g}+36 g}\\+(1+6 g)^2 \sqrt{12+\frac{1}{g}+36 g}
\end{array}\Bigg)
\left\{\begin{array}{ll}
-&g<\frac{1}{6}\\+&\frac{1}{6}<g<\half\end{array}\right. .
\end{align}
This line nicely fits into the diagram \ref{fig24phasenextsimple}, but is not shown to keep the figure clear. The fact that we have been able to find this line explicitly is due to a simple form of the multitrace term. Later, for more complicated models we will not be able to do so and the line could be found using the same method numerically, which we will however not do.

\podnadpis{A general $c_2^2$ model}

Similar model has been studied before. Namely the model with $F(c_2)=h c_2^2$ and a fixed $r=-1$ in \cite{shishanin}, i.e. the model given by the action
\be
S(M)=-\half \trl{M^2}+h \trl{M^2}^2+g \trl{M^4}\ .\label{actionshis}
\ee
The phase structure in the $(g,h)$ plane was then considered. It is very easy to analyze this structure using the outlined numerical approach.

\begin{figure}%[!tb]
\centering % \begin{center}/\end{center} takes some additional vertical space
\includegraphics[width=.7\textwidth]{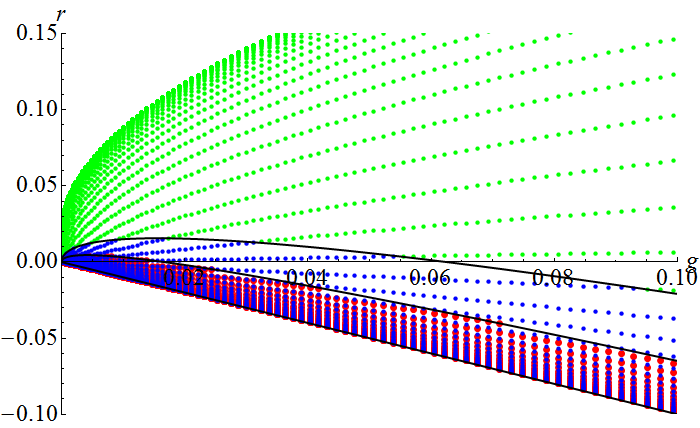}
% "\includegraphics" is very powerful; the graphicx package is already loaded
\caption{(Color online) The phase diagrams of model (\ref{actionshis}). The green dots denote the existence of the one cut solution, blue of the two cut solution and red of the asymmetric one cut solution.}\label{ob22_shish}
\end{figure}

The figure \ref{ob22_shish} shows the phase diagram of such model. There are three regions in the phase diagram, the green region where only the one cut solution exists, the blue region where only the two cut solution exists and the red region, where both the asymmetric one cut solution and the two cut solution exist. The three regions are separated by two parabolas, which are not difficult to compute. We take
$\reff=-1+4 h c_2$ and $\reff=-4\sqrt g$ or $\reff=-2\sqrt{15g}$ for the upper or the lower boundary respectively. This then yields the two boundary lines
\begin{align}
h(g)=&\frac{1}{4} \lr{\sqrt{g} - 4g}\ ,\\
h(g)=&\frac{3}{82} \lr{\sqrt{15} \sqrt{g} - 30 g}\ .
\end{align}
As we can see, both these expression precisely fit into the diagram \ref{ob22_shish}. The green region extends all the way up for positive $h$. But since for the second moment we obtain
\be
c_2=\frac{1}{4(g+h)}\ ,
\ee
if $h<-g$ the solution is not well defined and does not exist. Which is also seen in the diagram, as there is no solution under the $h=-g$ line.

Note that the expression for the second line computed in \cite{shishanin} is different and does not reproduce the line in the diagram. 

\subsubsection{The fourth moment multitrace models}

We will now proceed further with the discussion of the multitrace matrix models which include the powers of the fourth moment in their action. We will begin with two simple models and later consider a models which includes both the fourth and the second moments in the multitrace.

\podnadpis{Very simple fourth moment multitrace models}

To begin our study of the quartic multitrace matrix models, let us consider two very simple models given by the action
\be S(M)=\half r \trl{M^2}+g \trl{M^4}\pm c_4^2\ .\label{sec6_simple}\ee
This action leads to the saddle point equation
\be r \lambda_i+4 g \lambda_i^3\pm 8 c_4 \lambda_i^3=\frac{2}{N}\sum_{i\neq j}\frac{1}{\lambda_i-\lambda_j}\ .\ee
Looking at the formula for the corresponding effective coupling
\be \geff=g\pm2 c_4\ ,\ee
we know what to expect. Far from the $r$ axis, the value of $c_4$ is not substantial and we will get a small deformation in the phase diagram. Close to the positive $r$ axis, we still expect a finite shift, as the eigenvalue of the effective single trace model becomes the Wigner semicircle. This shift will be to he left for the positive sign and to the right for the negative sign.

\begin{figure}[!tb]
\centering % \begin{center}/\end{center} takes some additional vertical space
\includegraphics[width=.7\textwidth]{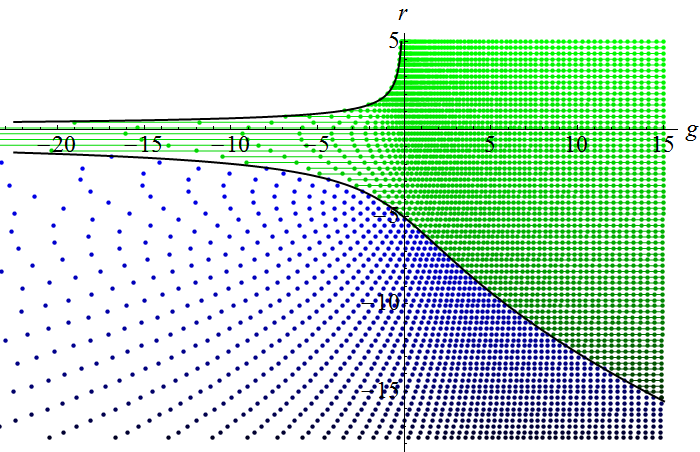}
\includegraphics[width=.7\textwidth]{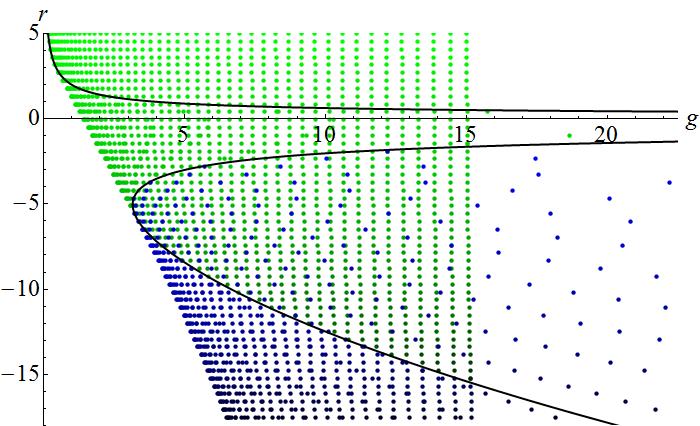}
% "\includegraphics" is very powerful; the graphicx package is already loaded
\caption{(Color online) The phase diagrams for the matrix model (\ref{sec6_simple}), with the positive sign on top and the negative sign on bottom. The green region denotes the one cut solution, the blue region denotes the two cut solution. See the text about the discussion of the boundary lines.}\label{ob62_simple}
\end{figure}

Close to the origin this shift will become more significant and as $r$ becomes negative, the fourth moment of the effective distribution diverges and the diagram is deformed to the infinity, in the same direction as before. This expectation is confirmed in the phase diagrams in the figure \ref{ob62_simple}.

The boundary lines are computed as follows. The upper line is given by $\geff=0$, with $c_4$ the corresponding expression of the fourth moment of the semicircle distribution 
\be c_4=\frac{1}{32}\delta^3 r\ee
and is given by
\be r(g)=\frac{2}{\sqrt{\mp g}}\ .\ee
The lower line is given by $r=-4\sqrt{\geff}$, with $c_4$ the corresponding expression of the fourth moment of two cut solution
\be c_4=D^2 \delta^2 \geff + \quater \delta^4 \geff\ee
at the phase transition. In the case of the positive sign it is given by the expression
\be r(g)=-2 \sqrt 2  \sqrt{g + \sqrt{10 + g^2}}\label{sec6_usualline}\ee
and in the case of the negative sign it is given by the two lines
\be r(g)=-2 \sqrt 2 \sqrt{g + \sqrt{-10 + g^2}}\ ,\ r(g)=-2 \sqrt 2 \sqrt{g - \sqrt{-10 + g^2}}\ .\label{sec6_usualline2}\ee

\begin{figure}
\centering % \begin{center}/\end{center} takes some additional vertical space
\includegraphics[width=.48\textwidth]{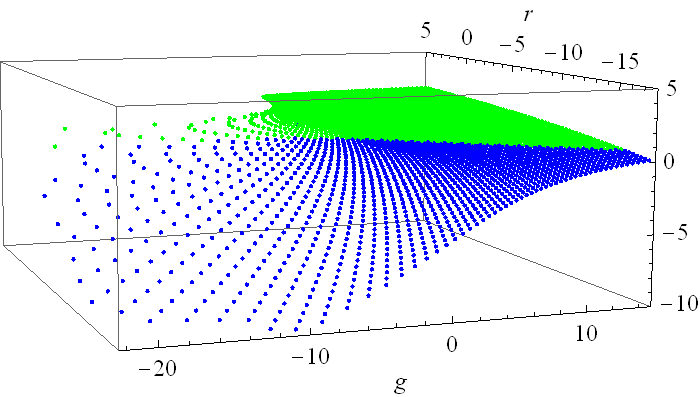}
\includegraphics[width=.48\textwidth]{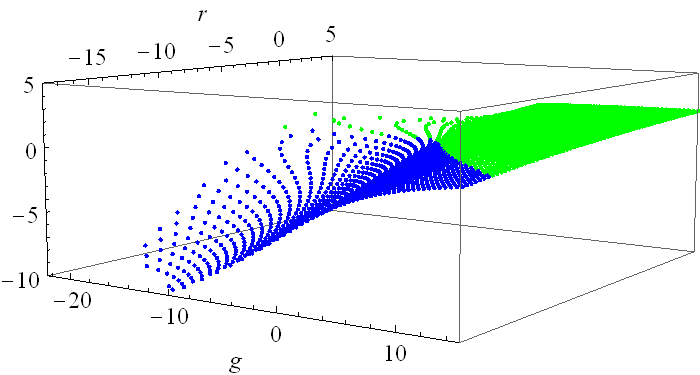}\\ 
\includegraphics[width=.48\textwidth]{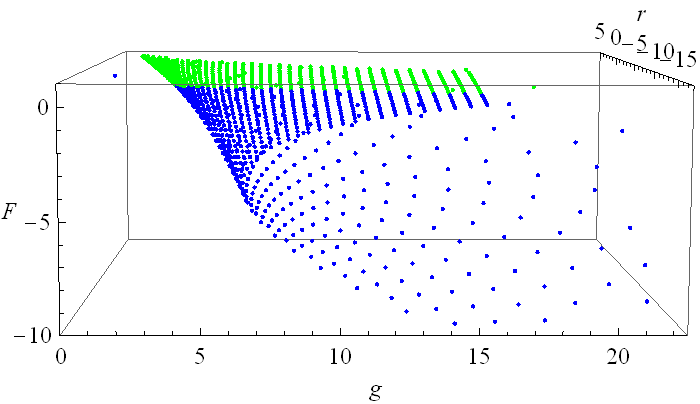}
\includegraphics[width=.48\textwidth]{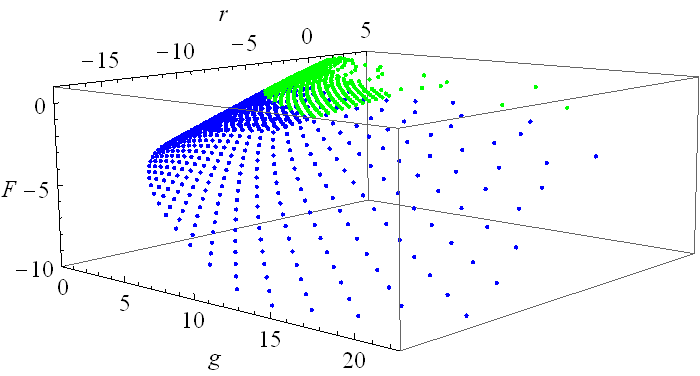}
% "\includegraphics" is very powerful; the graphicx package is already loaded
\caption{(Color online) The free energies of the the matrix models (\ref{sec6_simple}), with a positive sign in to top row and a negative sign in the bottom row.}\label{ob62_simplefree}
\end{figure}

The free energy graphs in the figure \ref{ob62_simplefree} show the following phase transitions of the models. Not much new for the model with the positive sign, the usual third order phase transition at the line (\ref{sec6_usualline}). However for the negative sign we get some new behavior. At the line $r=2/\sqrt{g}$ the models goes through a first order phase transition from a narrow to a wide one cut distribution. And then at the second line in (\ref{sec6_usualline2}) the third order phase transition, with some area missing in the phase diagram even for positive $g$.

\podnadpis{Not that simple fourth moment multitrace models}

As a second step, we will consider a slightly more complicated model given by
\be S(M)=\half r \trl{M^2}+g \trl{M^4}\pm \lr{c_4-2c_2^2}^2\ ,\label{sec6_more}\ee
motivated by the kinetic term effective action of the fuzzy field theory. Deriving the saddle point equation is now a simple exercise and from that we get the effective mass and coupling given by
\be \reff=r\mp16 c_2 (c_4 - 2 c_2^2) \ ,\ \geff=g\pm2 \lr{c_4-2c_2^2}\ .\label{morecomplicatedeffs}\ee
The extra term vanishes for the semicircle distribution and thus the points near the positive $r$ axis in the phase diagram will not be shifted at all. It however still diverges as we approach the negative $r$ axis and its vicinity will be shifted dramatically.

\begin{figure}[!tb]
\centering % \begin{center}/\end{center} takes some additional vertical space
\includegraphics[width=.7\textwidth]{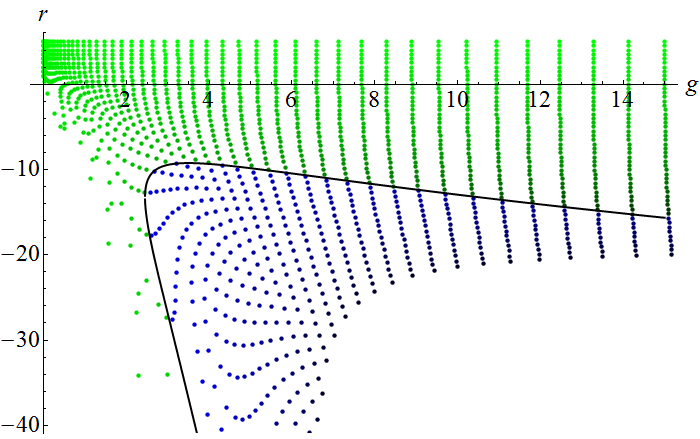}
\includegraphics[width=.7\textwidth]{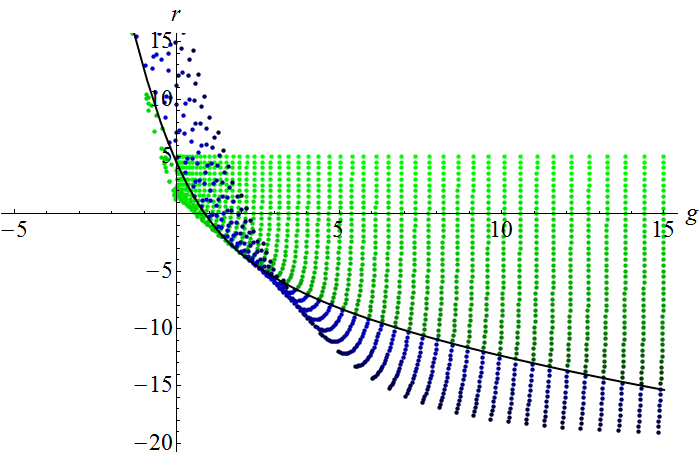}
% "\includegraphics" is very powerful; the graphicx package is already loaded
\caption{(Color online) The phase diagrams for the matrix model (\ref{sec6_more}), with a positive sign on the left and a negative sign on the right. The green region denotes the one cut solution, the blue region denotes the two cut solution. See the text about the discussion of the boundary lines.}\label{ob62_more}
\end{figure}

The phase diagram of both models is presented in the figure \ref{ob62_more}. The diagram confirms our expectation. For the positive sign we get a greatly extended region of existence of the one cut solution for negative $r$ and small $g$, but the system does not have a stable solution for negative values of $g$. Interestingly, for negative sign the system has a stable solution for negative $g$ as long as $r$ is positive and large enough. The two cut solution is however destabilized for a large part of the diagram and shifted under other solutions.

The phase transition lines in the figure \ref{ob62_more} have been computed as follows. Taking the moments of the two cut solution and plugging in the condition on the phase transition $\reff=-4\sqrt \geff$ we obtain
\be D=-\frac{4}{\reff}\ ,\ \delta=-\frac{4}{\reff}\ .\ee
This and (\ref{morecomplicatedeffs}) yields a pair of selfconsistency conditions
\be c_2=\frac{1}{\sqrt{\pm 2(c_4-2c_2^2)+g}}\ ,\ c_4=\frac{5}{\pm 8 (c_4-2c_2^2)+5 g}\ .\ee
For the positive sign these equations have two different solutions and we obtain two lines
\bse
\bal
r(g)=&-2 \sqrt{2} \sqrt{g - \sqrt{g^2-6}} - \frac{4}{\sqrt 3} \lr{g + \sqrt{g^2-6}}^{3/2}\\
r(g)=&-2 \sqrt{2} \sqrt{g - \sqrt{g^2-6}} - \frac{4}{\sqrt 3} \lr{g - \sqrt{g^2-6}}^{3/2}
\end{align}
\ese
For the negative sign in (\ref{sec6_more}), the selfconsistency conditions have only one solution and we obtain the phase transition line
\be
r(g)=-2 \sqrt{2} \sqrt{g + \sqrt{g^2+6}} + \frac{4}{\sqrt 3} \lr{\sqrt{g^2+6}-g}^{3/2}\ .
\ee

\begin{figure}
\centering % \begin{center}/\end{center} takes some additional vertical space
\includegraphics[width=.48\textwidth]{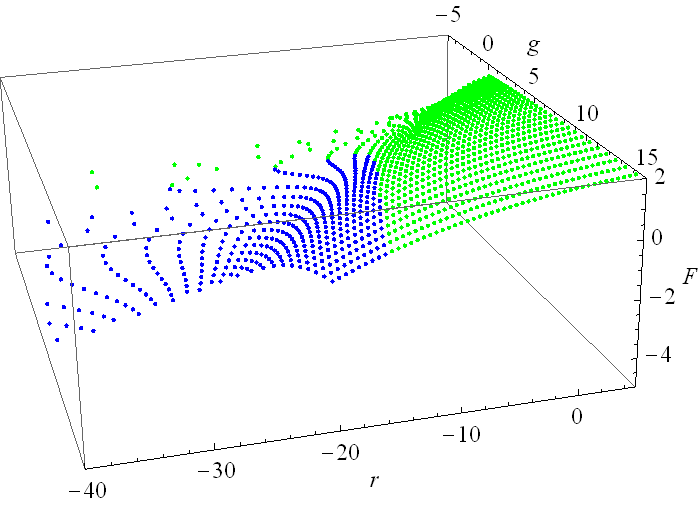}
\includegraphics[width=.48\textwidth]{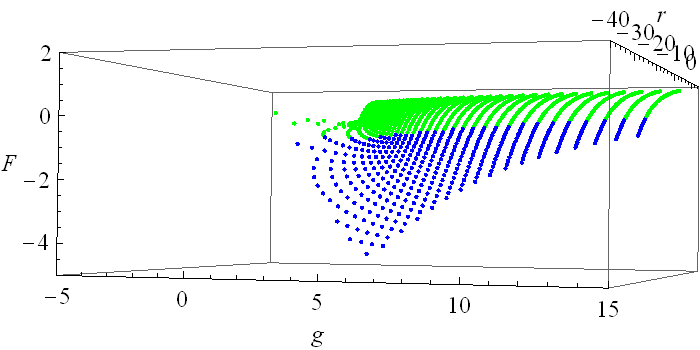}\\ 
\includegraphics[width=.48\textwidth]{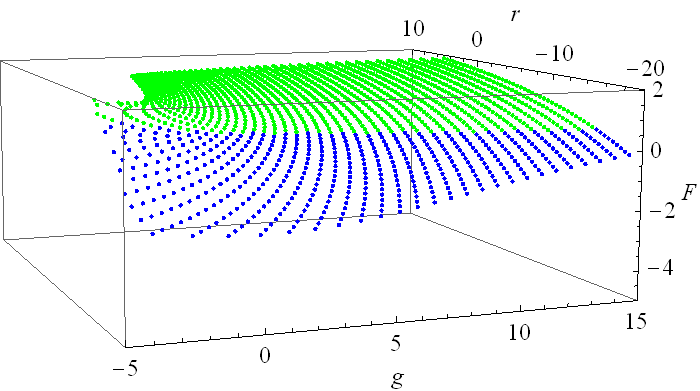}
\includegraphics[width=.48\textwidth]{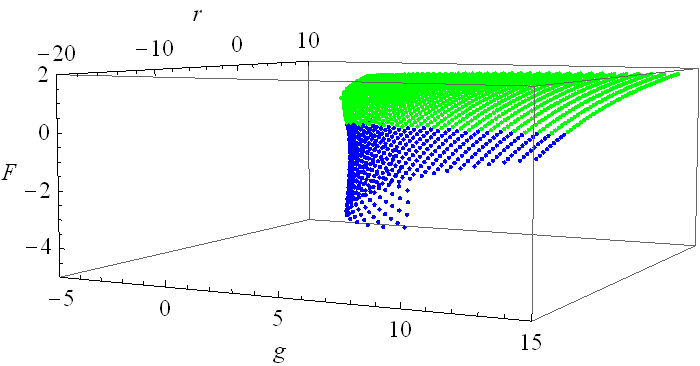}
% "\includegraphics" is very powerful; the graphicx package is already loaded
\caption{(Color online) The free energies of the matrix models (\ref{sec6_more}), with a positive sign on the left and a negative sign on the right.}\label{ob62_morefree}
\end{figure}

Looking at the phase energy diagrams in the figure (\ref{ob62_morefree}) we see the fold of the diagram in the case of the negative sign. It is however very difficult to compute the line, along which the diagram is folded. What can be checked more easily is that the folded part of the diagram extends all the way to infinity. Considering the transformation of the line $\reff=-x\sqrt{\geff}$ for large values of the parameter $x$ we see that in this limit, the line is mapped to the line $r=(x/2)^{3/4}$.

This means that there is a blue dot under every dot in the folded region and the ''folded'' two cut case is the preferred solution almost everywhere. Apart from the small region on the left of the black line in the top left part of the second diagram in the figure \ref{ob62_more}.

\podnadpis{Lessons learned}

We thus see, that the introduction of the multitrace terms into the matrix model action yields quite dramatic changes to the phase diagram, possibly introducing new phase transitions and regions of (in)existence. The key to this drama has been the large values of the moments of the single trace model distribution close to the negative $r$ axis, which diverge as $g\to0$.

To obtain a reasonable deformation of the phase diagram, we need to have a finite $\reff$ and $\geff$ even as the moments diverge. We will see one such model in the section \ref{sec33simplphase}.

But before that, let us look at a simple model which introduces an asymmetry into the multitrace action.

\subsubsection{Asymmetric multitrace model}\label{sec24asym}

As an example of an asymmetric multitrace model, let us solve a very simple model given by
\be\label{24simpleasym}
S(M)=-c_1^2+\half r \trl{M^2}+g\trl{M^4}\ ,
\ee
which leads by now straightforwardly to the effective single trace model
\be\label{24asymtrafo}
S_{eff}(M)=\trl{M}+\half \reff\trl{M^2}+\geff \trl{M^4}\ ,
\ee
with
\be\label{24asymeffrg}
\reff=\frac{r}{4c_1^2}\ ,\ \geff=\frac{g}{16c_1^4}\ .
\ee
But since we have rescaled the matrix, we need to be careful about the relation of the effective distribution moments and the moments of the true distribution. The scaling involved in changing (\ref{24simpleasym}) to (\ref{24asymtrafo}) was $M\to M/x_0=M/(-2c_1)$, which means on one hand
\be
\reff=\frac{r}{x_0^2}\ ,\ \geff=\frac{g}{x_0^4}\ .
\ee
and we also obtain a condition
\be
c_1=\frac{c_{1,eff}}{x_0}=\frac{c_{1,eff}}{-2c_1}\ ,
\ee
which can be easily solved
\be
c_1=-\sqrt{\half (-c_{1,eff})}\ \Rightarrow\ x_0=\sqrt{2(-c_{1,eff})}\ .
\ee

\begin{figure}[tb]
       \begin{center}
    \vspace*{-0.2cm}
      \includegraphics[width=0.6\textwidth]{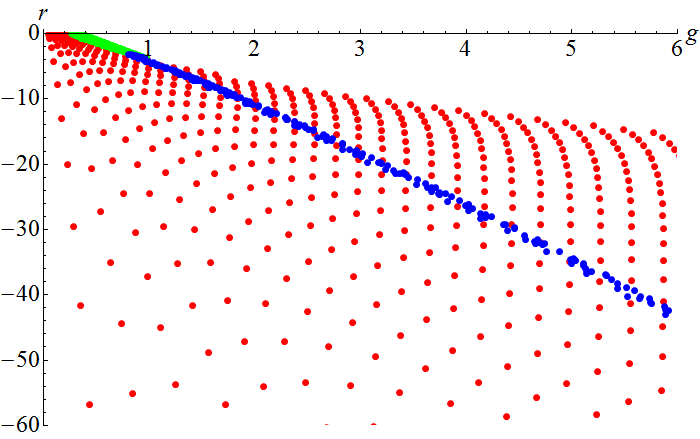}
%    \vspace*{-0.3cm}
	\caption{(Color online) The numerical phase diagram of the asymmetric multitrace model (\ref{24simpleasym}).}
       \label{fig24asymphase}
	\end{center}
	\end{figure}

\begin{figure}[!tb]
       \begin{center}
    \vspace*{-0.3cm}
      \includegraphics[width=0.8\textwidth]{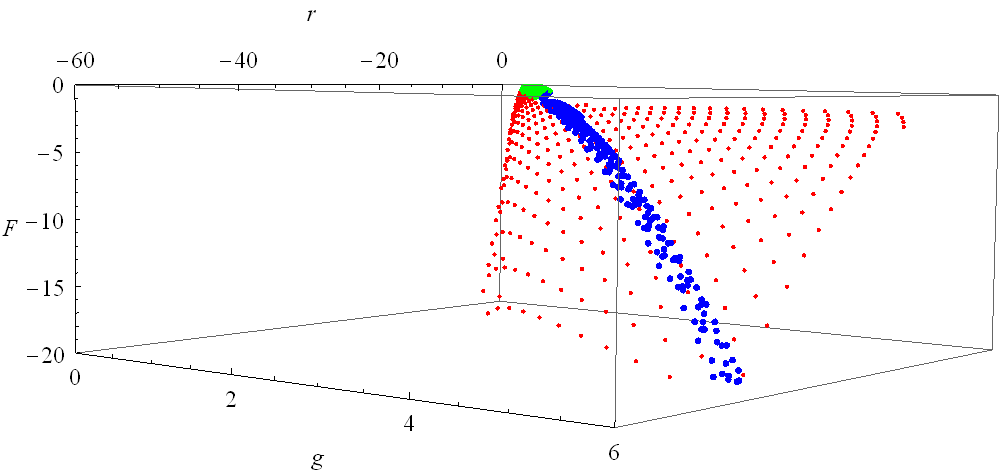}
			\includegraphics[width=0.8\textwidth]{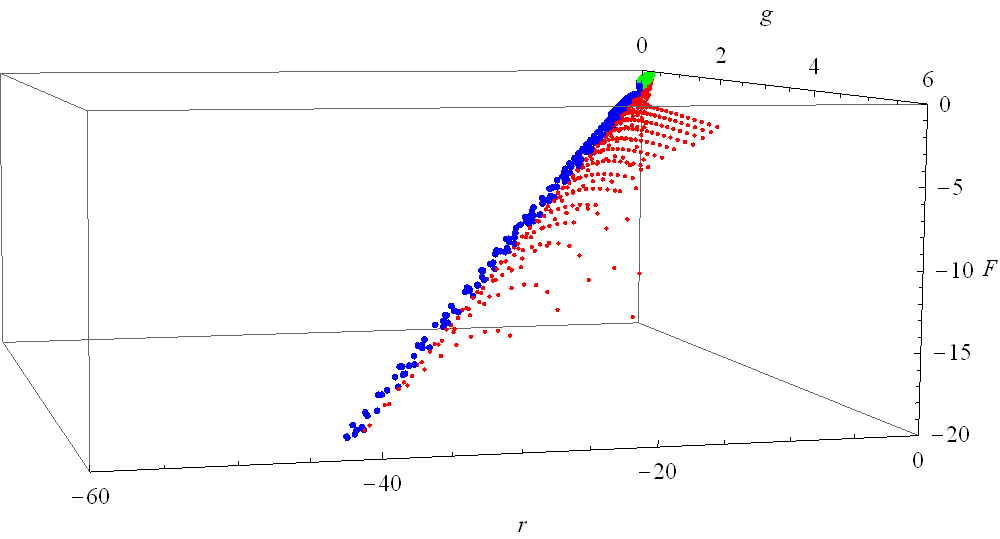}
    \vspace*{-0.2cm}
	\caption{(Color online) The numerical free energy diagram of the asymmetric multitrace model (\ref{24simpleasym}).}
       \label{fig24asymphasefree}
	\end{center}
	\end{figure}

Since we have no explicit expressions for the distributions, there is very little we can do further. Therefore, we will perform the same numerical procedure as before. We take the phase diagram of the effective single trace model in the figure \ref{fig22asymquartiphasediag} and by choosing particular values of $\reff$ and $\geff$ and inverting (\ref{24asymeffrg}) with a corresponding effective moment we obtain a point $(r,g)$ in the phase diagram of the original model. This way we obtain the phase diagram and the free energy diagram as in the figures \ref{fig24asymphase} and \ref{fig24asymphasefree}.

	
We have obtained an extremely drastic transformation of the phase diagram. The asymmetric one cut solution is deformed only slightly, but the other one cut solution have been shifted towards the origin due to a small value of $c_{1,eff}$.

Unfortunately there is very little we can do further. It would be interesting to analyze the model with a faster numerical procedure, especially to see the extent of the symmetric one cut and the two cut regions in the figure \ref{fig24asymphasefree}. Here, we just conclude with the observation, that the asymmetric solution of the model (\ref{24simpleasym}) is the preferred solution everywhere it exists and other solutions have a bleak chance of existing, apart from a small region close to the origin.

%Note, that for ${g=0}$ there is no stable solution. One can see it from the graph \ref{fig24asymphase} or directly trying to solve the equations determining the distribution together with the selfconsistency equations.

We leave it as an exercise for the reader to try to analyze the model (\ref{24simpleasym}) with a positive sign in the multitrace term.\footnote{A hint. There will be no stable asymmetric solution to this model. Why?} Both these models can be handled analytically in the case $g=0$ and the reader is invited to do it also.

%% file: 3_intro.tex
In this last section, we will put the introduced machinery of the matrix models to work in the particular case of the fuzzy field theory.

First, we will show why and how a fuzzy field theory can be described by a matrix model and then we will show that this is a multitrace matrix model. We will not be able to describe it completely, but we will arrive at different approximations to it.

We will then consider the phase structure of two rather different models, both approximating the scalar field theory on the fuzzy sphere in a very different way.

The first approximation will be nonperturbative\footnote{We will see later in what precise sense we mean this.}, but will contain only the multitraces of the first and second moments of the distribution. The second approximation will be consistently perturbative and contain multitraces of the first four orders. Interestingly, as we will see, the first approximation does a pretty decent job in explaining the numerical phase diagram of the section \ref{sec13}, opposing to the second one.

Before we proceed, let us comment on the connection between the phases of the fuzzy field theory from the section \ref{sec13} and the phases of the matrix model. Since the eigenvalues of the matrix correspond to the values of the field on the discretized noncommutative ''lattice'', and the eigenvalue distribution describes possible intervals for these values, it is natural to identify the symmetric one cut phase of the matrix model with the disorder phase of the field theory. And similarly the asymmetric one cut case with the uniform order phase and the two cut phase with the non-uniform order phase.

%% file: 3_1matrix.tex
The statement that the fuzzy field theory is equivalent to a particular matrix model is at this point rather self-evident. The field theory correlation functions (\ref{functional}) are matrix integrals (\ref{3.1}). And the large-$N$ limit of the matrix models corresponds to the commutative limit of the fuzzy theory, recalling (\ref{11tehta}).

The question is, what kind of matrix model describes the fuzzy field theory. It is not a simple Hermitian matrix model (\ref{3.3}) for some set of couplings $g_n$, since the kinetic term is not invariant under the diagonalization of the matrix $M\to U\Lambda U^\dagger$. It has a very non-trivial $U$ dependence and the angular integral cannot be straightforwardly computed. Assuming that the function $F(M)$ is invariant, we can write
\be
\avg{F}=\frac{1}{Z}\int \lr{\prod d\lambda_i}\,F(\Lambda)e^{-N^2\slr{\trl{V(M)}-\frac{2}{N^2}\sum_{i<j}\log|\lambda_i-\lambda_j|}}\int dU\,e^{-N^2\half\trl{M\K M}}\ ,
\ee
where
\be
Z=\int \lr{\prod d\lambda_i}\,e^{-N^2\slr{\trl{V(M)}-\frac{2}{N^2}\sum_{i<j}\log|\lambda_i-\lambda_j|}}\int dU\,e^{-N^2\half\trl{M\K M}}\ .
\ee
The computation of the angular integral will be the key point and we will denote it
\be\label{31anfintegral}
I=\int dU\,e^{-N^2\half\trl{M\K M}}\ .
\ee
Some of the results we obtain will be valid for a general kinetic term, some only for the standard double commutator kinetic term (\ref{12standkint}). We will do our best to comment on this every time a confusion could be possible. Both in the standard or in the more complicated cases, the kinetic terms couples the matrix $M$ to some external matrices, usually the $L_i$'s, in a way usually not considered in the literature.

The integral $I$ is only a function of the eigenvalues of the matrix $M$. So regardless of its structure, we can write it as
\be\label{31angeff}
\int dU\,e^{-N^2\half\trl{M\K M}}=e^{-N^2 S_{eff}(\Lambda)}
\ee
for some effective action $S_{eff}$ to be determined. We thus came the third time across a quantity we name effective, this time the kinetic term effective action, an effective interaction among the eigenvalues due to the coupling to the matrices $L_i$. We have reserved the name effective action for this object in section \ref{sec22}.

Notice, that at this point the large-$N$ behavior of the $dU$ integrand is not evident and we may have to introduce some different scaling of the couplings and the matrix to make all the terms contribute in the large-$N$limit than before. This should be possible, as we have three scalings to play with and three terms to fix. However, we expect this scaling to be unique, without any freedom we had before. We will return to this later.

\subsubsection{The free theory case and the persistence of the semicircle}

In this section, we will show that the eigenvalue distribution of the model given by the free theory
\be
S(M)=\half\trl{M\K M}+\half r \trl{M^2}\ ,\label{4.2}
\ee
is just a rescaled semicircle. This was first realized in \cite{stein05_1} and later in a different context in \cite{our1}. Here, we will follow the latter derivation.

We will be interested in the recursion rules for the moments of the distribution, derived using explicit Wick contractions. Knowing the moments, we will be able to reconstruct the eigenvalue distribution.

Looking at the distribution (\ref{4.2}) and the expansion of the random matrix $M$ in terms of the basis $T_{lm}$ (\ref{expansion}) we see that the correlation of two components of the field is
\begin{eqnarray}
	\contraction{}{c}{_{lm} }{c}c_{lm} c_{l'm'}=\delta_{ll'}\delta_{m m'} G(l),	
\end{eqnarray}
where $G(l)$ is the propagator given by the kinetic term $G(l)=1/[K(l)+r]$. As mentioned before, for the standard Laplacian kinetic term, we get $G(l)=1/[(l(l+1)+r]$. This then yields the two point function of the form
\begin{eqnarray}
	\avg{(M M)_{ij}}=\frac{1}{N}\sum_{l=0}^{N-1} (2l+1)G(l) \delta_{ij}\equiv f \delta_{ij},\label{4.4}
\end{eqnarray}
where $f$ is defined by this equation. We are now ready to investigate the moments of eigenvalue distribution of $M$. According to (\ref{3.30}), these are given by $\avg{\trl{M^{2m}}}$. We expect only the even moments to be non-vanishing due to the symmetry of the measure. We now single out one of the matrices in this trace and consider the contraction of this matrix with other matrix in the product, i.e.
\begin{eqnarray}
	\avg{(M^{2m-2-p})_{il}
	\contraction{}{M}{_{lr}(M^{p})_{rj}}{M}
	M_{lr}(M^{p})_{rj} M_{ji}}=c\ \delta_{li}\delta_{rj}\avg{(M^{2m-2-p})_{il}(M^{p})_{rj}},
\end{eqnarray}
where the two matrices that are being contracted are connected by the usual clip. Due to the planarity of the diagram, there are no contractions between the two groups of matrices. We can therefore write
\begin{eqnarray}
	\avg{(M^{2m-2-p})_{il}\contraction{}{M}{_{lr}(M^{p})_{rj}}{M} M_{lr}(M^{p})_{rj} M_{ji}}=\frac{f}{N} \avg{\trl{M^p}}\avg{\trl{M^{2m-2-p}}}.
\end{eqnarray}
Summing over all possible contractions, i.e. over all possible values of $p$ yields
\begin{eqnarray}
	\avg{\trl{M^{2m}}}=\frac{f}{N}\sum_{p=0}^{m-1}\avg{\trl{M^{2p}}}\avg{\trl{M^{2(m-1-p)}}} \ , \ m\geq1.
\end{eqnarray}
The condition on $m$ arises from the fact that we need at least one matrix $M$ to be able to consider any contractions. We now define the rescaled moments
\begin{eqnarray}
		F_{2m}=\frac{1}{N}\avg{\trsl{ \lr{\frac{M}{2\sqrt{f}}}^{2m}}},\label{4.12}
\end{eqnarray}
which are going to be finite and in the terms of which the recursion rule becomes
\begin{eqnarray}
	4F_{2m}=\sum_{p=0}^{m-1}F_{2p}F_{2(m-1-p)}.\label{4.13}
\end{eqnarray}
Note that $F_0$ enters here as an initial condition and is not given by this formula. But clearly $F_0=\avg{\Tr\ id}/N=1$. One could immediately recognize equation (\ref{4.13}) as a property of the Catalan numbers and identify $F_{2m}$ with a rescaled version of these. However, to illustrate better the method used in \cite{our1} for further calculations, we will do something different. We will define the function
\begin{eqnarray}
	\phi(t)=\sum_{m=0}^\infty t^{2m} F_{2m},
\end{eqnarray}
which is the moment generating function for the distribution $\rho(x)$, the distribution of the eigenvalues of the matrix $M$. If we now multiply (\ref{4.13}) by $t^{2m}$ and sum over all $m$, we obtain
\begin{align}
	4\sum_{m=1}^\infty t^{2m} F_{2m}=&t^2 \sum_{m=1}^\infty \sum_{p=0}^{m-1}t^{2p}F_{2p}t^{2(m-1-p)}F_{2(m-1-p)},\\
	4\big( \phi(t)-1\big)=&t^2\phi^2(t).\label{4.17}
\end{align}
This is an easy quadratic equation for $\phi(t)$. We chose the solution which has the correct $t\to 0$ limit and obtain
\begin{eqnarray}
	\phi(t)=2\frac{1-\sqrt{1-t^2}}{t^2}.\label{4.18}
\end{eqnarray}
Looking back at the definition of the resolvent (\ref{3.33}) we find
\begin{eqnarray}\label{resol}
	\omega(z)=\frac{1}{z} \phi(1/z)
\end{eqnarray}
and with the discontinuity equation (\ref{3.22}) we see that such generating function corresponds to the distribution
\begin{eqnarray}
	\rho(\lambda)=\frac{2}{\pi}\sqrt{1-\lambda^2}.
\end{eqnarray}
As advertised, this is nothing else than the normalized Wigner distribution, which has Catalan numbers as the moments.\footnote{More precisely the $2n$-th moment of the semicircle distribution of radius $R$ is $c_n (R/2)^{2n}$, with $c_n$ the $n$-th Catalan number.} From the definition of the moments (\ref{4.12}) we see that the radius of the distribution is 
\be\label{31radscaled}
R=2\sqrt f\ .
\ee
It is not difficult to compute the factor $f$ for the case of the usual kinetic term. We introduce $l=Nx$ and compute
\be
f=\frac{1}{N}\sum_{l=0}^{N-1}\frac{2l+1}{l(l+1)+r}=\int_0^1dx\frac{2Nx}{N^2 x^2+r}=\frac{1}{N}\log\lr{1+\frac{N^2}{r}}\ .\label{31efforSF}
\ee
We obtained
\be
R^2=\frac{4}{N}\log\lr{1+\frac{N^2}{r}}\ .
\ee
Clearly to obtain a value which includes $r$, we have to scale $r\to rN^2$. To obtain a finite value for the radius, we then need to scale $M\to MN^{-1/2}$, which can be checked to give a finite value for $\rho(\lambda)/N$. Note, that opposing to the case of the matrix models from the section \ref{sec22}, we did not have a freedom in scaling and both the scaling of $r$ and $M$ are now given. This is easily seen in terms of the action, which now includes one extra term, the kinetic term. We need to fix the scaling of this term to $N^2$ also, which uses up the freedom from before.

Note, that without any scaling, the eigenvalues would slowly come together in the large $N$ limit, even faster so if we enhance the mass term. This means that the kinetic terms introduces an attractive force among the eigenvalues, which overcomes the Vandermonde repulsion.

After everything is rescaled, comparing the formula for the radius of the Wigner distribution (\ref{22radwigner})
\be
R^2=\frac{4}{r}
\ee
with the new radius
\be
R^2=4\log\lr{1+\frac{1}{r}}\approx\frac{2}{r}-\frac{1}{r^2}+\ldots
\ee
we see, that the effect of the kinetic term is to shrink the eigenvalue distribution. We again see that the kinetic term thus corresponds to an attractive force among the eigenvalues, competing against the repulsive force of the Vandermonde determinant, however now tamed by the scaling to give a finite value for the radius in the large $N$ limit. As we have seen before, we can expect two new possible features of this new attraction.

First of all, we could obtain a stable solution for the values of parameters, where the pure matrix model does not allow such solution, as we have seen for for example in (\ref{24simplecorrection}). However, since the radius of the semicircle distribution diverges as $r\to0$, this is not going to happen also in the interacting theory, as we will see in the section \ref{sec33simplphase}.

Second, the asymmetric one cut solution of the interacting pure matrix model (\ref{22quarasym}) could become the energetically preferred solution. Recalling the results of the numerical simulations mentioned in the section \ref{sec13}, we expect this to be happen.

The same method of explicit Wick contractions was used in \cite{our1} also to compute the eigenvalue distribution of the matrix $B=\K M$ and the joint distribution of the matrices $M,B$. It was shown that $B$ follows also the semicircle law with radius given by $\avg{BB}$ and that the eigenvalues of $M$ and $B$ are correlated as follows
\be
\rho(x,y)=\rho(x)\rho(y)\frac{1-\gamma^2}{(1-\gamma^2)^2-4\gamma(1+\gamma^2)xy+4\gamma^2(x^2+y^2)}\ ,
	\label{4.46}
\ee
where
\be
\gamma=\frac{\avg{MB}}{\sqrt{\avg{BB}\avg{MB}}}\ .
\ee
In \cite{our2}, this result was extended to the joint distribution of three and four matrices.

In \cite{stein05_1,stein05_2}, it was proposed to take the rescaling of the radius of the distribution into account by describing the fuzzy field theory by a matrix model with the action
\be
\half \trl{M\K M}+\half r \trl{M^2}\ \to\ \half \frac{1}{f(r)}\trl{M^2}\ ,\label{3_1efpot}
\ee
or\footnote{Note, that this is not very satisfactory, since the integral $I$ should depend only on the matrix $M$. However, it is still a good starting point for the investigation of the consequences of the kinetic term.}
\be
I=e^{-\half \lr{\frac{1}{f(r)}-r}\trl{M^2}}\ .
\ee
Looking at (\ref{22radwigner}) and (\ref{31radscaled}), it is not difficult to see that the two models do have the same eigenvalue distribution. Adding the interaction term $g \trl{M^4}$ to (\ref{3_1efpot}), one can investigate consequences of the kinetic term to the interacting theory. The following prescription for the renormalization of the theory was proposed.

The range of the eigenvalue distribution of the model
\be
S_{eff}(M)=\half \frac{1}{f(r)}\trl{M^2}+g \trl{M^4} \label{3_1fulleff}
\ee
can be straightforwardly computed using (\ref{22quarticdelta}) to be
\be 
\delta=\frac{\sqrt{48g+\frac{1}{f^2(r)}}-\frac{1}{f(r)}}{6g}\ .
\ee
This radius can be used to define a correlation length. Since the correlation length is given by the mass parameter, one can look for a free theory of the same correlation length and identify its mass parameter $r_R$ as the renormalized mass. Matching the radii of the free and interacting distributions
we obtain
\bal
\frac{\sqrt{48g+\frac{1}{f^2(r)}}-\frac{1}{f(r)}}{6g}=4f(r_R)\ .
\end{align}
In \cite{stein05_1}, this formula was used in the context of regularizing the theories on non-compact $\R^{2d}_\theta$ and it has been shown that it reproduces the conventional one loop calculations in both two and four dimensions.

The phase structure of the model (\ref{3_1fulleff}) was further investigated. The conditions for the eigenvalue distribution
%\bal
%\ ,\\
%\ ,
%\end{align}
have solution only above some critical value $g_c$. It was then concluded that the deformed phase transition line (\ref{22basictrln}) terminates, reproducing the numerical results. We will however see, that this is not the case. When one considers a better approximation of the angular integral $I$, this critical point vanishes and the line extends all the way to origin. We will return to this point later.

%% file: 3_2multitrace.tex
Even though the considerations of the previous section could explain some properties of the field theory, they are not sufficient if one wants to treat the theory more precisely. We will have to find a way how to handle the integral (\ref{31anfintegral}). We will present three different way how this can be approached. They differ in their precision and scope and are useful in different settings.

The first two compute the integral perturbatively \cite{ocon,samann,samannnew}, order by order in the powers of kinetic term. At higher orders, the formulas become very involved and not very practical, but at low orders this method can capture all the features of the angular integral.

The third method approaches the integral little differently \cite{poly}. It is capable of capturing some part of the integral {\ref{31anfintegral}) non-perturbatively, however misses a class of other terms completely.

In the rest of this section we will describe the methods and their results, in the next section we will analyze the resulting phase structure.

\subsubsection{Perturbative angular integral via the character expansion computation}

The first attempt to consider the angular integral was made in \cite{ocon}, where a method was developed to compute the angular integral as a perturbative expansion in the powers of the kinetic term. This method was later extended in \cite{samann} to compute the terms of the first three orders.

Let us introduce a small parameter $\ep$
\be\label{32angint}
\int dU\,e^{-\half \ep \trl{M\K M}}=\int dU\,e^{-\half \ep \trl{M[L_i,[L_i, M]]}}\ ,\ee
where we have restricted ourselves to the the standard kinetic term (\ref{12standkint}).\footnote{The factor of $N^2$ is to reappear shortly.} Parameter $\ep$ can be thought of as an inverse temperature in the statistical model and we are thus performing a high temperature expansion.

If we introduce a collective index $\mu=0,\ldots,N^2-1$ for the polarization tensors $T_{l,m}$ and write $M=c_\mu T_\mu$ with $c_\mu=\trl{M T_\mu}$, we can write
\be
\half\trl{M[L_i,[L_i, M]]}=c_\mu c_\nu \half\trl{T_\mu[L_i,[L_i, T_\nu]]}=c_\mu c_\nu K_{\mu\nu}\ .
\ee
Note, that the matrix $K$ is constant and does not depend on the matrix $M$. Since the kinetic term vanishes for the identity matrix, $\mu=0$ term can be omitted. In this notation, the argument the integral (\ref{32angint}) can be written as a series in $\ep$
\bal
\int dU\,e^{-\half \ep \trl{M\K M}}=&1-\ep\int dU\,K_{mn}c_m c_n+\ep^2\half \int dU\,(K_{mn}c_m c_n)^2+\no&+\ep^3 \frac{1}{6}\int dU\,(K_{mn}c_m c_n)^3=1-\ep I_1+\ep^2\half I_2-\ep^3\frac{1}{6}I_3\ ,
\end{align}
where $m,n=1,\ldots,N^2-1$. Once we know the integrals $I_{1,2,3}$, it is straightforward to reexponiate and obtain
\be
I=e^{-\lr{\ep I_1+\ep^2\half\lr{I_1^2-I_2}+\ep^3\frac{1}{6}\lr{I_3-3I_1 I_2+2 I_1^3}}}\ .
\ee

The integrals $I_{1,2,3}$ are computed using several group theoretic methods and identities. After the separation of $M$ into eigenvalues and angular degrees of freedom $M=U\Lambda U^\dagger$, instead of explicitly computing integrals of the form
\be
\int dU\,U_{p_1q_1}\ldots U^\dagger_{r_1s_1}\ldots
\ee
one writes e.g. for $I_1$
\be\label{32haarint}
I_1=K_{m_1n_1}\int dU\,\trl{(U\otimes U)(\Lambda\otimes \Lambda)(U^\dagger\otimes U^\dagger)(T_{m_1}\otimes T_{n_1})}\,
\ee
and similarly for the other two cases. We have thus rewritten the integral of product of traces as an integral of a single trace in a (tensor) product of the fundamental representation. This product can be decomposed into irreducible representations of $SU(N)$ and an orthogonality relation for the Haar measure \cite{ocon} can be used to compute integrals like (\ref{32haarint}). We then get
\be
I_1=K_{m_1n_1}\sum_{\rho}\frac{1}{\textrm{dim}\,\rho}\textrm{Tr}_\rho\lr{\Lambda\otimes \Lambda}\textrm{Tr}_\rho\lr{T_{m_1}\otimes T_{n_1}}\ ,
\ee
where the sum goes over all the irreducible representations contained in the 2 (or 4 or 6) fold tensor product of the fundamental representation of $SU(N)$. The traces are taken in the particular representation $\rho$, and the first trace of $\Lambda$ is simply the character $\chi_\rho(\Lambda)$. The computation of the other trace is more involved and will be omitted here. Interested reader is refereed to the original work \cite{ocon,samann}. As a result, one obtains rather complicated formulas for the integrals $I_{1,2,3}$ involving traces of powers of the matrix $K_{mn}$, which are valid for any value of $N$. Here, we will present only their large-$N$ limit, which was computed in \cite{samann} for the cases of ${\CPn,n=1,2,3}$.
\begin{subequations}\label{sam}
\bal
S_{eff}^{(n=1)}=&N^2\half\Bigg[\ep \half\lr{c_2-\frac{1}{N}c_1^2}-\ep^2 \frac{1}{24}\lr{c_2-\frac{1}{N}c_1^2}^2-\label{sam1}\\
&-\ep^3\frac{1}{216}N\lr{c_3-3\frac{1}{N} c_1 c_2+2\frac{1}{N^2}c_1^3}^2\Bigg],\no
S_{eff}^{(n=2)}=&N^2\half\Bigg[\ep \frac{4}{3}\frac{1}{N}\lr{c_2-\frac{1}{N}c_1^2}-\ep^2 \frac{1}{9N^2}\lr{c_2-\frac{1}{N}c_1^2}^2-\label{sam2}\\
&-\ep^3\frac{1}{405N^3}\lr{8\lr{c_2-\frac{1}{N}c_1^2}^3+5N\lr{c_3-3\frac{1}{N} c_1 c_2+2\frac{1}{N^2}c_1^3}^2}\Bigg],\no
S_{eff}^{(n=3)}=&N^2\Bigg[ \ep \frac{6^{2/3}3}{4}\frac{1}{N^{4/3}}\lr{c_2-\frac{1}{N}c_1^2}-\ep^2 \frac{ 6^{4/3}3}{160}\frac{1}{N^{8/3}}\lr{c_2-\frac{1}{N}c_1^2}^2-\label{sam3}\\
&-\ep^3\frac{3}{400}\frac{1}{N^4}\lr{10\lr{c_2-\frac{1}{N}c_1^2}^3+3N\lr{c_3-3\frac{1}{N} c_1 c_2+2\frac{1}{N^2}c_1^3}^2}\Bigg].\nonumber
\end{align}
\end{subequations}
For all the terms to contribute, we have to rescale the matrix $M$, or equivalently the eigenvalue $x$ such that the terms in the parentheses do not contain any factors of $N$. If we write ${M=\tilde M N^{\theta x}}$, we obtain
\be
c_1\sim N^{1+\theta_x}\ ,\ c_2\sim N^{1+2\theta_x}\ .
\ee
Note that this means that opposing to the section \ref{sec2}, we are defining the moments without the normalization factor $1/N$. The finite value of the moments in the large $N$ limit will be guaranteed by the scaling of the matrix $M$. This condition fixes the scaling of the matrix to
\bse\label{32firstscaling}
\be
\theta_{x,1}=-\half\ , \ \theta_{x,2}=0\ ,\ \theta_{x,3}=\frac{1}{6}\ .
\ee
This the fixes the scaling of the mass and the coupling ${r=\tilde r N^{\theta_r},g=\tilde g N^{\theta_g}}$, such that the corresponding terms scale as $N^2$
\be
\theta_{r,1}=2\ , \ \theta_{r,2}=1\ ,\ \theta_{r,3}=\frac{2}{3}\ ,
\ee
and
\be
\theta_{g,1}=3\ , \ \theta_{g,2}=1\ ,\ \theta_{g,3}=\frac{1}{3}\ .
\ee
\ese

Let us also mention that the original work \cite{samann} uses rather different notation than the presented formulas. We will now present a short dictionary between the two notations, the quantities as in \cite{samann} denoted by a bar. The action has been
\bal
\bar S=&\beta\,\trl{\bar M [\bar L_i,[\bar L_i,\bar M]]+\bar r \bar M^2+\bar g \bar M^4}\ ,\\
& [\bar L_i,[\bar L_i,T^l_m]]=2l(l+n)T^l_m\ .
\end{align}
To bring the kinetic term into the same form, we need to write $M= \bar M\sqrt{4 \beta}$. This then means that $r=\half\bar r$ and $g= \frac{\bar g}{16\beta} $. For the different scaling we obtain
\be
\theta_x=\frac{\bar \theta_\lambda+\half \bar \theta_\beta}{n} \ , \ \theta_r=\frac{\bar \theta _r}{n} \ , \ \theta_g=\frac{\bar \theta_g - \bar \theta_\beta}{n}\ ,
\ee
where the extra factor of $n$ is simply due to the fact that in \cite{samann}, quantities were scaled with $L$. And finally, since the moments were not normalized, we have
\be c_m=\frac{l^n}{n!}(4\beta)^{m/2}\bar c_m\ .\ee

\subsubsection{Perturbative angular integral via the bootstrap}

Recently a different method has been developed to compute the perturbative expansion of the integral (\ref{31anfintegral}) \cite{samannnew}. The expressions one encounters are easier to manipulate and authors of \cite{samannnew} have been able to push the expansion one order further. This will be particularly interesting, as we will encounter multitrace terms of quartic order. We will now briefly describe the idea of the method and then state the results.

The idea is the following. We write the kinetic term effective action as a general multitrace expression
\bal
S_{eff}=&a_2 \trl{M^2}+a_{1,1}\trl{M}\trl{M}+a_4\trl{M^4}+a_{3,1}\trl{M^3}\trl{M}+\no
&+a_{2,2}\trl{M^2}^2+a_{2,1,1}\trl{M^2}\trl{M}^2+\ldots\ .\label{32generalmult}
\end{align}
If we now find a differential operator, for which
\be De^{-S_{kin}(M)}=O(\Lambda)e^{-S_{kin}(M)}\ ,\label{32difkin}\ee
where $O(\Lambda)$ is an expression invariant under ${M\to U M U^\dagger}$, we clearly obtain
\be D\int dU\,e^{-S_{kin}(U MU^\dagger)}=O(\Lambda)\int dU\,e^{-S_{kin}(UMU^\dagger )}\ .\ee
This then yields
\be De^{-S_{eff}(M)}=O(\Lambda)e^{-S_{eff}(\Lambda)}\ .\ee
So if we find an operator such that (\ref{32difkin}), we get a constraint on the coefficients in (\ref{32generalmult}).

One such operator is given by the translational invariance of the kinetic term and is
\be D_1=\sum_a \pd{}{M_{aa}}\ .\ee
Since $D_1e^{-S_{kin}}=0$ we get $D_1e^{-S_{eff}}=0$. This yields constraints
\begin{align}\label{32thanksto}
a_{1,1}=-\frac{1}{N}a_2\ ,& \ a_{3,1}=-\frac{1}{N}4a_4\ ,\no
a_{1,1,1,1}=-\frac{1}{2N}a_{2,1,1}\ ,&\ a_{2,1,1}=-\frac{1}{2n}\lr{3 a_{3,1}+4a_{2,2}}\ ,\no
a_{1,1,1,1,1,1}=-\frac{1}{3N}a_{2,1,1,1,1}\ ,&\ a_{4,1,1}=\frac{1}{2N}(5a_{5,1}+2 a_{4,2})\ ,\\
\vdots&\nonumber
\end{align}
Further, considering higher order differential operators of the form
\begin{align}
\left.\pd{}{M_{ab}}\pd{}{M_{ba}}\right|_{M=0}\ ,& \ 
\left.\pd{}{M_{aa}}\pd{}{M_{bb}}\right|_{M=0}\ ,\no
\left.\pd{}{M_{ab}}\pd{}{M_{bc}}\pd{}{M_{cd}}\pd{}{M_{da}}\right|_{M=0}\ ,&\ 
\left.\pd{}{M_{ab}}\pd{}{M_{bc}}\pd{}{M_{ca}}\pd{}{M_{dd}}\right|_{M=0}\ ,\no
\left.\pd{}{M_{ab}}\pd{}{M_{ba}}\pd{}{M_{cd}}\pd{}{M_{dc}}\right|_{M=0}\ ,&\ 
\left.\pd{}{M_{ab}}\pd{}{M_{ba}}\pd{}{M_{cc}}\pd{}{M_{dd}}\right|_{M=0}\ ,\no
\left.\pd{}{M_{aa}}\pd{}{M_{bb}}\pd{}{M_{cc}}\pd{}{M_{dd}}\right|_{M=0}\ &
\end{align}
yields remaining conditions up to fourth order. For higher orders, similar higher order differential operators need to be considered.

This way, one obtains algebraic equations for the coefficients $a$ in terms of traces of products of matrices $L_i$. The general formulas valid at a general $N$ and a general $\CPn$ are again rather complicated. The first three expressions are, in the large $N$ limit,
\bal
a_2=&\frac{\Sigma_1}{2N^2}\ ,\\
a_4=&\frac{\beta\Sigma_1(2n(n+1)(n+2)N^4+N^2(4-n(n+2))\Sigma_1)}{32(n+2)N^{9}}\ ,\\
a_{2,2}=&-\frac{\beta\Sigma_1 4n(n+1)(n+2)N^6}{16n(n+2)N^{12}}+\frac{\beta \Sigma^2(-N^6)}{16n(n+2)N^{12}}\ ,
\end{align}
where $\Sigma_1$ is trace over the eigenvalues of the quadratic Casimir given by (in the large-$N$ limit)
\be \Sigma_1=\frac{2L^{2n+2}}{(n+1)!(n-1)!}\ .\ee
Thus
\bse\label{32samana222}
\bal
a_2=&\frac{n}{2(n+1)}\lr{n!N}^{2/n}\ ,\\
a_4=&\frac{(4-n(n+2))n^2\lr{n!}^{4/n}N^{\frac{4}{n}-3}}{8(n+1)^2(n+2)}\ ,\\
a_{2,2}=&-\frac{n}{4(n+1)^2(n+2)}\lr{n!}^{4/n}N^{\frac{4}{n}-2}\ .
\end{align}
\ese

For the case of the fuzzy sphere, after rescaling, which turns out to be the same as (\ref{32firstscaling}), the effective action becomes in the large $N$ limit
\bal
S_{eff}=&N^2\Bigg[ \frac{1}{4}t_2-\frac{1}{3\times16}t_2^2-\frac{34}{135\times16^2}t_2^4-\no&-
\frac{4}{27\times 64}t_3^2-\frac{10}{135\times 16^2}t_4^2+\frac{40}{135\times 16^2}t_4 t_2^2\Bigg]\label{32raw}\ ,
\end{align}
where $t$'s are the systematized moments
\be
t_n=\trl{M-\frac{1}{N}\trl{M}}\ ,
\ee
or explicitly
\begin{subequations}
\begin{align}
t_2=&\trl{M^2}-\frac{1}{N}\trl{M}^2\ ,\\
t_3=&\trl{M^3}-3\frac{1}{N} \trl{M}\trl{M^2} + 2\frac{1}{N^2} \trl{M}^3\ ,\\
t_4=&\trl{M^4}- 4 \frac{1}{N}\trl{M^3} \trl{M} + 6 \trl{M^2} \trl{M}^2 - 3\frac{1}{N^3} \trl{M}^4\ .
\end{align}
\end{subequations}
Note, that the effective action could be rewritten in the terms of $t$'s rather than $c$'s thanks to the conditions (\ref{32thanksto}). (\ref{32raw}) can then be rearranged into
\bal
S_{eff}=N^2\Bigg[\half \lr{\half t_2-\frac{1}{24}t_2^2+\frac{1}{2880}t_2^4}-\frac{1}{432}t_3^2-\frac{1}{3456}\lr{t_4-2t_2^2}^2\Bigg]\ .\label{32rearr}
\end{align}

\subsubsection{Non-perturbative angular integral via the boostrap}\label{polyapproach}

We will now present the third method to handle the angular integral. It has been developed in \cite{poly} for the fuzzy sphere and later used in \cite{mine1} for the higher projective spaces.

The starting point is the fact from the section \ref{sec31} that the introduction of the kinetic term into the action of the free matrix model with ${g=0}$ does not change the shape of the eigenvalue distribution, but only rescales its radius.

We write (\ref{31angeff}) again
\be
\int dU\,e^{-\half \trl{M\K M}}=e^{-S_{eff}(\Lambda)}\ ,
\ee
the effective action being function of the eigenvalues of the matrix $M$ and therefor can be written in the terms of the trace invariants $c_n=\trl{M^n}$. Moreover, since we have assumed the identity matrix to have a vanishing kinetic term, the effective action is a function of the transitionally invariant traces
\be
t_n=\trl{M-\frac{1}{N}\trl{M}}^n\ .
\ee
We obtained the same result as the consequence of the translational invariance not too long ago in (\ref{32raw}).

The fact, that for the free theory the solution is a semicircle again is incorporated by writing the effective action in the following way
\be\label{32sefffull}
S_{eff}=\half F(t_2)+a_1 t_3^2+(b_1+b_2 t_2)(t_4-2t_2^2)^2+c_1(t_6-5 t_2^3)(t_4-2 t_2^2)\ldots=\half F(c_2)+\mathcal R\ .
\ee
Here, $F(t_2)$ is a function to be determined and $a,b,c$'s are some numerical factors. The remainder term $\mathcal R$ contains products of
\be\label{32difsemi}
t_{2n}-C_n t_2^n=t_{2n}-\frac{(2n)!}{n!(n+1)!} t_2^n\ ,\ t_{2n+1}\ ,
\ee
$C_n$ the $n$-th Catalan number, and captures the difference between the distribution given by $t_n$ and the Wigner distribution. The effective action must always contain a product of at least two such terms. This can be seen as follows.

The saddle point equation given by the action $S_{eff}+\half z \trl{M^2}$ is
\be\label{32effsadle}
(z+F'(c_2))\lambda_i+\pd{\mathcal R}{\lambda_i}=(z+F'(c_2))\lambda_i+\sum_{m}\pd{\mathcal R}{t_m}\pd{t_m}{\lambda_i}=2\sum_{i<j}\frac{1}{\lambda_i-\lambda_j}\ .
\ee
The semicircle distribution has to be solution of this equation. The first term rescales it's radius, all the terms in the sum have to vanish when $t_m$ are the moments of the semicircle distribution, as they would produce higher order terms in the equation. So $\mathcal R$ must contain product of at least two terms which vanish when (\ref{32difsemi}) vanishes.

We can see that the perturbative solution from the previous section (\ref{32rearr}) has precisely the form given by (\ref{32sefffull}).

We will denote ${t_2=t}$. The function $F(t)$ is then determined by (\ref{31radscaled}). For a semicircle distribution $\mathcal R$ vanishes and equation (\ref{32effsadle}) yields a solution with radius
\be
R^2=\frac{4N}{z+F'(t)}
\ee
and the second moment
\be
t=\frac{N^2}{z+F'(t)}\ .
\ee
Matching these with (\ref{31radscaled}) yields after a little of algebra
\begin{subequations}\label{32effact}
\bal
F'(t)+z=&\frac{N^2}{t}\label{effact1}\ ,\\
t=&f(z)\ .\label{effact2}
\end{align}
\end{subequations}
Using the expression (\ref{31efforSF}) for $f(z)$ for the case of the standard kinetic term on the fuzzy sphere then yields
\be
F_{S^2_F}(t)=N^2\log\lr{\frac{t}{1-e^{-t}}}\label{32FS2}\ .
\ee
The power series expansion of this expression is
\be
F(t)=\half t - \frac{1}{24}t^2 + \frac{1}{2880}t^4+\ldots=\half (c_2-c_1^2) - \frac{1}{24}(c_2-c_1^2)^2 + \frac{1}{2880}(c_2-c_1^2)^4+\ldots\ ,
\ee
matching the first orders of the perturbative solution (\ref{32rearr}).

From (\ref{32FS2}) we see, that we require $t$ not to scale with $N$ to obtain a consistent large-$N$ limit. This then yields the scaling of the matrix to be $\theta_x=-1/2$, in agreement with the fuzzy sphere case of (\ref{32firstscaling}).

The path from here to the phase diagram is straightforward. This is a particular case of the multitrace matrix model from the section \ref{sec24} and we will study its phase structure in the next section. But before we do that, let us conclude this section with the presentation of the method for a general kinetic term and then for the case of the the fuzzy $\CPn$.

For a general kinetic term, let the large $l$ dependence of $K(l)$ be
\be K(l)\sim l^\alpha \ .\ee
We then straightforwardly obtain
\be f(z)=N^{2-\alpha} \bar f(zN^{-\alpha})\ ,\ee
where
\be \bar f(x)=\int \frac{d y}{y^{\alpha/2}+x}\ .\ee
If we denote the inverse of $\bar f$ as $g$, we then get
\be F'(t)=\frac{N^2}{t}-N^\alpha g(tN^{\alpha-2})\ .\ee
Thus $F$ has the right scaling properties if ${t\sim N^{2-\alpha}}$ which yields general scaling of the matrix
\be\label{32thetaalpha}
\theta_{x,\alpha}=1-\frac{\alpha}{2}\ .
\ee

Computing the function $f(z)$ for the case of the fuzzy $\CPn$ is straightforward. We just need to recall the section \ref{secCPn} and the formulas (\ref{11CPnspectrum}) for the eigenvalues of the standard Laplacian and for the multiplicity. We then obtain
\be
f(z)=\sum_{l=0}^L \frac{\textrm{dim}(n,l)}{l(l+n)+z}\approx 2n N^2 \int_0^1 dx \frac{x^{2n-1}}{L^2 x^2+z}\ ,
\ee
where we have rescaled $l= L x$ and used the mentioned formulas
\be
\textrm{dim}(n,l)\sim\frac{l^{2n-1}}{(n-1)!^2n} \ , \ N\sim\frac{L^n}{n!}\ .
\ee
The result is then
\be
f(z)=\frac{N^2}{z}\tFo\lr{1,n;n+1,-\frac{(n!N)^{2/n}}{z}}\,
\ee
where $\tFo$ is the hypergeometric function. It is a little anticlimactic, as one immediately realizes, that we will not be able to obtain a closed formula for the effective action $F(t)$. The equations involved in solving (\ref{32effact}) would be transcendental. The only thing we can do is a perturbative calculation of $F(t)$ in powers of $t$, which can be carried out to an arbitrary order. We simply expand both $f$ and $F$ in (\ref{32effact}) and require the equations to hold order by order.

This yields up to the seventh order
\bal
\hspace*{-0.4cm}
F(t)=&N^2\bigg[\frac{n(n!)^{2/n}}{n+1}\lr{N^{2/n-2}t}
-\frac{n(n!)^{4/n}}{2(n+1)^2(2+n)}\lr{N^{2/n-2}t}^2\no
&\hspace*{-0.3cm}-\frac{2(n-1)n(n!)^{6/n}}{3(n+1)^3(6+5n+n^2)}\lr{N^{2/n-2}t}^3\no
&\hspace*{-0.3cm}-\frac{n(12-24n+n^2+7n^3)(n!)^{8/n}}{4(n+1)^4(n+2)^2(12+7n+n^2)}\lr{N^{2/n-2}t}^4\no
&\hspace*{-0.3cm}-\frac{2n(-24+92n-77n^2-8n^3+17n^4)(n!)^{10/n}}{5(n+1)^5(n+2)^2(60+47n+12n^2+n^3)}\lr{N^{2/n-2}t}^5\no
&\hspace*{-0.3cm}-\frac{n\lr{\begin{array}{lr}720-3660n+3854n^2+1524 n^3-\\-2429 n^4-439 n^5+423 n^6+10n^7\end{array}}(n!)^{12/n}}{3(n+1)^6(n+2)^3(n+3)^2(120+74n+15 n^2+n^3)}\lr{N^{2/n-2}t}^6\no
&\hspace*{-0.3cm}-\frac{12n\lr{\begin{array}{lr}-720+5304 n - 10960 n^2+5426 n^3+4521 n^4 - \\3282 n^5-884 n^6+472 n^7+128n^8\end{array}}(n!)^{14/n}}{7(n+1)^7(n+2)^3(n+2)^2(840+638n+179 n^2+22 n^3+n^4)}\lr{N^{2/n-2}t}^7\no&\hspace*{-0.3cm}+\ldots\bigg]\ .\label{32effactionCPn}
\end{align}
This shows that we can rescale the matrix $M$ such that all the terms are of the same order and contribute. We namely get similarly to the previous sections
\be
\theta_x=\half-\frac{1}{n}\ , \ \theta_r=\frac{2}{n}\ ,\ \theta_g=\frac{4}{n}-1\ .
\ee
These scalings agree with the scalings (\ref{32firstscaling}) from \cite{samann} shown in the previous section, as well as with the $\CPn$ laplacian $\alpha=2/n$ in (\ref{32thetaalpha}). Moreover, the first terms agree with the perturbnative results (\ref{32samana222}) of \cite{samannnew}.

As we will see in the next section, the perturbative expressions like (\ref{32effactionCPn}) are not sufficient to study the phase structure of the theory near the origin of the parameter space. To overcome our inability to find a closed expression for the effective action, we will approximate (\ref{32effactionCPn}) using the method of Pade approximants \cite{pade}.

The behavior of $F(t)$ for large and small values of $t$ can be determined universaly \cite{poly}. Small values of $t$ correspond to large $z$ and from (\ref{32effact}) we obtain
\be F(t)=\lr{\int_0^1 dx\,2x K(N x)}t\ee
and we expect ${F\sim t}$. Large values of $t$ correspond to small $z$ and from (\ref{effact1}) one expects ${F\sim \log t}$. This allows us to search for the effective action in the form
\bse\label{32aproxmeth}
\be F(t)=\log\lr{1+t h(t)}\ .\label{32padeFh}\ee
We now find $h(t)$ in a form of ratio of two polynomials
\be\label{32padeFh2}
h(t)=\frac{a_0+a_1 t^1+\ldots+a_{M_a} t^{M_a}}{1+b_1 t+\ldots+b_{M_b} t^{M_b}}
\ee
\ese
such that the expansion of (\ref{32padeFh}) in powers of $t$ reproduces (\ref{32effactionCPn}). Since we know the expansion of $F(t)$ up to seventh order in $t$, we can take ${M_a=M_b=3}$. If we need more precision, we need to find more terms in the expansion of $F(t)$, as the known seven terms would not determine the coefficients uniquely. At the end of the day, we obtain the effective action in the form of a logarithm of the following expressions, ${F(t)=\log\lr{\ldots}}$
\begin{subequations}
\begin{align}
\hspace*{-0.4cm}
n=2 \ &,\ 
\frac{\lr{\begin{array}{lr}-11789184402450 + 93978027360 t + 5998294160460 t^2\\ + 2932050372000 t^3 + 497642752529t^4\end{array}}}{30(-392972813415+
527096352132t - 197206475949 t^2 + 39981049252 t^3)}\ ,\\
\hspace*{-0.4cm}
n=3 \ &,\ 
\frac{\lr{\begin{array}{lr}-824984552000 + 272764119600 6^{2/3} t + 1454433162840\,6^{1/3} t^2\\ + 1945660952760 t^3 + 167744125503\, 6^{2/3}
t^4\end{array}}}{560(-1473186700+ 1591968810\, 6^{2/3} t - 2246388516\, 6^{1/3} t^2 + 1411751691\, t^3)}\ ,\\
\hspace*{-0.4cm}
n=4 \ &,\   
\frac{\lr{\begin{array}{lr}148500768040625 + 25279264566220000\sqrt6 t + 150340268216376000 t^2\\ + 60165687487957760\sqrt6 t^3 + 60941670673759896 t^4\end{array}}}{5\lr{\begin{array}{lr}29700153608125 + 5008332667471000\sqrt6 t\\ - 18230533095002200 t^2 + 4236389020569472\sqrt6 t^3\end{array}}}\ ,\\
\hspace*{-0.4cm}
n=5 \ &,\ 
\frac{\lr{\begin{array}{lr}
159606423231255\, 2^{4/5} 15^{3/5} + 9098272445481450\, t\\ + 6433230719486025\, 2^{1/5} 15^{2/5} t^2
 + 2256395876247656\, 2^{2/5} 15^{4/5} t^3\\ + 5382149790792650\, 2^{3/5} 15^{1/5} t^4\end{array}}}{\lr{\begin{array}{lr}
159606423231255\, 2^{4/5} 15^{3/5} + 1117951283918700 t -\\
1890281408310225\, 2^{1/5} 15^{2/5} t^2 + 562828122041281\, 2^{2/5} 15^{4/5} t^3\end{array}}}\ .
\end{align}
\end{subequations}

%% file: 3_3phasesimple.tex
In this section, we will start with the description of the phase structure of the matrix models representing the fuzzy scalar field theories. The following two sections are based on \cite{mine1,mine2}.

In this section, we will investigate the phase structure of the second moment multitrace model (\ref{32FS2}). As it has been mentioned several times, this model includes multitrace terms of only the second moment, but includes it in all orders and packed in a nonperturbative way.

We will first study the symmetric regime with $c_1=0$ and consider the full model afterward.

\subsubsection{Perturbative and nonperturbative symmetric regime}

The phase structure of the model
\be\label{33simplesym}
S=\half F(c_2)+\half r \trl{M^2}+g \trl{M^4}
\ee
with $F(c_2)$ given by (\ref{32FS2}) is now a simple exercise. It is a particular multitrace matrix model, with the phase structure described in section \ref{sec24}. 

The set of conditions (\ref{22quartcond1},\ref{33cond2cut}) determining the eigenvalue distribution of the model (\ref{33simplesym}) in the one cut and two cut cases is
\bse\label{33firstconditions}
\bal
\frac{3}{4} \delta^2 g + \frac{1}{4}\delta\lr{r+\frac{1}{c_2}+\frac{1}{1-e^{c_2}}}=&1\ ,\\
\delta^2=\frac{1}{g}\ ,&\ 4Dg+r+\frac{1}{c_2}+\frac{1}{1-e^{c_2}}=0\ .
\end{align}
together with the conditions on the second moment for the one cut and two cut solution (\ref{2momentsphi4a}, \ref{22moments2cuta})
\be
c_2=\frac{\delta}{4}+\frac{\delta^3 g}{16}\ , \ c_2=D\ .
\ee
\ese
Unfortunately this is a set of transcendental equations and cannot be solved analytically, not even for the two cut case. We will approach these numerically shortly, but first we will look only at the phase transition line. At the phase transition, the condition on the second model simplifies as before
\be
c_2=\frac{1}{\sqrt g}\ .
\ee
The phase transition line between the one cut and the two cut distribution given by $\reff=-4\sqrt{g}$ becomes
\be
r(g)=-5\sqrt g - \frac{1}{1-e^{1/\sqrt g}}\ ,\label{12boundSF2}
\ee
first obtained in \cite{poly}. Note that this formula is also general and does not rely on any particular form of $F$ and can be used for any kinetic term and we get for the transition line in the general case
\be
r(g)=-4 \sqrt g-F'(1/\sqrt{g})\ .
\ee

\begin{figure}[tb]
\centering % \begin{center}/\end{center} takes some additional vertical space
\includegraphics[width=.8\textwidth]{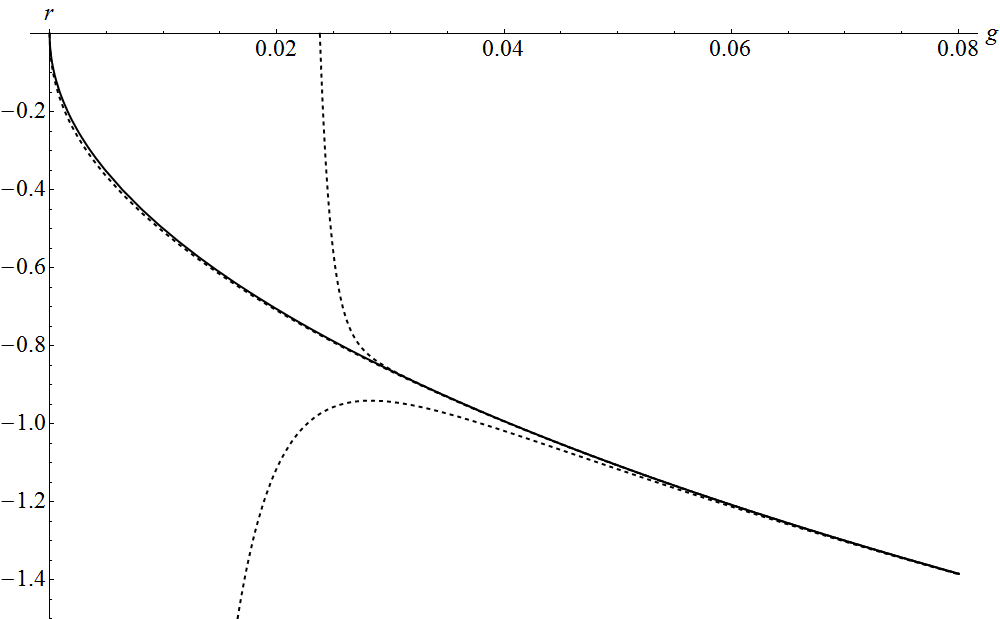}
% "\includegraphics" is very powerful; the graphicx package is already loaded
\caption{Comparison of the different results for the phase transition lines of the model (\ref{33simplesym}). The solid line denotes the exact transition line (\ref{12boundSF2}), the downward dashed line is the transition line computed from the first seven terms of the expansion of $F(t)$, namely
\[r(g)=-\half -\frac{1}{1209600 g^{7/2}} + \frac{1}{30240 g^{5/2}}- \frac{1}{720 g^{3/2}} + \frac{1}{12 \sqrt{g}} - 4 \sqrt{g}\ .\] The upward dashed line is a similar formula with $50$ terms used. The third dashed line is obtained using an approximation $F(t)$ using the Pade approximation method from the first seven terms of the perturbative series.}\label{ob33_differetlines}
\end{figure}

A striking feature of this formula is the essential singularity at $g=0$. Since the expansion of $F(t)$ in $t$ turns into the expansion of the transition line in $1/\sqrt{g}$, this renders any perturbative series an asymptotic one and unreliable near the origin of the parameter space. One has to be careful when interpreting the results with only first few terms of the expansion known. The figure \ref{ob33_differetlines} compares the exact, perturbative and approximate formulas for the transition line. Clearly a reasonable treatment of the raw perturbative data, such as the approximation method (\ref{32aproxmeth}),  is needed to obtain an useful phase transition line.

Note that if we work with a finite term approximation to $F$ to compute the phase transition lines, this result is an exact, nonperturbative line of that model. The only incomplete part is the knowledge of $F$.

\begin{figure}[tb]
\vspace*{-0.2cm}
\centering % \begin{center}/\end{center} takes some additional vertical space
\includegraphics[width=.7\textwidth]{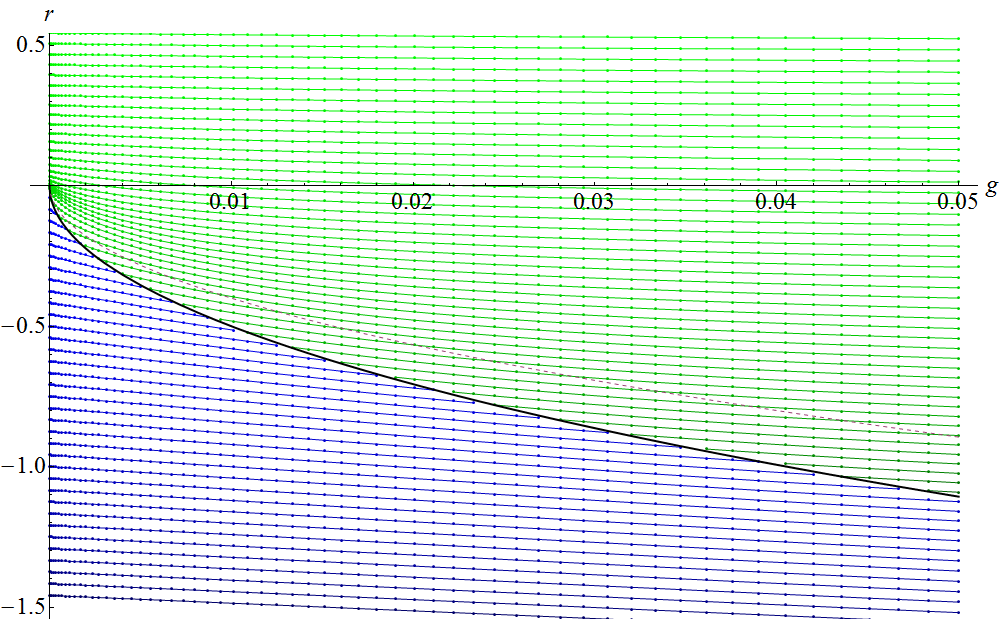}
\includegraphics[width=.7\textwidth]{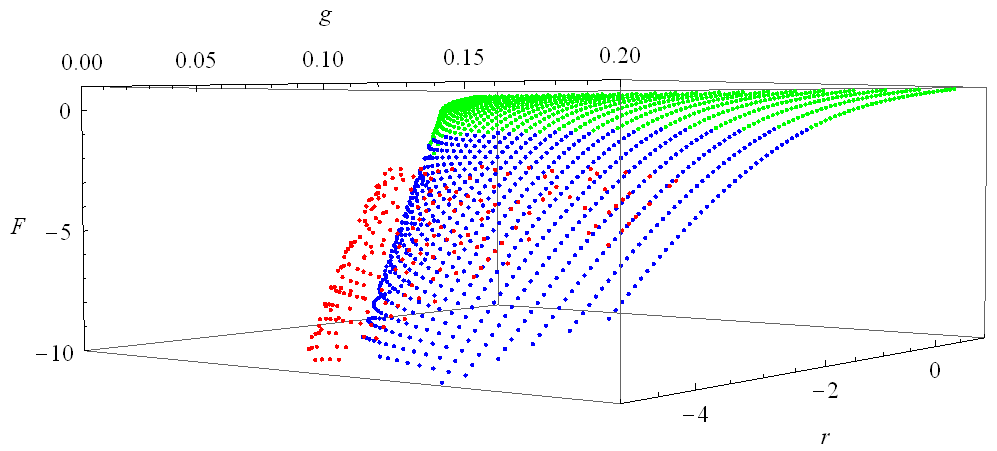}
% "\includegraphics" is very powerful; the graphicx package is already loaded
\vspace*{-0.2cm}
\caption{(Color online) The phase diagram and the free energy diagram of the model (\ref{33simplesym}) obtained numerically. The green region denotes the one cut solution, the blue region the two cut solution and the red region denotes the asymmetric one cut solution, which is easily seen to have higher free energy than the two cut solution. The green lines represent the symmetric one cut/asymmetric one cut phase transition, the red lines the symmetric one cut/two cut phase transition and the blue lines the two cut/asymmetric one cut phase transition. The blue line is the analytic transition line (\ref{12boundSF2}), the dashed line denotes the transition line of the original model.}\label{ob33symdiag}
\vspace*{-0.2cm}
\end{figure}

To get results away from the transition line, we have to retreat to numerical solution of (\ref{33firstconditions}). Using the method described in section \ref{sec22} we get a phase diagram and the free energy diagram of the theory show in the figure \ref{ob33symdiag}. We show also the asymmetric one cut solution of the model, but we can see that it has higher free energy than the other solutions and we will not discuss further.

Several authors have considered some perturbative results for the model \cite{ocon,samann,ydrinew1}. The phase transition lines the authors obtain are the perturbative lines similar to the dashed lines in the figure \ref{ob33_differetlines}, which have little hope to accurately describe the theory close to the origin. Moreover since no odd terms have been considered in the action, the asymmetric solutions should have a higher free energy. A region of parameters space where this is not true has been found in \cite{samann}, but this is only a relict of the perturbative treatment.

Most recently, perturbative phase structure of the model corresponding to the fuzzy disc scalar field theory has been considered \cite{samanfuzzydisc}, including the odd terms. Due to the perturbative treatment only, we expect the results to be changed in nonperturbative considerations.

\subsubsection{Perturbative asymmetric regime}

The first attempts to include the asymmetric solution of the matrix model corresponding to the uniform order phase of the field theory was presented in \cite{samann}. However the authors considered only the symmetric terms in the action, which is not satisfactory for the asymmetric regime.

In \cite{mine1} a model which included asymmetric terms was analyzed perturbatively. A simple model was considered, with asymmetry introduced by considering the complete asymmetric version of $t_2= c_2-c_1^2$ in the perturbative expression for $F(t)$. This yields a model
\bal
S_{eff}=&\half F(c_2 -c_1^2)+\half r \trl{M^2}+g \trl{M^4}\no\approx& \half A_1 (c_2 -c_1^2)+ \half A_2 (c_2 -c_1^2)^2+\half r \trl{M^2}+g \trl{M^4}\ .\label{33themodel}
\end{align}
The equations determining the asymmetric one cut solution (\ref{22generalonecut}) are (\ref{23kuk},\ref{2momentseffectiveasymmetric})
\begin{subequations}\label{mostimportant}
\bal
0=&-\half c_1 \lr{A_1+2A_2(c_2-c_1^2)}+2 D^3 g + 3 D \delta g \no&+\half D\big(r+A_1+2A_2(c_2-c_1^2)\big)\ ,\\
1=&3 D^2 \delta g +\frac{3}{4}\delta^2 g \quater\delta\big(r+A_1+2A_2(c_2-c_1^2)\big)\ ,\\
c_1=&3 D^3 \delta g + \frac{3}{2}D\delta^2 g + \quater D \delta\big(r+A_1+2A_2(c_2-c_1^2)\big)\ ,\\
c_2=&3 D^4 \delta g + 3 D^2 \delta ^2 g + \quater \delta^3 g + \quater D^2 \delta \big(r+A_1+2A_2(c_2-c_1^2)\big)+\no&+\frac{1}{16}\delta^2 \big(r+A_1+2A_2(c_2-c_1^2)\big)\ .
\end{align}
These are supplemented by the phase transition condition (\ref{23asymphasecondition})
\be
4 D^2 g + 2 \delta g + r + 4 D g x + 4 g x^2 + A_1+2A_2(c_2-c_162)=0\ .
\ee
\end{subequations}
As we have seen in the section \ref{sec23}, these equations are beyond any hope to solve analytically. The procedure adopted in \cite{mine1} was thus to solve them perturbatively in the kinetic term contribution. Technically, this meant replacing $A_1\to \ep A_1,A_2\to \ep^2 A_2$ and looking for solution of the equations (\ref{mostimportant}) as a power series in $\ep$. Obviously, if we include more terms of the expansion of $F(t)$ in (\ref{33themodel}), one can go further also in the expansion of the solution. The following, rather lengthy formula, for the transition line has been obtained for the seventh order approximation of $F$
\begin{align}
&\hspace*{-0.6cm}r(g)=
-2\sqrt 15 g+\frac{4 A_1}{5}+\lr{\frac{135 A_1^2+848 A_2}{1500 \sqrt 15}}\frac{1}{\sqrt g}
\label{33mega}
+\\&
+\frac{1}{g}\lr{\frac{2025 A_1^3+11 236 A_3+10170 A_1 A_2}{562500}}
+\no&
+\frac{1}{(\sqrt g)^3}\frac{
\lr{\begin{array}{lr}5740875 A_1^4 + 32464800 A_1^2 A_2 + 19805040 A_2^2 \\
+29707560 A_1 A_3 + 19056256 A_4\end{array}}}{2025000000 \sqrt 15}
+\no&
+\frac{1}{g^2}\frac{
\lr{\begin{array}{lr}19683000 A_1^5 + 108172665 A_1^2 A_3 + 70685676 A_1 A_4 + 31561924 A_5\\
 + 126372150 A_1^3 A_2 + 106028514 A_3 A_2 + 144230220 A_1 A_2^2
\end{array}}}{113906250000}
+\no&\hspace*{-0.44cm}
+\frac{1}{(\sqrt g)^5}\frac{
\lr{\begin{array}{lr}237487696875 A_1^6 + 1709468550000 A_1^4 A_2 + 2801052738000 A_1^2 A_2^2 \\
+  609290035200 A_2^3 + 1400526369000 A_1^3 A_3 + 2741805158400 A_1 A_2 A_3 \\
+ 423995740920 A_3^2 + 913935052800 A_1^2 A_4 + 753770206080 A_2 A_4 \\
+ 471106378800 A_1 A_5 + 26764511552 A_6
\end{array}}}{1366875000000000 \sqrt 15}
+\no&\hspace*{-0.44cm}
+\frac{1}{g^3}\frac{
\lr{\begin{array}{lr}
2391484500000 A_1^7 + 15074840643750 A_1^4 A_3 + 9480160743525 A_1 A_3^2  \\
+ 9749822053500 A_1^3 A_4 + 4512250464660 A_3 A_4 + 5266755968625 A_1^2 A_5  \\
 +376020872055 A_1 A_6 + 620602111612 A_7 + 19049945512500 A_1^5 A_2 \\
+ 43874199240750 A_1^2 A_3 A_2 + 16853619099600 A_1 A_4 A_2 \\+  3760208720550 A_5 A_2
+40199575050000 A_1^3 A_2^2 \\+  12640214324700 A_3 A_2^2 + 19499644107000 A_1 A_2^3
\end{array}}}{192216796875000000}
+\no&\hspace*{-0.44cm}
+\frac{1}{(\sqrt g)^7}\frac{
\lr{\begin{array}{lr}
51778255111171875 A_1^8 + 451444780593000000 A1^6 A_2  \\
 +1156207911142500000 A_1^4 A_2^2 + 886545939600000000 A_1^2 A_2^3  \\
 +97520988528864000 A_2^4 + 346862373342750000 A_1^5 A_3  \\
 +1329818909400000000 A_1^3 A_2 A_3 + 877688896759776000 A_1 A_2^2 A_3  \\
 +329133336284916000 A_1^2 A_3^2 + 170459137959667200 A_2 A_3^2  \\
 +221636484900000000 A_1^4 A_4 + 585125931173184000 A_1^2 A_2 A_4  \\
 +151519233741926400 A_2^2 A_4 + 227278850612889600 A_1 A_3 A_4 \\
 +23876085465308160 A_4^2 + 121901235661080000 A_1^3 A_5  \\
 +189399042177408000 A_1 A_2 A_5 + 44767660247452800 A_3 A_5  \\
 +9469952108870400 A_1^2 A_6 + 5969021366327040 A_2 A_6  \\
 +20891574782144640 A_1 A_7 + 4811616828772352 A_8\end{array}}}{3690562500000000000000\sqrt 15}\ .\nonumber
\end{align}
Note that this formula is also general and does not rely on any particular form of $F$ and can be used for any kinetic term.

It is important to stress that this is not truly a phase transition line yet. It is just a boundary of the region of existence of the asymmetric one cut solution. It is however not guaranteed that such solution will be realized, as it is not clear where, if at all the free energy of such solution is lower than of the symmetric solutions.

Moreover, the character of the expansion is not clear at the moment either. Assuming that the series in $\frac{1}{\sqrt{g}}$ is convergent, one can use the formula (\ref{33mega}) to obtain an approximate location of the triple point of the theory. The table \ref{table1} shows the location of the triple point for different orders of the perturbative approximation for the fuzzy sphere.

\begin{table}[tb]
\centering
\begin{tabular}{|c|ll|}
\hline
order & \multicolumn{1}{c}{$g_c$} & \multicolumn{1}{c|}{$r_c$} \\
\hline 
2 & 0.02091 & -0.7221\\
3 & 0.02127 & -0.7282\\
4 & 0.0215 & -0.7326\\
5 & 0.02171 & -0.7356\\
6 & 0.02185 & -0.7379\\
7 & 0.02195 & -0.7397\\
8 & 0.02204 & -0.7412\\
\hline
\end{tabular}
\caption{Comparison of the location of the triple point in the case of the fuzzy sphere for different perturbative orders.}\label{table1}
\end{table}

\begin{table}[tb]
\centering
\begin{tabular}{|c|ll|}
\hline
n & \multicolumn{1}{c}{$g_c$} & \multicolumn{1}{c|}{$r_c$} \\
\hline
1 & 0.02160 & -0.7369\\
2 & 0.1714 & -2.048\\
3 & 0.9320 & -5.333\\
4 & 2.890 & -9.784\\
5 & 6.713 & -15.20\\
\hline
\end{tabular}
\caption{Numerical values for the $\CPFn$ triple points. The value for $n=1$, i.e. the fuzzy sphere, was obtained using the approximate phase boundary and not using the exact expression.}\label{tabulka2}
\end{table}

The results for the first few $\CPn$'s are given in the table \ref{tabulka2}. If however the expansion is asymptotic, it is not clear how well or if at all these values approximate the location of the triple point. Clearly, a non-perturbative tool is needed.

\subsubsection{Non-perturbative asymmetric regime}

\begin{figure}[tb]
\centering % \begin{center}/\end{center} takes some additional vertical space
\includegraphics[width=.7\textwidth]{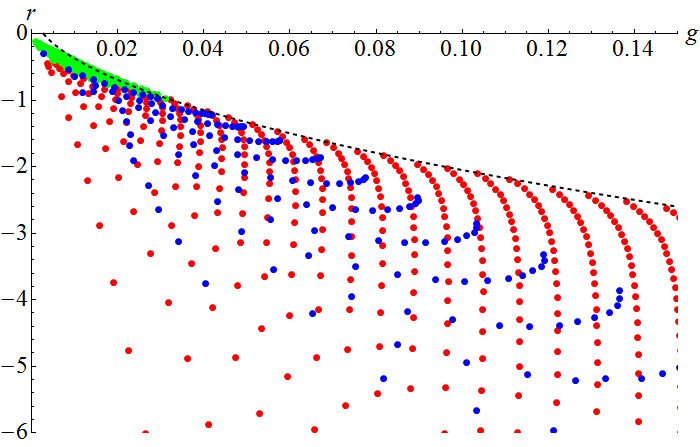}
\caption{(Color online) The phase diagram the matrix model (\ref{33simpleasymmodel}) obtained numerically. Note the difference in the scales of the two free energy diagrams}\label{33diagsympleasymmmm}
\end{figure}

\begin{figure}[tb]
\centering
\vspace*{-0.3cm}
\includegraphics[width=.8\textwidth]{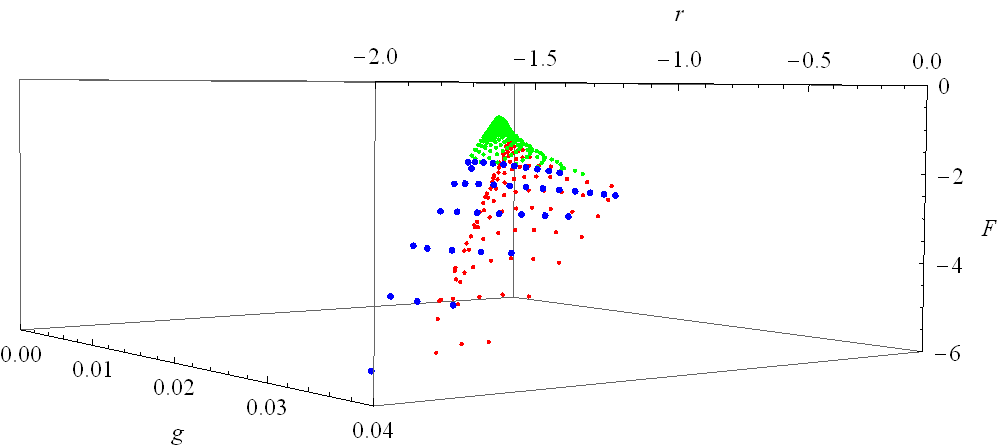}
\includegraphics[width=.8\textwidth]{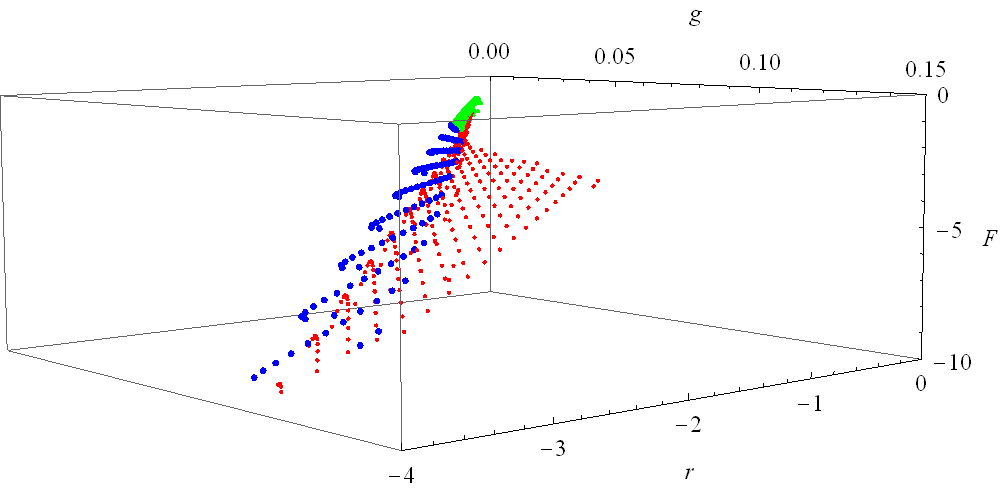}
% "\includegraphics" is very powerful; the graphicx package is already loaded
\vspace*{-0.2cm}
\caption{(Color online) The free energy of the the matrix model (\ref{33simpleasymmodel}) obtained numerically. Note the difference in the scales of the two diagrams}\label{33diagsympleasymnum}
\end{figure}

To obtain the free energy diagram as well as a phase transition line reliable for all values of $g$, we treat the problem numerically. Starting from the action
\be\label{33simpleasymmodel}
S=\half F(c_2 -c_1^2)+\half r \trl{M^2}+g \trl{M^4}
\ee
we obtain the saddle point equation
\be
-c_1F'(c_2-c_1^2)+\lr{r+F'(c_2-c_1^2)}\lambda_i+4 g \lambda_i^3=\frac{2}{N}\sum_{i\neq j}\frac{1}{\lambda_i-\lambda_j}\ .
\ee
This corresponds to an asymmetric effective single trace model
\be
S_{eff}=a_{eff} \trl{M}+\half \reff \trl{M^2}+g\trl{M^4}\ .
\ee
To connect it with the model (\ref{23asymqrt}) we have studied in the section \ref{sec23}, we rescale the matrix (or eigenvalues) by $x_0=-c_1F'(c_2-c_1^2)$. Now the model is in the desired form
\be
S_{eff}=\trl{M}+\half \reff \trl{M^2}+\geff\trl{M^4}
\ee
with
\be
\reff=\frac{r+F'(c_2-c_1^2)}{x_0^2}\ ,\ \geff=\frac{g}{x_0^4}\ .
\ee
A similar, yet simpler model was studied in section \ref{sec24asym} and it was shown there how to relate the original model to this solution. Let us stress that the moments of the effective single trace model differ from the true moments by a corresponding factor of $x_0$. Also the eigenvalue distribution is related to the effective one by
\be
\rho(\lambda)=x_0\rho_{eff}(x_0 \lambda)\ .
\ee
Following that discussion, with a more complicated relation of $\reff$ and $x_0$ to the moments, we obtain the phase diagram and the free energy diagram as shown in the figures \ref{33diagsympleasymmmm}, \ref{33diagsympleasymnum}. There are several interesting features of these diagrams.

\begin{figure} [!htb]
\centering % \begin{center}/\end{center} takes some additional vertical space
\vspace*{-0.3cm}
\includegraphics[width=.7\textwidth]{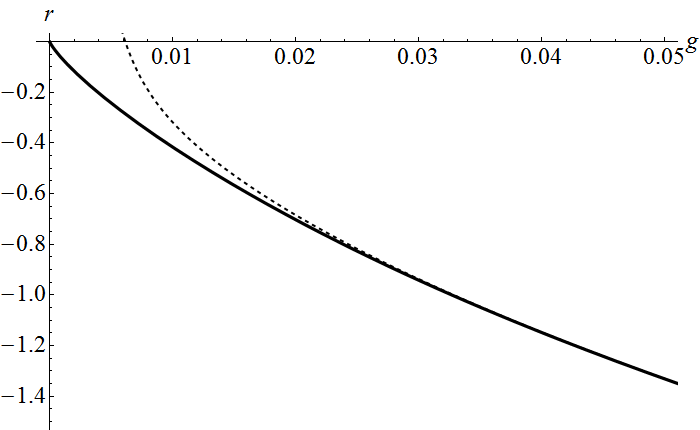}
% "\includegraphics" is very powerful; the graphicx package is already loaded
\vspace*{-0.2cm}
\caption{The boundary of the asymmetric one cut solution region from the figure \ref{33diagsympleasymnum}, shown in solid, compared with the perturbative line (\ref{33mega}), shown in dashed.}\label{33boundline}
\end{figure}

First, we see that there is a big part of the phase diagram where the solution does not exist at all. This does seem a little strange, since opposing to examples like (\ref{24simplecorrection}), we have added an attractive force so there is no reason for no stable solution to exist. However, as before we realize that what does not exist is an asymmetric solution. The symmetric solution is still there and what we found is that for those values of parameters, the attraction is not strong enough to stabilize an asymmetric distribution.

Second, before we removed them, there were a lot of strangely placed points around the $g=0$ in the free energy diagram. When we looked closer at these points, we found out the the values of $g$ are very small numbers of order $10^{-10}$ and lower. Looking even closer, we see that this is a numerical artifact, as the true solution of the equations turns out to be at $x_0=0$. This corresponds either to an unstable solution $c_2-c_1^2\to\infty$ or to a symmetric case $c_1=0$. Either way, that is a solution we are not interested in and these points will be thrown away.

And finally we see, that the asymmetric one cut solution is the preferred solution everywhere it exists. This means that once the eigenvalues can go and decide to go asymmetric, it goes as asymmetric as it gets.

If we now look just for the boundary of the asymmetric one cut region, we find the line shown in the figure \ref{33boundline}. As we can see, for larger values of $g$, the perturbative line (\ref{33mega}) reconstructs the solution nicely. But for small values the difference is significant. And gets more significant as we include more terms in the perturbative expansion (\ref{33mega}). We thus conclude that the expansion is asymptotic in for small values of $g$.

We now have all the ingredients needed to obtain the full phase diagram of the theory.

\subsubsection{Interplay of the symmetric and the asymmetric regime}

\begin{figure} [tb]
\centering % \begin{center}/\end{center} takes some additional vertical space
\includegraphics[width=.9\textwidth]{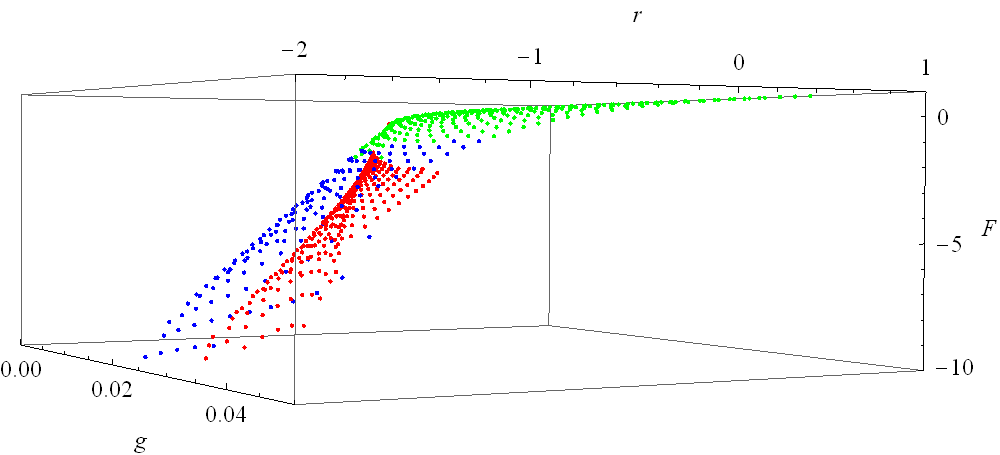}
% "\includegraphics" is very powerful; the graphicx package is already loaded
\caption{(Color online) Comparison of the symmetric and the asymmetric free energies of the model (\ref{33simpleasymmodel}).}\label{33comparefreediag}
\end{figure}

\begin{figure} [tb]
\centering % \begin{center}/\end{center} takes some additional vertical space
\includegraphics[width=.7\textwidth]{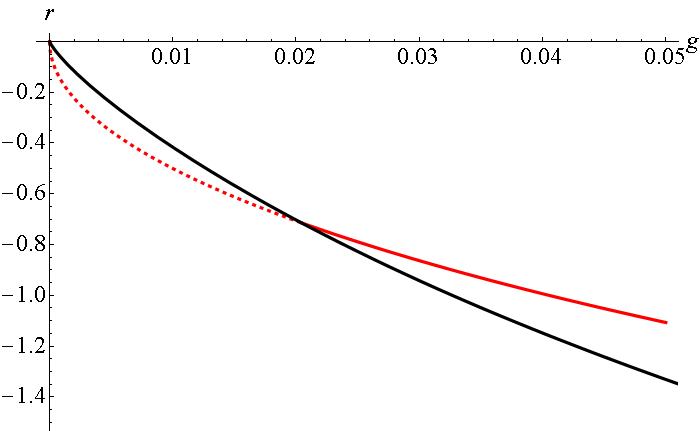}
% "\includegraphics" is very powerful; the graphicx package is already loaded
\caption{(Color online) The complete phase diagram of the model (\ref{33simpleasymmodel}). The black line is the numerically obtained boundary of existence of the asymmetric one cut solution from the figure \ref{33boundline}. The red line is the analytic boundary of existence of the symmetric one cut solution (\ref{12boundSF2}).)}\label{33completesimplephase}
\end{figure}

To study the interplay of the symmetric and asymmetric regimes, we need to put the free energy diagrams \ref{ob33symdiag} and \ref{33diagsympleasymnum} on top of each other, which is done in the figure \ref{33comparefreediag}. And quite to our surprise, we see that the asymmetric regime is the preferred wherever it is possible. It is a surprise, as the perturbative treatment in \cite{samann} suggested otherwise. We thus conclude, that this was has probably been a relict of the perturbative approach not present in the full nonperturbative results. And we obtain the phase diagram of the theory as in the figure \ref{33completesimplephase}.

\begin{figure}[!htb]
\centering
\includegraphics[width=.7\textwidth]{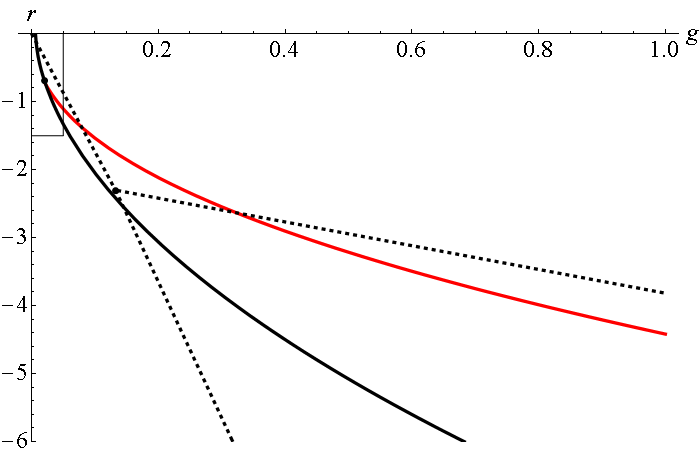}
\caption{(Color online) The phase transition lines, shown in dashed, which are linear fits to the data points, from the numerical work \cite{denjoenum2} compared with the phase transition lines presented here. The small rectangle in the top left corner denotes the area covered by the figure \ref{33completesimplephase}.}\label{fignum}
\end{figure}

The intersection of the two lines occurs at a critical coupling equal to approximately
\be
g_c\approx 0.02\ .
\ee
We see that this is a little lower than the value in the table \ref{table1}. This is clearly the result of the asymptotic character of the perturbative expansion (\ref{33mega}).

And the value is smaller roughly by a factor of $7$ from the numerical values (\ref{13numericalptriple}) of the numerical works \cite{denjoenum2,num14}. To compare better, we draw our phase diagram and the numerical phase diagram in the same picture in the figure \ref{fignum}.

We see that the diagrams, together with the location of the triple point are in in a qualitative agreement, but to claim they agree quantitatively would be a rather bold statement.

One possible source of the discrepancy is clearly the remainder term $\mathcal R$ in (\ref{32sefffull}) which will deform the transition lines. However there is a possible problem also on the part of the numerics. The lines in the phase diagrams are linear extrapolation of data. And looking at the original figure \ref{plotnum}, we see that the data points are at values in the range $r\in(0.4,0.6)$. This is a little too far away from the triple point and the linear extrapolation does not take into account the turn the transition lines make for small values of $g$. As long as this turn is not drastically changed by the higher terms deformation, numerical data do overestimate the triple point. After some preliminary numerical analysis closer to the origin of the phase diagram, this seems to be the case, however it is not clear to what extent \cite{denjoeprivate}.

%% file: 3_4phasefull.tex
In this last section, we will study the phase structure of a model with the perturbative effective action (\ref{32rearr}). Moreover, as very little changes if we consider the first part nonperturbatively, we will use the expression for $F(t)$ rather than the perturbative part. The action includes quartic multitrace terms, but as we will shortly see, does not explain the phase structure of the field theory.

Again, we will first study the symmetric case and the asymmetric case afterward.

\subsubsection{The symmetric regime}

\begin{figure}[!tb]
\centering % \begin{center}/\end{center} takes some additional vertical space
\includegraphics[width=.7\textwidth]{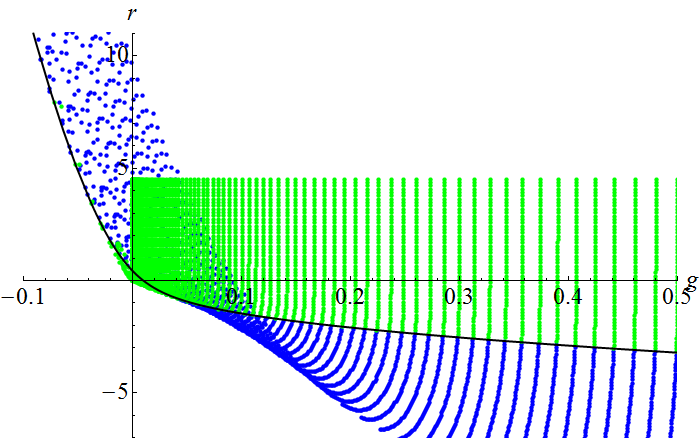}
\includegraphics[width=.7\textwidth]{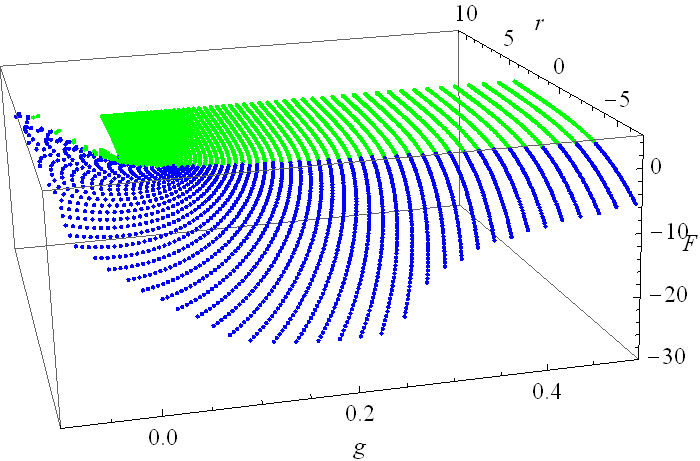}
% "\includegraphics" is very powerful; the graphicx package is already loaded
\caption{(Color online) The phase diagram and the free energy of the the matrix models (\ref{sec6_sphere}). The green region denotes the one cut solution, the blue region denotes the two cut solution. The blue line is the exact transition line (\ref{34fullsymtrafo}).}\label{ob62_shere}
\end{figure}

The full action for the symmetric regime is 
\be S=\frac{1}{2}F(c_2)-\frac{1}{3456}\lr{c_4-2c_2^2}^2+\half r \trl{M^2}+g \trl{M^4}\ ,\label{sec6_sphere}\ee
which is a combination of the nonperturbative part (\ref{32FS2}) obtained in \cite{poly} and the perturbative term obtained in \cite{samannnew}. The effective parameters are given by
\be \reff=r+F'(c_2)+\frac{1}{216}c_2(c_4-2c_2^2) \ ,\ \geff=g-\frac{1}{1728}(c_4-2c_2^2)\ .\label{34symeff}\ee
Approaching the problem numerically, we can find the phase diagram and the free energy diagram of this model by now familiar method and obtain diagrams show in the figure \ref{ob62_shere}.

The free energy diagram has clearly features which we do not expect from the diagram of the fuzzy field theory. It allows for a solution for negative $g$, for some values of $g>0$ it has no solution and it has more than one solution elsewhere. This is the first clue that there is something wrong with the interpretation of the perturbative effective action (\ref{32rearr}) as the action of the scalar field on the fuzzy sphere. But at the moment, let us follow the discussion.

If we are interested only in the phase transition line, the expressions simplify again and we are able to find the line analytically. Plugging (\ref{34symeff}) in (\ref{22moments2cut}) gives
\bse
\bal
c_2=&-\frac{r+F'(c_2)+\frac{1}{216}c_2(c_4-2c_2^2)}{4\lr{g-\frac{1}{1728}(c_4-2c_2^2)}}\ ,\\
c_4=&c_2+\frac{1}{4\lr{g-\frac{1}{1728}(c_4-2c_2^2)}}\ .
\end{align}
\ese
which together with ${\reff=-4\sqrt{\geff}}$ yields after some algebra
\be\label{34fullsymtrafo}
r(g)=-\frac{1}{1 - e^{\frac{4 \sqrt 3}{\sqrt{24 g + \sqrt{1 + 576 g^2}}}}} 
 +\frac{2}{\sqrt 3 \lr{24 g + \sqrt{1 + 576 g^2}}^{3/2}}
 - \frac{5 \sqrt{24 g + \sqrt{1 + 576 g^2}}}{4 \sqrt 3}\ ,
\ee
the line shown in the figure \ref{ob62_shere} together with the transition line (\ref{12boundSF2}) for the model without the quartic multitrace term. It is important to point out, that this phase transition is exact. Starting from the action (\ref{sec6_sphere}), no perturbative considerations have been made. So as long as this models is viewed this formula is complete.

But if the model  is viewed as a representation of the fuzzy field theory, the action is a perturbative expansion and the given formula is just a perturbative approximation of the true transition line. And we see again that it is not a good approximation to the true transition line, which is expected to go through the origin. This is another clue, that the perturbative effective action (\ref{sec6_sphere}), even if the part $F(t)$ is taken nonperturbatively, is not a good starting point to the analysis of phase structure of the fuzzy field theory.

For completeness though, let us continue with the discussion of the consequences of the perturbative effective action to its bitter end with the asymmetric regime.

To compare with the asymmetric results we will obtain shortly, we shall compute the transition line also perturbatively. We will take the expansion of $F$ up to the fourth order (to be consistent with the fourth order of the other term) and obtain the transition line
\be\label{34symperttrafo}
r=- 4 \sqrt{g}-\half + \frac{1}{12 \sqrt{g}}  +\frac{7}{5760 g^{3/2}}+\frac{29}{1935360 g^{5/2}}\ .
\ee
Note, that this formula agrees with the first terms in the $1/\sqrt{g}$ expansion of (\ref{34fullsymtrafo}).

\subsubsection{The asymmetric regime}

The full action for the asymmetric regime is again the combination of the perturbative part (\ref{32rearr}) and the non-perturbative part (\ref{32FS2}), but now with all the odd terms included
\begin{align}
S=&\frac{1}{2}F(c_2-c_1^2)-\frac{1}{3456}\Big[(c_4 - 4 c_3 c_1 + 6 c_2 c_1^2 - 3 c_1^4)-2(c_2-c_1^2)^2\Big]^2-\nonumber\\&-\frac{1}{432}\lr{c_3 - 3 c_1 c_2 + 2 c_1^3}^2+\half r\, \trl{M^2}+g\, \trl{M^4}\ .\label{sec6_sphereAS}
\end{align}
This action leads to the following saddle point equation
\begin{align}
\bigg[\frac{1}{432} \Big(-25 c_1^7 + 3 c_1^5 (25 c_2-8 )
 - 25 c_1^4 c_3
+6 c_1^2 (5 c_2-2 ) c_3 
+ c_3 (6 c_2 - 2 c_2^2 + c_4)\no
+ c_1^3 (48 c_2 - 60 c_2^2 + 5 c_4) 
+c_1 (10 c_2^3-18 c_2^2 - 4 c_3^2 - 5 c_2 c_4)\Big)
-c_1 F'(c_2 - c_1^2)\bigg]
+\no+
\bigg[\frac{1}{432} \Big(25 c_1^6 + 
   c_1^4 (24 - 60 c_2)
 - 4 c_2^3 
+ 20 c_1^3 c_3 
+ 4 c_1 (3 - 2 c_2) c_3 
+   \no+c_1^2 (30 c_2^2-36 c_2 - 5 c_4) 
+ 2 c_2 c_4\Big)
+F'(c_2 - c_1^2)+r\bigg]\lambda_i+\no+
\bigg[\frac{1}{144} \Big(-5 c_1^5 + 
   2 c_1^3 (5 c_2-2 ) 
	- 2 c_3 
	- 4 c_1^2 c_3 
	+   c_1 (6 c_2 - 2 c_2^2 + c_4)\Big)\bigg]\lambda_i^2+\no
+\bigg[\frac{ 1}{432} \Big(5 c_1^4 
- 10 c_1^2 c_2 
+ 2 c_2^2 
+   4 c_1 c_3 
- c_4\Big)+
4g\bigg]\lambda_i^3
=\frac{2}{N}\sum_{i\neq j}\frac{1}{\lambda_i-\lambda_j}\ ,
\end{align}
which we will, for the brevity of the equations, write as
\be X_0 +(r+X_1)\lambda_i+X_2 \lambda_i^2+(4g+X_3)\lambda_i^3=\frac{2}{N}\sum_{i\neq j}\frac{1}{\lambda_i-\lambda_j}\ .\ee
This equation leads to an effective single trace matrix model which is not of the form (\ref{23asymqrt}), since the $\lambda^2$ term leads to a $\trl{M^3}$ term in the action of the effective model. We can however get rid of such term by shifting the matrix $M$. By changing $M\to (M-a\,\textrm{id})/x_0$, where
\begin{subequations}\label{31x01}
\begin{align}
x_0=&X_0 + \frac{2 X_2^3}{27 (4 g + X_3)^2} - \frac{(r + X_1) X_2}{3 (4 g + X_3)}\ ,\\
a=&\frac{X_2}{3 (4 g + X_3)}\ ,
\end{align}
\end{subequations}
and after dropping the constant term, we obtain the effective single trace matrix model in the desired form (\ref{23asymqrt})
\be
S=\trl{M}+\half\reff\trl{M^2}+\geff \trl{M^4}\ ,
\ee
with
\begin{subequations}\label{34reffgeff}
\begin{align}
\reff=&\frac{12 g (r + X_1) - X_2^2 + 3 (r + X_1) X_3}{3 (4 g + X_3)x_0^2}\ ,\\
\geff=&\frac{4g+X_3}{4x_0^4}\ .
\end{align}
\end{subequations}
Moreover, the moments of the distribution $c_n$, which are to be used when evaluating $X$'s are related to the moments of the effective single trace matrix model distribution (\ref{2momentseffectiveasymmetric}) by
\begin{subequations}\label{31c1azc4}
\begin{align}
c_1 =& \frac{c_{1,eff}}{x_0} - \frac{a}{x_0}\ ,\\
c_2 =& \frac{c_{2,eff}}{x_0^2} - \frac{2 c_{1,eff} a}{x_0^2} + \frac{a^2}{x_0^2}\ ,\\
c_3 =& \frac{c_{3,eff}}{x_0^3} - \frac{3 c_{2,eff} a}{x_0^3} + \frac{3 c_{1,eff} a^2}{x_0^3} - \frac{a^3}{x_0^3}\ ,\\
c_4 =& \frac{c_{4,eff}}{x_0^4} - \frac{4 c_{3,eff} a}{x_0^4} + \frac{6 c_{2,eff} a^2}{x_0^4} - \frac{
     4 c_{1,eff} a^3}{x_0^4} + \frac{a^4}{x_0^4}\ .
\end{align}
\end{subequations}

Unfortunately there is little we can do with our approach in this case. If we choose particular values of $\reff$ and $\geff$, the moments of the effective single trace model are known and the equations to be solved numerically are (\ref{31x01}) plugged into (\ref{31c1azc4}) together with (\ref{34reffgeff}). This system is way too complicated to be handled. So to include the effect of the new terms, we retreat to the perturbative calculation and repeat the procedure of \cite{mine1}.

This means we reintroduce the parameter $\ep$ and solve the equations determining the solution as a series in powers of $\ep$, later setting $\ep=1$. The final result we obtain is
\be r(g)= - 2 \sqrt{15} \sqrt{g}\frac{2}{5} 
- \frac{19}{ 18000 \sqrt{15} \sqrt{g}} 
+\frac{29}{1125000 g}
- \frac{7886183}{4374000000000 \sqrt{15} g^{3/2}}\ .\label{34asympert}\ee
We stress that this formula is a perturbative one even if the model (\ref{sec6_sphereAS}) is viewed as a complete action. We cannot do much more here, so we continue to the comparison of the two regimes.

\subsubsection{The interplay of the perturbative symmetric and asymmetric regimes}

\begin{figure}[tb]
\begin{center}
\includegraphics[width=.7\textwidth]{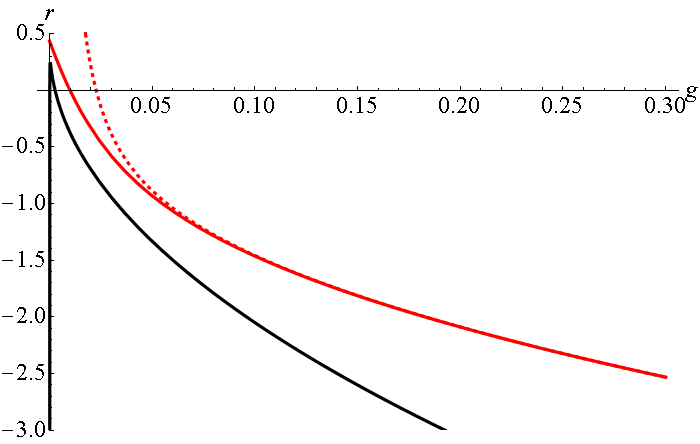}
% "\includegraphics" is very powerful; the graphicx package is already loaded
\caption{(Color online) The phase transition lines (\ref{34fullsymtrafo}) - the red solid line,(\ref{34symperttrafo}) - the red dashed line and (\ref{34asympert}) - the black line in one figure, showing the incompleteness of the perturbative description.}
\end{center} 
\label{ob62x}
\end{figure}

The figure \ref{ob62x} shows a little sad plot of the symmetric transition lines (\ref{34symperttrafo}) and (\ref{34fullsymtrafo}) with the asymmetric transition line (\ref{34asympert}). These transition lines do not intersect and the perturbative model as it stands does not have a triple point. Also, recalling the free energy diagram of the symmetric regime in the figure \ref{ob62_shere}, the phase structure of such model is not clear at all and surely does not represent the phase structure discussed in the section \ref{sec13}.

We thus conclude, that the phase structure of the fuzzy field theory is very sensitive to the terms of the perturbative expansion. To be able to make any reasonable statements about the properties of the phase diagram of the field theory, once we include some multitrace term, we need to include terms of all orders in that moment. A similar result has been obtained recently in a Monte-Carlo study of the perturbative multitrace model (\ref{sec6_sphereAS}) in \cite{ydrisupernew}, where it is shown that the perturbative version of the model (\ref{sec6_sphere}) does not reconstruct the phase structure of the fuzzy field theory. It would be interesting to see, how our results connect to this work.

%% file: 4_concl.tex
We have presented an application of the matrix model techniques to the study of the phase structure of the scalar field theory on the fuzzy sphere. Our goal was to reproduce the available results of various numerical simulations. We have analyzed two different models which approximate the field theory in two different ways.

One of the models did a decent job, reproducing the expected qualitative features of the diagram and giving a reasonable quantitative values of the parameters of the diagram. The second model did much worse and have been neither in qualitative nor quantitative agreement with the expectations. As we have seen, the key property distinguishing these two models and leading to the success or the failure was the fact that the approximation used in the first model was a nonperturbative. And the fact, that the second approximation was in some sense more precise, as it included several extra terms, played little role.

So the main lesson we have learned is that if one uses a matrix model to approximate the fuzzy field theory, it is very important that the terms are arranged in a nonperturbative packing. Perturbative approximations are of little use if we aim to explain the phase structure of the theory.

We will now conclude our journey with several open questions or interesting related problems.
We start with a short remainder, that we have left two loose ends on the numerical side of our work. One should try to improve the algorithm for the treatment of the two cut solution of the asymmetric quartic model and come up with a better numerical treatment of the second, perturbative approximation matrix model, capable of handling the $t_3$ term.

\podnadpis{The kinetic term effective action}

We do not have to go too far for further problems. Some better, perhaps a complete treatment of the angular integral (\ref{31anfintegral}) is desirable. Unfortunately, there are very few hints on how to approach this problem further, as the perturbative treatment seems to be way too complicated \cite{samannnew} and there is no hint how to generalize the nonperturbative treatment of \cite{poly} beyond the second moment. The reason might be that we are on a wrong track. The moments, in which we try to express the kinetic term effective action, might not be the most useful variable. The effective action is a function of the eigenvalues and the moments are perhaps a bad way how to express this dependence. For example if we tried to express the repulsive Vandermonde term in terms of the moments, we would very soon run into difficulties. So stopping for a while and rethinking the whole problem over might bring some new insight.

\podnadpis{Nonperturbative approximation of perturbative approximation}

We have presented a possible method how to pack a perturbative expansion into a nonperturbative form, yielding a nonperturbative approximation of perturbative approximation. Even though still an approximation, it might be useful for computations. We have very little information of about the remainder term $\mathcal R$ (\ref{32sefffull}) so the method based on Pade approximants is not very useful, but we still can cook up an approximation by
\bal
\mathcal R=&-\frac{1}{432}t_3^2-\frac{1}{3456}\lr{t_4-2t_2^2}^2\approx \log\slr{h\lr{\frac{1}{432},t_3}h\lr{\frac{1}{3456},t_4-2t_2^2}}\ ,\\
\textrm{where}&\no
&h(\alpha,x)\sim\begin{array}{ll}
-\alpha x^2 & x\to0\\
\log x & x\to\infty
\end{array}
\end{align}
for example
\bal
h(\alpha,x)=&\frac{1}{1+\lr{1+\alpha}x^2}+\frac{x^2}{1+x}\ ,\\
h(\alpha,x)=&e^{-\lr{1+\alpha}x^2}+x\lr{1-e^{-x}}\ .
\end{align}
It would definitely be interesting to study a matrix model with such multitrace term.

\podnadpis{Other noncommutative spaces}

As mentioned in the text, there are numerical results available for several other spaces, the fuzzy disc \cite{fuzzydiscnum}, the three dimensional case of ${\mathbb R\times S^2_F}$ \cite{rsfnum}, the noncommutative plane \cite{num14panero2} to name some. Moreover, for the first there is a perturbative analysis available also \cite{rsfper} and for the second an perturbative approximation similar to the one studied here \cite{samanfuzzydisc}. The fuzzy disc effective action includes the linear $c_4$ term and also does not vanish for the semicircle. It would be very interesting to generalize the nonperturbative treatment of the section \ref{polyapproach} to these spaces and see how these results compare.

Also, we could clearly do computations for different spaces, where no previous results are available, such as $\mathbb CP^2,\mathbb CP^3$, a general $\CPn$ or the fuzzy foursphere $S^4$ \cite{s4}. Moreover, if one considers a general $\phi^{2k}$ theory and repeat the procedure of the section, one could look for the aspects of non-renormalizability of certain theories, like $\phi^6$ on $\mathbb R^4$ or $\phi^4$ on $\mathbb R^6$, that are reflected in the regularized matrix versions. This was suggested in \cite{ocon} and \cite{samann}, but was not elaborated on neither there, nor here. Some very interesting structure of $M^6$ matrix model is revealed in \cite{multiband} and the straightforward connection of this model to the $\phi^6$ theory conjectured here makes those results relevant.

\podnadpis{Phase structure of the UV/IR free theories}

And finally, the most interesting problem. As mentioned in the text, there are several modifications of the naive versions of the scalar field theory which remove the UV/IR mixing phenomenon. Both for the fuzzy sphere and for Euclidean spaces \cite{uvir,gw,nouvir2}. The phase structure of such spaces is of prime interest, as the non-uniform order phase should be removed together with the UV/IR mixing. This remains to be shown and perhaps one could learn something new from the way the extra phase is removed. 

\podnadpis{Beyond noncommutative field theory}

Undoubtedly, there is a lot of possibility to apply the results for the asymmetric quartic model and the multitrace models beyond the phase structure of the noncommutative field theories, as the idea has been successfully used before \cite{novy1,novy2,novy3}. Interestingly, a similar matrix model appears also in the study of the random geometry on the fuzzy sphere \cite{glaser}.

%% file: appendix2.tex
In this section, we will present the algorithm used to numerically solve the equations (\ref{23conditions}, \ref{ap1condition}) for the asymmetric two cut solution of the model (\ref{23asymqrt}). Trying to use numerical methods of the computer algebra program directly on the equation (\ref{ap1condition}) yields little success, so we were forced to come up with something else.

The algorithm starts with choosing some particular values of $r$ and $g$ as before. Then we take the value of $D_2$ to be $-D_1(1+x)$, where $x$ is our new parameter. We chose a particular value of $x$ and solve the equations (\ref{23conditions}) numerically using standard features of the computer algebra program. Again, navigating our way through handful of irrelevant solutions\footnote{These are imaginary solutions, solutions with negative $\delta$'s and solutions where the two intervals overlap.}, we check for the value of (\ref{ap1condition}) for the relevant solution (if it exists).

Doing this, we scan through a reasonable interval of $x$ observing when does the value change sign. When we find an interval of $x$ with different sign of (\ref{ap1condition}) at the endpoint, we zoom to this area and look further. We again split the interval into parts and look for the change of sign. When we find such interval, we take a linear approximation to the function given by the values at the endpoints and take the root of this linear fit as the solution for $x$ and thus for all the parameters $D_{1,2}$ and $\delta_{1,2}$.

We have used the following values for our search. Interval of $x$ has been $(0,0.3)$, which was first split into $500,300,100$ or $50$ parts, depending on the values of $r$ and $g$. Generally, for small $g$ larger number was needed to find the solution. If the solution has not been found for smaller value of the number, the algorithm ran again with the value of $500$ to make sure there is no solution. This way we achieved a balance between precision and speed. The part where the value of (\ref{ap1condition}) has changed sign was divided further into $50$ parts. We then chose for our value of $x$ the midpoint of the small interval where (\ref{ap1condition}) changes sign. The program had trouble when the edge of the first interval was too close to the zero, so if the value of (\ref{ap1condition}) was too close to zero (absolute value smaller than 0.5) in our program, we have narrowed down the second interval.

The biggest problem of the algorithm was the exact location of the transition line between the asymmetric one cut and the two cut solution. We did our best to tune the parameters of the algorithm to achieve good precision while having a reasonable speed, but there still might be points around the boundary which we have missed. Clearly, there is also a lot of room for improvement of the speed of this algorithm. However for the current purposes this version did serve quite well.

%% file: acta_tekel.bbl
\addcontentsline{toc}{section}{References}
\begin{thebibliography}{999}
\bibitem{stein05_1}
H. Steinacker, 
``A Non-perturbative approach to non-commutative scalar field theory,''
\JHEP {\bf 0503}, 075 (2005), 
[hep-th/0501174].

\bibitem{stein05_2}
H. Steinacker, 
``Quantization and eigenvalue distribution of noncommutative scalar field theory,''
[hep-th/0511076].

\bibitem{stein04_1}
H. Steinacker, 
``Quantized gauge theory on the fuzzy sphere as random matrix model,''
Nucl.\ Phys.\  B {\bf 679}, 66 (2004),
[hep-th/0307075].

\bibitem{our1}
V.P.~Nair, A.P.~Polychronakos, J.~Tekel,
``Fuzzy spaces and new random matrix ensembles,''
\PRD {\bf 85}, 045021 (2012)
[1109.3349 [hep-th]].

\bibitem{ocon}
D.~O'Connor, C.~S\"{a}mann,
``Fuzzy Scalar Field Theory as a Multitrace Matrix Model,''
\JHEP {\bf 0708}, 066 (2007)
[0706.2493 [hep-th]].

\bibitem{connes1}
A. Connes,
``Noncommutative Geometry,''
(Academic Press, 1994).

\bibitem{landi}
G. Landi,
``An Introduction to Non-commutative Spaces and their Geometry,''
(Springer-Verlag, 1997).

\bibitem{snyder}
H.S. Snyder,
``Quantized Spacetime,''
Phys. Rev. {\bf 71} (1947) 38.

\bibitem{doplicher}
S. Doplicher, K. Fredenhagen, J.E. Roberts,
``The quantum structure of spacetime at the Planck scale and quantum fields,''
Comm. Math. Phys. {\bf 172} (1995), no. 1, 187-220,
[hep-th/0303037].

\bibitem{ncsm1}
A. Connes, J. Lott,
``Particle models and non-commutative geometry,''
Nucl. Phys. (Proc. Suppl.) {\bf 18B} (1991) 29.

\bibitem{ncsm2}
D. Kastler, T. Schucker,
``A detailed account of Alain Connes' version of the standard model,''
Rev. Math. Phys. {\bf 8} (1996) 205,
[hep-th/9501077].

\bibitem{ncsm3}
A. Connes, M. Marcolli
``Noncommutative Geometry, Quantum Fields and Motives,''
AMS Colloquium Publications Vol. 55 (2008).

\bibitem{walter}
W. van Suijlekom,
``Noncommutative Geometry and Particle Physics,''
Springer Netherlands, 2015.

\bibitem{grav1}
A. Chamseddine, G. Felder and J. Frohlich,
``Gravity in non-commutative geometry,''
Commun. Math. Phys. {\bf 155} (1993) 205,
[hep-th/9209044].

\bibitem{grav2}
G. Landi and C. Rovelli,
``Gravity from Dirac eigenvalues,''
Phys. Rev. Lett. {\bf 78} (1997) 3051,
[gr-qc/9708041].

\bibitem{qhe1}
D. Karabali and V. P. Nair,
``Quantum Hall effect in higher dimensions,''
Nucl. Phys. B {\bf 641} (2002) 533,
[hep-th/0203264].

\bibitem{qhe2}
D. Karabali and V.P. Nair,
``Quantum Hall Effect in Higher Dimensions, Matrix Models and Fuzzy Geometry,''
J. Phys. A: Math. Gen. {\bf 39} (2006) 12735,
[hep-th/0606161].

\bibitem{qhe3}
B. Morariu, A.P. Polychronakos,
``Quantum mechanics on non-commutative Riemann surfaces,''
Nucl. Phys. B {\bf 634} (2002) 326,
[hep-th/0201070].

\bibitem{fuzzygauge}
D. Karabali, V.P. Nair, A.P. Polychronakos,
``Spectrum of Schrodinger field in a non-commutative magnetic monopole,''
Nucl.Phys. B{\bf 627} (2002) 565,
[hep-th/0111249].

\bibitem{ppsug}
V. Schomerus,
``String Theory and Noncommutative Geometry,''
\JHEP {\bf 09} (1999) 032,
[hep-th/9903205].

\bibitem{witten}
N. Seiberg, E. Witten,
``D-branes and Deformation Quantization,''
\JHEP {\bf 06} (1999) 030
[hep-th/9908142].

\bibitem{noncom1}
M.R. Douglas and N.A. Nekrasov,
``Non-commutative field theory,''
\RMP ~{\bf 73} (2001) 977,
[hep-th/0106048].

\bibitem{noncom2}
R.J. Szabo
``Quantum Field Theory on Noncommutative Spaces,''
Phys. Rept. {\bf 378} (2003) 207
[hep-th/0109162].

\bibitem{branekabat}
D. Kabat, W. Taylor,
``Linearized supergravity from Matrix theory,''
Phys.Lett.B {\bf 426} (1998) 297,
[hep-th/9712185].

\bibitem{branenair}
V.P. Nair and S. Randjbar-Daemi, 
``On Brane Solutions in M(atrix) Theory,''
Nucl. Phys. B {\bf 533} (1998) 333,
[hep-th/9802187].

\bibitem{taylor}
W. Taylor,
``M(atrix) Theory: Matrix Quantum Mechanics as a Fundamental Theory,''
\RMP ~{\bf 73} (2001) 419,
[hep-th/0101126].

\bibitem{abedis}
Y. Abe,
``Construction of Fuzzy Spaces and Their Applications to Matrix Models,''
Ph. D. thesis, Physics Department, City College of New York (2001),
[1002.4937].

\bibitem{fuzzy3}
M.~Panero, 
``Quantum Field Theory in a Non-Commutative Space: Theoretical Predictions and Numerical Results on the Fuzzy Sphere,''
SIGMA {\bf 2} (2006) 08,
[hep-th/0609205].

\bibitem{num14panero1}
W.~Bietenholz, F.~Hofheinz, H.~Mej\'{i}a-D\'{i}az, M.~Panero,
``Scalar fields on a non-commutative space,''
J. Phys.: Conf. Ser. {\bf 651} 012003 (2015),
[hep-th/1402.4420].

\bibitem{revmat}
``Applications of random matrix theory to condensed matter and optical physics,''
\textit{The Oxford Handbook of Random Matrix Theory}, edited by G. Akemann,J. Baik, P. Di Francesco (Oxford University Press, 2011).

\bibitem{wigner1}
E. Wigner,
``Characteristic Vectors of Bordered Matrices with Infinite Dimensions,''
Ann. Math. {\bf 62} (1955) 548.

\bibitem{wigner2}
E. Wigner,
``On the Distribution of the Roots of Certain Symmetric Matrices,''
Ann. Math. {\bf 67}(1958) 325.

\bibitem{metha}
M.L. Mehta
``Random Matrices,''
3d edition (Elsevier 2004).

\bibitem{thooft}
G. 't Hooft,
``A planar diagram theory for strong interactions,''
Nucl. Phys. B {\bf 72} (1974) 461.

\bibitem{gaume}
L. Alvarez-Gaum\'{e},
``Random surfaces, statistical mechanics and string theory,''
Lausane lectures (1990).

\bibitem{matrix2}
M.~Mari\~{n}o,
``Les Houches lectures on matrix models and topological strings,''
[hep-th/0410165].

\bibitem{matrix3}
P.~Di Francesco,
``2D Quantum Gravity, Matrix Models and Graph Combinatorics,''
[math-ph/0406013].

\bibitem{string}
P. Ginsparg and G. Moore,
''Lectures on 2D gravity and 2D string theory'',
TASI lectures (1992),
[hep-th/9304011].

\bibitem{revmat1}
C.W.J. Beenakker,
``Applications of random matrix theory to condensed matter and optical physics,''
\textit{The Oxford Handbook of Random Matrix Theory}, edited by G. Akemann,J. Baik, P. Di Francesco (Oxford University Press, 20011).

\bibitem{revmat2}
T. Guhr, A. M\"uller-Groeling and H.A. Weidenm\"uller,
``Random Matrix Theories in Quantum Physics: Common Concepts,''
Phys. Rep. {\bf 299}, 189 (1998),
[cond-mat/9707301].

\bibitem{revmat3}
Les Houches, ``Chaos and quantum systems,'' (Elsevier 1991).

\bibitem{chaos}
P. Bourgade and J.P. Keating, 
``Quantum chaos, random matrix theory, and the Riemann $\zeta$-function,''
Seminaire Poincar\'e XIV, 115 (2010).

\bibitem{rieman}
A.M Odlyzko
``The distribution of spacings between zeros of the zeta function,''
Math. Comput. {\bf 48} (1987) 273.

\bibitem{rieman2}
J. Keating, N.C. Snaith,
``Random matrix theory and $\zeta(1/2+i t)$,''
Comm. Math. Phys. {\bf 214} (2000) 57.

\bibitem{ncgeo}
J.~Madore,
``An Introduction to Non-commutative Differential Geometry and its Physical Applications,''
(Cambridge University Press, Cambridge - 1995).

\bibitem{sf21}
J. Madore,
``The Fuzzy sphere,''
Class.Quant.Grav. 9 (1992) 69.

\bibitem{sf22}
H. Grosse, C. Klim\v{c}\'{i}k, P. Pre\v{s}najder,
``Towards Finite Quantum Field Theory in Non-Commutative Geometry,''
Int.J.Theor.Phys. 35 (1996) 231,
[hep-th/9505175].

\bibitem{sf23}
H. Grosse, C. Klim\v{c}\'{i}k, P. Pre\v{s}najder,
``Topologically nontrivial field configurations in noncommutative geometry,''
Commun.Math.Phys. 178 (1996) 507,
[hep-th/9510083].

\bibitem{abe1}
Y. Abe,
``Construction of fuzzy $S^4$,''
Phys. Rev. D {\bf 70} (2004) 126004,
[hep-th/0406135].

\bibitem{s4}
J. Medina, D. O'Connor,
``Scalar Field Theory on Fuzzy $S^4$,''
\JHEP {\bf 11} (2003) 051
[hep-th/0212170].

\bibitem{cpnnair1}
D. Karabali, V.P. Nair and R. Randjbar-Daemi,
``Fuzzy spaces, the M(atrix) model and the quantum Hall effect,''
in \textit{From Fields to Strings: Circumnavigating Theoretical Physics}, Ian Kogan Memorial Collection, M. Shifman, A. Vainshtein and J. Wheater (eds.) (World Scientific, 2004)
[hep-th/0407007].

\bibitem{cpnnair2}
V.P. Nair,
``Noncommutative mechanics, Landau levels, twistors and Yang-Mills amplitudes,''
Lecture Notes in Physics {\bf 698} (Springer-Verlag, Berlin, Heidelberg, 2006),
[hep-th/0506120].
  
\bibitem{blowcpn}
H. Grosse, H. Steinacker,
``Finite Gauge Theory on Fuzzy $CP^2$,''
 Nucl. Phys {\bf B 707} (2005) 145-198,
[hep-th/0407089].

\bibitem{cpnoriginal}
G. Alexanian, A.P. Balachandran, G. Immirzi, B. Ydri,
``Fuzzy CP**2,''
J. Geom, Phys, {\bf 42} (2002) 28,
[hep-th/0103023].

\bibitem{coherent}
A.M. Perelomov
``Generalized Coherent States and Their Applications,''
(Springer-Verlag, 1996).

\bibitem{star1}
A.P. Balachandran, B.P. Dolan, J. Lee, X. Martin, D. O'Connor,
``Fuzzy complex projective spaces and their star products,''
J. Geom, Phys, {\bf 43} (2002) 184,
[hep-th/0107099].

\bibitem{star3}
V.P. Nair,
``Gravitational fields on a non-commutative space,''
Nucl. Phys B, {\bf 651} (2003) 313,
[hep-th/0112114].

\bibitem{star2}
D. Karabali, V.P. Nair
``The effective action for edge states in higher dimensional quantum Hall systems,''
Nucl.Phys. B {\bf 679} (2004) 427,
[hep-th/0307281].

\bibitem{stenposi}
H. Steinacker,
``Non-commutative geometry and matrix models,''
Proceedings of Science (QGQGS 2011) 004,
[1109.5521 [hep-th]]

\bibitem{pois1}
M. Bordemann, E. Meinrenken and M. Schlichenmaier,
``Toeplitz quantization of Kahler manifolds and $gl(N),N\to\infty$ limits,''
Commun. Math. Phys. {\bf 165} (1994) 281,
[hep-th/9309134].

\bibitem{pois2}
M. Kontsevich,
``Deformation quantization of Poisson manifolds, I,''
Lett. Math. Phys. {\bf 66} (2003) 157,
[q-alg/9709040].

\bibitem{fuzzy2}
A.P. Balachandran, S. Kurkcuoglu and S. Vaidya,
``Lectures on Fuzzy and Fuzzy SUSY Physics,''
[hep-th/0511114].

\bibitem{uvir2}
S. Minwalla, M. Van Rammsdonk, N. Seinberg,
``Non-commutative perturbative dynamics,''
\JHEP {\bf 2} (2000) 020
[hep-th/9912072].

\bibitem{uvir1}
C.S. Chu, J. Madore, H. Steinacker,
``Scaling limits  of the fuzzy sphere at one loop,''
\JHEP {\bf 3} (2005) 075,
[hep-th/0106205].

\bibitem{uvirkuk}
S. Vaidya and B. Ydri,
``On the Origin of the UV-IR Mixing in Noncommutative Matrix Geometry,''
Nucl. Phys. B {\bf 671} (2003) 4,
[hep-th/0305201].

\bibitem{uvir}
B.P.~Dolan, D.~O'Connor, and P.~Pre\v{s}najder,
``Matrix $\phi^4$ Models of the Fuzzy Sphere and their Continuum Limits,''
\JHEP {\bf 03} (2002) 013
[hep-th/0109084].

\bibitem{gw}
H.~Grosse, R.~Wulkenhaar,
``Renormalization of $\phi^4$-theory on noncommutative $\mathbb R^4$ in the matrix base,''
Commun.Math.Phys. {\bf 256} (2005) 305
[hep-th/0401128].

\bibitem{phaseover}
T.R. Govindarajan, S. Digal, K.S. Gupta, X. Martin,
``Phase strucutres in fuzzy geometries,''
Proceedings of Science (CORFU 2011) 058,
[1204.6165 [hep-th]].

\bibitem{noncomphase1}
G.S. Gubser, S.L. Sondhi,
``Phase structure of non-commutative scalar field theories,''
Nucl. Phys. B {\bf 605} (2001) 395,
[hep-th/0006119].

\bibitem{medina}
J. Medina,
``Fuzzy Scalar Field Theories: Numerical and Analytical Investigations,''
Ph. D. thesis, Physics Department, Instituto Politecnico Nacional, Mexico City (2006),
[0801.1284 [hep-th]].

\bibitem{comr2}
J. Glimm, A.M. Jaffe, 
``Phase transition for $\phi^4_2$ quantum fields,''
Comm. Math. Phys. {\bf 45} (1975), 203-216. 

\bibitem{comr2num}
W.~Loinaz, R.S.~Willey,
``Monte Carlo Simulation Calculation of Critical Coupling Constant for Continuum $\phi^4_2$,''
\PRD {\bf 58} (1998) 076003
[hep-lat/9712008].

\bibitem{noncomphase2}
A.P. Balachandran, T.R. Govindarajan, B. Ydri,
``The Fermion Doubling Problem and Non-commutative Geometry,''
Mod. Phys. Lett. {\bf A15} (2000) 1279,
[hep-th/9911087].

\bibitem{noncomphase3}
G.H Chen, Y.S. Wu,
``Renormalization group equations and the Lifshitz point in non-commutative Landau-Ginsburg theory,''
Nucl. Phys. B {\bf 6022} (2002) 189,
[hep-th/0110134].

\bibitem{rsfnum}
J. Medina, W. Bietenholz, D. O'Connor,
``Probing the fuzzy sphere regularization in simulations of the $3d\, \lambda\phi^4$ model,''
\JHEP {\bf 04} (2008) 041
[0712.3366 [hep-th]].

\bibitem{num14panero2}
H.~Mej\'{i}a-D\'{i}az, W.~Bietenholz, M.~Panero,
``The Continuum Phase Diagram of the 2d Non-Commutative $\lambda \phi^4$ Model,''
\JHEP {\bf 1410} (2014) 56.
[1403.3318 [hep-lat]].

\bibitem{fuzzydiscnum}
F. Lizzi, B. Spisso,
``Noncommutative Field Theory: Numerical Analysis with the Fuzzy Disc,''
Int.J.Mod.Phys. A27 (2012) 1250137,
[1207.4998 [hep-th]].

\bibitem{nummartin}
X. Martin,
``A matrix phase for the $\phi^4$ scalar ﬁeld on the fuzzy sphere,''
\JHEP {\bf 04} (2004) 77,
[hep-th/0402230].

\bibitem{paneronum}
M.~Panero, 
``Numerical simulations of a non-commutative theory: the scalar model on the fuzzy sphere,''
\JHEP {\bf 05}, 082 (2007)
[hep-th/0608202].

\bibitem{denjoenum1}
F.~Garc\'{i}a-Flores, X.~Martin, D.~O'Connor, 
``Simulation of a scalar field on a fuzzy sphere,''
Proceedings of Science {\bf LAT 2005} (2006) 262
[hep-lat/0601012].

\bibitem{denjoenum2}
F.~Garc\'{i}a-Flores, X.~Martin, D.~O'Connor, 
``Simulation of a scalar field on a fuzzy sphere,''
Internat. J. Modern Phys. A {\bf 24} (2009), 3917$-$3944
[0903.1986 [hep-lat]].

\bibitem{num14}
B.~Ydri, 
``New Algorithm and Phase Diagram of Noncommutative $\Phi^4$ on the Fuzzy Sphere,''
\JHEP {\bf 03}, 065 (2014)
[1401.1529 [hep-th]].

\bibitem{matrix1}
B. Eynard,
``An introduction to Random Matrices,''
Saclay Lecture Notes (2001).

\bibitem{matrix11}
B. Eynard, T. Kimura, S. Ribault
``Random matrices,''
[1510.04430 [math-ph]].

\bibitem{rndmath}
A. Guionnet,
``On Random Matrices,''
Lecture Notes in Mathematics (Springer, New York 2008).

\bibitem{matrixnum}
A. Edelman, N. Raj Rao,
``Random Matrix Theory,''
Acta Numerica (2005), 1.
  
\bibitem{brezin}
E.~Brezin, C.~Itzykson, G.~Parisi, J.B.~Zuber,
``Planar diagrams,''
Commun.Math.Phys. 59 (1978).

\bibitem{shimishimi}
Y. Shimamune,
``On the Phase Structure of Large N Matrix Models and Gauge Models,''
Phys.Lett. B 108 (1982) 407.

\bibitem{newcritical}
S.R. Das, A. Dhar, A.M. Sengupta, S.R. Wadia,
``New Critical Behavior in $d=0$ Large $N$ Matrix Models,''
Mod.Phys.Lett. A {\bf 5} (1990) 1041-1056.

\bibitem{itzub}
C. Itzykson, J.B. Zuber,
``The planar approximation. II,''
J. Math. Phys. {\bf 21} (1980) 411.

\bibitem{shishanin}
A.O. Shishanin,
``Phases of the Goldstone multitrace matrix model in the large-N limit,''
Theor.Math.Phys. {\bf 152} (2007) 1258-1265.

\bibitem{our2}
J.~Tekel,
``Random matrix approach to scalar field on fuzzy spaces,''
\PRD {\bf 87}, 085015 (2013)
[1301.2154 [hep-th]].

\bibitem{samann}
C.~S\"{a}mann,
``The Multitrace Matrix Model of Scalar Field Theory on Fuzzy $\CPn$,''
SIGMA {\bf6} (2010)
[1003.4683 [hep-th]].

\bibitem{samannnew}
C.~S\"{a}mann,
``Bootstrapping Fuzzy Scalar Field Theory,''
\JHEP {\bf 04} (2015) 044
[1412.6255 [hep-th]].

\bibitem{poly}
A.P.~Polychronakos,
``Effective action and phase transitions of scalar field on the fuzzy sphere,''
\PRD {\bf 88}, 065010 (2013)
[1306.6645 [hep-th]].

\bibitem{mine1}
J.~Tekel,
``Uniform order phase and phase diagram of scalar field theory on fuzzy $\mathbb C P^n$,''
\JHEP {\bf 1410} (2014) 144
[1407.4061 [hep-th]].

\bibitem{pade}
G.A.~Baker, P.~Graves-Morris,
``Pad\'{e} Approximants,''
Cambridge University Press,  1996.

\bibitem{mine2}
J.~Tekel,
``Matrix model approximations of fuzzy scalar field theories and their phase diagrams,''
submitted to JHEP, [1510.07496 [hep-th]].

\bibitem{ydrinew1}
B.~Ydri, 
``A Multitrace Approach to Noncommutative $\Phi^4_2$,''
[1410.4881 [hep-th]].

\bibitem{samanfuzzydisc}
S. Rea, Ch. S\"{a}mann,
``The Phase Diagram of Scalar Field Theory on the Fuzzy Disc,''
\JHEP {\bf 11} (2015) 115, [1507.05978 [hep-th]].

\bibitem{denjoeprivate}
D.~O'Connor, private communiation.

\bibitem{ydrisupernew}
B. Ydri, K. Ramda, A. Rouag,
``Phase diagrams of the multitrae quartic matrix models of noncommutative $\Phi^4$,''
[1509.03726 [hep-th]].

\bibitem{rsfper}
M. Ihl, Ch. Sachse, C.~S\"{a}mann, 
``Fuzzy Scalar Field Theory as Matrix Quantum Mechanics,''
\JHEP {\bf 03} (2011) 091
[1012.3568 [hep-th]].

\bibitem{multiband}
K. Demeterfi, N. Deo, S. Jain, Ch.I. Tan,
``Multiband structure and critical behavior of matrix models,''
\PRD {\bf 42} (1990) 4105.

\bibitem{nouvir2}
R. Gurau, J. Magnen, V. Rivasseau, A. Tanasa,
``A translation-invariant renormalizable non-commutative scalar model'',
Commun.Math.Phys. {\bf 287} (2009) 275
[0802.0791 [math-ph]].

\bibitem{novy1}
I. Andrić, L. Jonke, D. Jurman, H.B. Nielsen,
``Homolumo Gap from Dynamical Energy Levels'',
Phys. Rev. D {\bf 77} (2008) 127701,
[0909.2346 [hep-th]].

\bibitem{novy2}
N. Deo,
``Glassy random matrix models,''
Phys. Rev. E {\bf 65} (2002) 056115,
[cond-mat/0204072].

\bibitem{novy3}
G.S. Krishnaswami,
``Phase transition in matrix model with logarithmic action: toy-model for gluons in baryons,''
\JHEP {\bf 03} (2006) 067,
[hep-th/0601216].

\bibitem{glaser}
J.W. Barrett, L. Glaser,
``Monte Carlo simulations of random non-commutative geometries,''
[1510.01377 [gr-qc]].

%\bibitem{stenposi} !!!
%H. Steinacker,
%``Non-commutative geometry and matrix models,''
%Proceedings of Science (QGQGS 2011) 004
%[1109.5521 [hep-th]].

%\bibitem{sakita}
%A. Jevicki, B. Sakita,
%``The quantum collective field method and its application to the planar limit,''
%Nuclear Physics B {\bf 3} (1980), 511-527.

%\bibitem{fuzzy1}
%A.P. Balachandran and S. Kurkcuoglu,
%``Topology change for fuzzy physics: Fuzzy spaces as Hopf algebras,''
%\IJMP ~{\bf A19} (2004) 3395;

%\bibitem{fuzzy4}
%B. Ydri,
%``Fuzzy physics,''
%Ph. D. thesis, Physics Department, Syracuse University (2001)

%\bibitem{oneovern}
%R.C. Brower, N. Deo, S. Jain, Ch.I. Tan
%``Symmetry Breaking in the Double-Well Hermitian Matrix Models,''
%Nucl. Phys. B {\bf 405} (1993) 166.

%\bibitem{vanid}
%S. Roman,
%``The Umbral Calculus,''
%(Academic Press, New York, 1984)  p. 29.

%\bibitem{fuzzygauge2}
%H. Steinacker,
%``Gauge theory on fuzzy spaces,''
%J. Phys.: Conf. Ser. {\bf 53} (2006) 872.

%\bibitem{sphtor}
%B. P. Dolan, D. O'Connor,
%``A Fuzzy Three Sphere and Fuzzy Tori,''
%\JHEP {\bf 0305} (2003) 018

%\bibitem{hch}
%Harish-Chandra,
%``Spherical functions on a semisimple Lie group. I,''
%Amer. J. Math. 80 (1958) 241.




\end{thebibliography}
